\documentclass[11pt]{JHEP3}
\usepackage{amsmath,amssymb,amsfonts}
\usepackage[multiple,stable]{footmisc}
\usepackage[amsmath,amsthm,thmmarks]{ntheorem}
\allowdisplaybreaks[4]
\usepackage[noconfig]{refstyle}

\numberwithin{equation}{section}

\let\eqref=\relax
\newref{eq}{name={eq.~},Name={Eq.~},names={eqs.~},Names={Eqs.~},rngtxt={-},refcmd=(\ref{#1})}
\newref{th}{name={Theorem~},Name={Theorem~},names={Theorems~},Names={Theorems~}}
\newref{l}{name={Lemma~},Name={Lemma~},names={Lemmas~},Names={Lemmas~}}
\newref{cor}{name={Corollary~},Name={Corollary~},names={Corollaries~},Names={Corollaries~}}
\newref{def}{name={Definition~},Name={Definition~},names={Definitions~},Names={Definitions~}}
\newref{f}{name={footnote~},Name={Footnote~},names={footnotes~},Names={Footnotes~}}
\newref{app}{name={Appendix~},Name={Appendix~},names={Appendixes~},Names={Appendixes~}}
\newref{tab}{name={Table~},Name={Table~},names={Tables~},Names={Tables~}}

\newcommand{\be}{\begin{equation}}
\newcommand{\ee}{\end{equation}}
\newcommand{\bea}{\begin{equation}\begin{aligned}}	
\newcommand{\eea}{\end{aligned}\end{equation}}		
\newcommand{\UA}{{\underline{A}}}					
\newcommand{\UB}{{\underline{B}}}
\newcommand{\UZ}{{\underline{0}}}
\newcommand{\Uth}{{\underline{\theta}}}
\newcommand{\BUth}{{\bar{\underline{\theta}}}}
\newcommand{\cc}{\text{c.c.}}

\newcommand{\field}[1]{\mathbb{#1}}
\newcommand{\oneon}[1]{\frac{1}{#1}}
\newcommand{\manifold}{\mathcal{M}}

\newcommand{\curvtwofrm}{\mathcal{R}}
\newcommand{\susyno}{\mathcal{N}}
\newcommand{\sderiv}{\mathcal{D}}
\newcommand{\tr}{\mathrm{tr}}
\newcommand{\R}{\field{R}}

\newcommand{\al}{\alpha}
\DeclareMathOperator{\diag}{diag}

\DeclareMathOperator{\SO}{SO}

\providecommand{\id}{\leavevmode\hbox{\small$\mathrm{1}$\kern-3.8pt\normalsize$\mathrm{1}$}}

\title{M-theory on Calabi-Yau Five-Folds}
\author{Alexander~S.~Haupt$^{a,b,c}$, Andre~Lukas$^{d}$, K.~S.~Stelle$^{a,c,e}$\\
$^a$Theoretical Physics Group, Imperial College London,\\ Prince Consort Road, London SW7 2AZ, U.K.\\
$^b$Institute for Mathematical Sciences, Imperial College London,\\ 53 Prince's Gate, London SW7 2PG, U.K.\\
$^c$Max-Planck-Institut f\"ur Gravitationsphysik (Albert-Einstein-Institut),\\ Am M\"uhlenberg 1, D-14476 Potsdam, Germany\\
$^d$Rudolf Peierls Centre for Theoretical Physics, Oxford University,\\ 1 Keble Road, Oxford, OX1 3NP, U.K.\\
$^e$Theory Division, Physics Department, CERN, CH-1211 Geneva 23, Switzerland\\ E-mail:
\email{a.haupt@imperial.ac.uk}, \email{lukas@physics.ox.ac.uk}, \email{k.stelle@imperial.ac.uk}}

\abstract{We study the compactification of M-theory on Calabi-Yau five-folds and the resulting ${\cal N}=2$ super-mechanics theories. By explicit reduction from 11 dimensions, including both bosonic and fermionic terms, we calculate the one-dimensional effective action and show that it can be derived from an ${\cal N}=2$ super-space action. We find that the K\"ahler and complex structure moduli of the five-fold reside in $2a$ and $2b$ super-multiplets, respectively. Constrained $2a$ super-multiplets arise from zero-modes of the M-theory three-form and lead to cross-couplings between $2a$ and $2b$ multiplets. Fermionic zero modes which arise from the $(1,3)$ sector of the 11-dimensional gravitino do not have bosonic super-partners and have to be described by purely fermionic super-multiplets in one dimension. We also study the inclusion of flux and discuss the consistency of the scalar potential with one-dimensional ${\cal N}=2$ supersymmetry and how it can be described in terms of a superpotential. This superpotential can also be obtained from a Gukov-type formula which we present. Supersymmetric vacua, obtained by solving the F-term equations, always have vanishing vacuum energy due to the form of this scalar potential. We show that such supersymmetric solutions exist for particular examples. Two substantial appendices develop the topology and geometry of Calabi-Yau five-folds and the structure of one-dimensional ${\cal N}=2$ supersymmetry and supergravity to the level of generality required for our purposes.}

\preprint{Imperial/TP/08/KSS/02; CERN-PH-TH/2008-201; arXiv:0810.2685 [hep-th]}

\begin{document}

\section{Introduction}
The technique of compactification has connected string- and M-theory to a wealth of supergravity theories in diverse dimensions and has led to important insights into both theoretical and phenomenological aspects of the theory. Ever since the seminal work~\cite{Candelas:1985en}, compactifications on Calabi-Yau spaces and related constructions have played a central r\^{o}le in this context. While most of this work has concentrated on Calabi-Yau three-folds, primarily in order to connect string theory to four-dimensional physics, Calabi-Yau four-folds have been used, for example in F-theory compactifications~\cite{Brunner:1996pk}, and compactification on K3 has played an important r\^{o}le in uncovering elementary duality relations~\cite{Hull:1994ys,Witten:1995ex}. Calabi-Yau four-folds have also appeared in string-/M-theory compactifications to two and three dimensions~\cite{Becker:1996gj,Haack:2002tu}. To the best of our knowledge, the first time Calabi-Yau five-folds have appeared in the physics literature was in Ref.~\cite{Kumar:1996zx} where subclasses of those manifolds feature in the discussion of certain vacuum constructions of F-theory and thirteen dimensional S-theory leading to supersymmetric two dimensional $\susyno=(1,1)$ and three dimensional $\susyno=2$ theories, respectively, and then again more detailed later in Ref.~\cite{Curio:1998bv} in a similar but more general context.

The main purpose of the present paper is to close an apparent gap in the scheme of M-theory compactifications by considering 11-dimensional supergravity on Calabi-Yau five-folds. Eleven-dimensional supergravity is the only one of the six ``known'' limits of M-theory with a sufficient number of physical spatial dimensions to allow for such compactifications (although, Calabi-Yau five-folds can, of course, be used for F-theory compactifications to two dimensions). M-theory backgrounds based on Calabi-Yau five-folds and their corrections induced by higher-order curvature terms have been considered in Ref.~\cite{Lu:2004ng}. Here, we will be concerned with the actual compactifications on such backgrounds and the resulting one-dimensional (super-)mechanics theories. Calabi-Yau five-folds reduce supersymmetry by a factor of $1/16$ and, given the eleven-dimensional theory has $32$ real supercharges, one expects one-dimensional theories with ${\cal N}=2$ supersymmetry from such reductions. 

Specifically, we will derive the general form of this one-dimensional ${\cal N}=2$ super-mechanics theory and analyse its relation to the underlying topology and moduli-space geometry of the five-folds. The necessary mathematical details regarding the topology and geometry of five-folds are, to a large extend, analogous to the the well-established three-fold case, and will be systematically developed as a preparation for our reduction. Another vital ingredient in our discussion is the structure of one-dimensional ${\cal N}=2$ supersymmetric theories~\cite{Coles:1990hr}. Although gravity is non-dynamical in one dimension, the component fields of the one-dimensional gravity supermultiplet (the lapse function and the gravitino) generate constraint equations which cannot be ignored. Therefore, we have to consider local one-dimensional ${\cal N}=2$ supersymmetry. Moreover, it turns out that the structure of the one-dimensional theories obtained from M-theory reduction is more general than the super-mechanics theories usually considered in the literature. In the present paper, we, therefore, invest considerable work in order to develop one-dimensional ${\cal N}=2$ supergravity to a sufficiently general level. 

Our work is motivated by a number of general considerations.
Reductions of M-theory to one dimension have played some r\^{o}le in the attempts to understand quantum M-theory~\cite{Julia:1980gr,Damour:2002cu} and we hope the results of the present paper may prove useful in this context. Arguments from topological string theory suggest a mini-superspace description of quantum string cosmology~\cite{Ooguri:2005vr} along the lines of ``traditional'' quantum cosmology~\cite{Hartle:1983ai}. Mini-superspace quantisation may be applied to the one-dimensional effective theories derived in this paper, hoping that this will describe some aspects of quantum M-theory on Calabi-Yau moduli spaces. In the present paper, we will not pursue this explicitly but possible applications in this direction are currently under investigation. A further motivation is related to the general problem of string vacuum selection and its possible interplay with cosmology. One aspect of the string vacuum degeneracy, which is often overlooked, is the ambiguous split of space-time into a number of internal, usually compact dimensions and four external dimensions. One might speculate that a more plausible geometry for an ``initial'' state in the early universe is one where all spatial dimension are treated on an equal footing. In the context of M-theory, such ``democratic'' backgrounds are given by 10-dimensional compact Ricci-flat spaces (neglecting flux for the time being) and, hence, Calabi-Yau five-folds provide a natural arena for this discussion. Assuming sufficiently slow, adiabatic time evolution, the problem of how three large spatial dimensions emerge from such a background can then be addressed by studying dynamics on the five-fold moduli space. This dynamics is, of course, described by the one-dimensional effective actions we will be deriving in the present paper.

As a low-energy effective description of M-theory, 11-dimensional supergravity is corrected by an infinite series of higher-order terms which are organised by their associated power of $\beta\sim\kappa_{11}^{4/3}$, where $\kappa_{11}$ is the 11-dimensional Newton constant. Let us first consider the situation at zeroth order in $\beta$, that is for 11-dimensional supergravity in its standard form. A background with vanishing flux, that is with zero anti-symmetric four-form tensor field $G=dA$, and an 11-dimensional metric which consists of a direct product of a Ricci-flat Calabi-Yau metric and time, clearly solves the 11-dimensional equations of motion at this lowest order. However, at linear order in $\beta$ the anomaly cancellation term $-\beta\int A\wedge X_8$, where $X_8$ is the well-known quartic in the curvature two-form, has to be added to the action. It has been observed in Ref.~\cite{Lu:2004ng} that $X_8$ can be non-zero when evaluated on Calabi-Yau five-folds backgrounds. In fact, here we will show that it is proportional to $c_4(X)$, the fourth Chern class of the five-fold $X$. At order $\beta$, the equation of motion for $G$ is accordingly corrected by a term $\beta X_8$ and is, hence, no longer necessarily satisfied for $G=0$. A further contribution to the $A$ equation of motion can arise from membranes wrapping a holomorphic curve $C$ with cohomology class $W=[C]$ in the Calabi-Yau five-fold. Taking into account these contributions, we show the three-form equation of motion leads to a topological consistency condition, required for a solution at order $\beta$ to exist. It states (modulo factors) that the cohomology class $[G\wedge G]$ plus the membrane class $W$ must be proportional to the fourth Chern class, $c_4(X)$. Here, we will consider several ways of solving this consistency equation. First, for vanishing flux, $G=0$, and no membranes, the five-folds $X$ needs to have vanishing fourth Chern class $c_4(X)$ and we will show that such five-folds indeed exist. Alternatively, for five-folds with $c_4(X)\neq 0$ a compensating non-zero flux and/or membrane is required. By means of a number of simple examples we will demonstrate that this can indeed frequently be achieved. In particular, we show that the consistency condition can be satisfied for the Calabi-Yau five-fold defined by the zero locus of a septic polynomial in $\mathbb{P}^6$. The ``septic'' is arguably the simplest  five-fold and the analogue of the quintic three-fold in $\mathbb{P}^4$.

The one-dimensional effective action will be calculated as an expansion in powers of $\beta$. As a first step we consider the situation at zeroth order in $\beta$. Effects from flux or membranes only come in at order $\beta$ and are, therefore, not relevant at this stage. In particular, we clarify the relation between Calabi-Yau topology/geometry and the structure of the one-dimensional supermechanics induced by M-theory at this lowest order in $\beta$. Many aspects of this relation are analogous to what happens for compactifications on lower-dimensional Calabi-Yau manifolds, others, as we will see, are perhaps less expected.  The topology of a Calabi-Yau five-fold $X$ is characterised by six a priori independent Hodge numbers, namely $h^{1,1}(X)$, $h^{1,2}(X)$, $h^{1,3}(X)$, $h^{2,2}(X)$, $h^{1,4}(X)$ and $h^{2,3}(X)$. In analogy with the four-fold case~\cite{Sethi:1996es}, an index theorem calculation together with the Calabi-Yau condition $c_1(X)=0$, leads to one relation between those six numbers. The moduli space of a Calabi-Yau manifold consists (locally) of a direct product of a K\"ahler and a complex structure moduli space~\cite{Candelas:1990pi}. For Calabi-Yau five-folds, these two parts of the moduli space are associated with the $(1,1)$ and the $(1,4)$ sectors, respectively. As we will see, the associated K\"ahler and complex structure moduli are part of $2a$ and $2b$ multiplets~\cite{Gibbons:1997iy} of one-dimensional ${\cal N}=2$ supersymmetry. A further set of bosonic zero modes originates from the M-theory three form $A$ in the $(2,1)$ sector. We will show that these modes become part of constrained $2a$ multiplets. This exhausts the list of bosonic zero modes. Expanding the 11-dimensional gravitino leads to fermionic zero modes in the sectors $(1,q)$ where $q=1,2,3,4$. For $q=1,2,4$ these pair up into super-multiplets with the aforementioned bosons but the $(1,3)$ fermions have no bosonic zero mode partners. We will show that this apparent contradiction can be resolved by the introduction of fermionic $2b$ multiplets, that is $2b$ multiplets with a fermion as their lowest component. With this assignment of zero modes to super-multiplets, the one-dimensional effective theory is an ${\cal N}=2$ sigma model which we present both in its component and superspace form. Some of its features are worth mentioning. For example, the sigma model metric for the $2a$ multiplets in the $(1,1)$ sector is not the standard Calabi-Yau K\"ahler moduli space metric~\cite{Candelas:1990pi}, as is usually the case for three-fold compactifications. However, the physical sigma model metric and the standard Calabi-Yau metric are related in a simple way. Also, it turns out that the sigma model metrics in the $(2,1)$ and $(1,3)$ sector depend inter alia on the K\"ahler moduli, so that we require a coupling of $2a$ and $2b$ multiplets. As far as we know such a coupling between $2a$ and $2b$ multiplets has not been studied in the context of one-dimensional ${\cal N}=2$ supersymmetry before. 

Then, we proceed to include the order $\beta$ effects from flux and membranes. We calculate the scalar potential, including four-form flux, membrane effects and effects from the non-zero Calabi-Yau curvature tensor. The latter requires evaluating the non-topological $R^4$ terms of M-theory on a five-fold background and we show that these terms can be expressed in terms of the fourth Chern class, $c_4(X)$. Our results indicate that the part of the scalar potential induced by the $(1,3)$-component of the four-form flux breaks one-dimensional ${\cal N}=2$ supersymmetry. Setting the $(1,3)$-part of the four-form flux to zero to maintain full supersymmetry induces an implicit potential for the complex structure moduli. It is not known whether this potential can be calculated explicitly and we have thus restricted our attention to Calabi-Yau five-folds for which this potential vanishes. This is equivalent to demanding that all $(2,2)$-forms can be obtained from the product of two $(1,1)$-forms. All the explicit examples of Calabi-Yau five-folds presented in this paper are of this type. The $(2,2)$-part of the scalar potential is compatible with one-dimensional supersymmetry and can be written in terms of a superpotential ${\cal W}$. As we will show, this superpotential can be obtained from the Gukov-type formula ${\cal W}\sim \int_XG_{\rm flux}\wedge J^3$, where $J$ is the K\"ahler form of the Calabi-Yau five-fold. We also present the explicit superpotential and scalar potential for a number of particular examples, including the septic in $\mathbb{P}^6$, and discuss implications for moduli stabilisation and dynamics.

The plan of the paper is as follows. In Section~\ref{sec:M} we review some basic facts about 11-dimensional supergravity. Some general results on the topology and moduli space geometry of Calabi-Yau five-folds are collected in Section~\ref{sec:5-folds}. In this section, we also present several explicit examples of five-fold backgrounds which solve the M-theory consistency condition. More details on this and derivations of some of the results are given in \appref{cy5}. In Section~\ref{sec:reduction}, we perform the reduction of M-theory on such backgrounds at zeroth order in $\beta$, starting with the bosonic action and then including terms bilinear in fermions. Section~\ref{sec:N=2} shows that the one-dimensional effective action obtained in this way has indeed two local supersymmetries and can be written in superspace form. Many of the necessary details and technical results on one-dimensional ${\cal N}=2$ supersymmetry and supergravity are collected in \appref{superspace}. In Section~\ref{sec:flux}, we derive the order $\beta$ corrections to the effective action and calculate the scalar potential and superpotential. We conclude in Section~\ref{sec:concl}. Conventions and notation used throughout this paper are summarised in \appref{conv}.

\section{The M-theory low energy effective action}\label{sec:M}

In this section, we review a number of results on 11-dimensional supergravity and its higher-derivative corrections, focusing on the aspects that will be important for the reduction on Calabi-Yau five-folds. More detailed reviews on the subject can, for example, be found in Refs.~\cite{Bilal:2003es,Miemiec:2005ry}.

The field content of 11-dimensional supergravity consists of the 11-dimensional space-time metric $g_{MN}$, the anti-symmetric  three form tensor field $A_{MNP}$ with field strength $G=dA$ and the gravitino $\Psi_M$, an 11-dimensional Majorana spinor. Here, we denote 11-dimensional curved indices by $M,N,\ldots = 0,1,\ldots ,10$ and their flat, tangent-space counterparts by $\underline{M},\underline{N},\ldots$. Where possible, we will use differential forms to keep our notation concise. Our conventions largely follow Ref.~\cite{Bilal:2003es} and are summarised in Appendix~\ref{app:conv}. 

We split the 11-dimensional action into four parts as
\be\eqlabel{action11}
	S_{11} = S_{11,\rm B} + S_{11,\rm F} + S_{11,\rm GS} + S_{11,R^4}+\ldots .
\ee
Here, the first and second terms are the bosonic and fermionic parts of 11-dimensional supergravity~\cite{Cremmer:1978km}, respectively, $S_{\rm GS}$ is the Green-Schwarz term related to the cancellation of the M5-brane world-volume anomaly~\cite{Duff:1995wd}, $S_{R^4}$ are the non-topological $R^4$ terms~\cite{Green:1997di,Green:1997as,Russo:1997mk} and the dots indicate additional higher order contributions, which we will not need for our purposes.

The bosonic part of the action reads~\cite{Cremmer:1978km}
\be\eqlabel{sugra11}
	S_{11,\rm B} = \oneon{2\kappa^2_{11}} 
		\int_\manifold \left\{ R \ast 1 - \oneon{2} G\wedge\ast G - \oneon{6} G\wedge G \wedge A \right\} ,
\ee
where $\kappa_{11}$ is the 11-dimensional gravitational constant, $R$ is the Ricci scalar of the 11-dimensional metric $g$ and ${\cal M}$ is the space-time manifold. The equations of motion from this bosonic action are given by
\begin{align}\eqlabel{geom}
	 R_{MN} &= \oneon{12} G_{M M_2\ldots M_4}{G_N}^{M_2\ldots M_4} - \oneon{144} g_{MN} G_{M_1\ldots M_4}G^{M_1\ldots M_4} , \\
	d \ast G  &= - \oneon{2} G\wedge G \;  .\eqlabel{Geom_class}
\end{align}
The gravitino dynamics is encoded in the fermionic action
\begin{multline}\eqlabel{sugra11_fermions}
	S_{11,\rm F} = - \oneon{2\kappa^2_{11}} \int_\manifold d^{11}x \sqrt{-g} \Big\{
		\bar{\Psi}_M \Gamma^{MNP} D_N(\omega) \Psi_P \\ \qquad\qquad +
		\oneon{96} \left( \bar{\Psi}_M \Gamma^{MNPQRS} \Psi_S + 12 \bar{\Psi}^N \Gamma^{PQ} \Psi^R \right) G_{NPQR}
		+ (\text{fermi})^4
	\Big\} ,
\end{multline}
where $\bar{\Psi}_M = i \Psi_M^\dagger \Gamma^\UZ$. Here and in much of what follows, we omit four-fermi terms. The covariant derivative $D_M$ is defined by
\begin{equation}
D_N (\omega) \Psi_P = (\partial_N + \oneon{4} {\omega_N}^{\underline{Q}\underline{R}} \Gamma_{\underline{Q}\underline{R}}) \Psi_P\; , \eqlabel{covder}
\end{equation}
with the spin connection $ {\omega_N}^{\underline{Q}\underline{R}}$. The corresponding equation of motion for $\Psi_M$ reads
\be\eqlabel{sugra11_gravitino_eom}
	\Gamma^{MNP} D_N(\omega) \Psi_P + \oneon{96} \left( \Gamma^{MNPQRS} \Psi_S 
			+ 12 g^{MN} \Gamma^{PQ} \Psi^R \right) G_{NPQR} + (\text{fermi})^3 = 0 \; .
\ee
The action $S_{11,{\rm B}}+S_{11,{\rm F}}$ for 11-dimensional supergravity is invariant under the supersymmetry transformations
\bea\eqlabel{sugra11_susy_transf}
	\delta_\epsilon g_{MN}  &= 2 \bar{\epsilon} \Gamma_{\left(M\right.} \Psi_{\left.N\right)} , \\
	\delta_\epsilon A_{MNP} &= -3 \bar{\epsilon} \Gamma_{\left[MN\right.} \Psi_{\left.P\right]} , \\
	\delta_\epsilon \Psi_M  &= 
			2D_M (\omega) \epsilon + \oneon{144} (\Gamma_M {}^{NPQR} - 8 \delta_M^N \Gamma^{PQR}) \epsilon G_{NPQR} + (\text{fermi})^3 \; ,
\eea
which are parameterised by an 11-dimensional Majorana spinor $\epsilon$. 

In its r\^{o}le as the low-energy effective theory of M-theory the action $S_{11,\rm B}+S_{11,\rm F}$ receives an infinite series of higher-order derivative corrections which are organised by integer powers of the quantity
\begin{equation}
 \beta =\frac{1}{(2\pi )^2}\left(\frac{\kappa_{11}^2}{2\pi^2}\right)^{2/3}\; .
\end{equation} 
One such correction which appears at order $\beta$ is the Green-Schwarz term~\footnote{Care has to be taken in order to obtain the correct sign of the Green-Schwarz term relative to the $GGA$ Chern-Simons term in the action~\eqref*{sugra11} and different versions exist in the literature~\cite{Bilal:2003es,deAlwis:1996ez,deAlwis:1996hr}. In general, the sign of the Chern-Simons term is fixed by supersymmetry and the relative sign is fixed by the anomaly cancellation condition on the five-brane world volume~\cite{Duff:1995wd}. In Ref.~\cite{Bilal:2003es}, several different arguments are presented for why the relative sign must be positive (in our conventions) and we adopt this result in the present paper.}
\be\eqlabel{sugra11_gs}
	S_{11,\rm GS} = - (2\pi )^4\frac{\beta}{2\kappa_{11}^2} \int_\manifold A \wedge X_8 ,
\ee
where $X_8$ is a quartic polynomial in the curvature two-form $\curvtwofrm$. It can be conveniently expressed in terms of the first and second Pontrjagin classes $p_1(T\manifold )$ and $p_2(T\manifold )$ of the tangent bundle $T\manifold$ of $\manifold$ as
\bea\eqlabel{def_x8}
	X_8 &= \oneon{48} \left(\left(\frac{p_1}{2}\right)^2 - p_2\right), \\
	p_1(T\manifold)  &= - \oneon{2} \left(\oneon{2\pi}\right)^2 \tr \, \curvtwofrm^2, \\
	p_2(T\manifold)  &=   \oneon{8} \left(\oneon{2\pi}\right)^4 \left((\tr\,\curvtwofrm^2)^2 - 2\,\tr\,\curvtwofrm^4\right) \, .
\eea
This Green-Schwarz term leads to a correction to the equation of motion~\eqref*{Geom_class} for $A$, which now reads
\be\eqlabel{sugra11_3form_eom}
	d \ast G =- \oneon{2} G\wedge G - (2\pi)^4 \beta X_8\; .
\ee
We note that the exactness of $d\ast G$ implies that the eight-form $\frac{1}{2}G\wedge G+(2\pi )^4\beta X_8$ must be cohomologically trivial on $\manifold$. This integrability condition will play an important r\^{o}le for compactifications on Calabi-Yau five-folds, as we will see. There is also a non-topological $R^4$ term at order $\beta$ which is related to the Green-Schwarz term~\eqref*{sugra11_gs} by supersymmetry. This term which we will need for our discussion of flux and scalar potentials in the one-dimensional effective theory is given by~\cite{Green:1997di,Green:1997as,Russo:1997mk}
\be
 S_{11,R^4}=\frac{\beta}{2\kappa_{11}^2}\frac{1}{9\cdot 2^{11}}\int_{\cal M} d^{11} x \sqrt{-g}\, t_8^{M_1\dots M_8}t_8^{N_1\dots N_8}R_{M_1M_2N_1N_2}\dots R_{M_7M_8N_7N_8}\; , \eqlabel{R4}
\ee
with the famous rank eight tensor $t_8$ which has been defined in Ref.~\cite{Schwarz:1982jn}.

\vskip 0.4cm

Equations of motion for anti-symmetric tensor fields can receive contributions from electrically charged objects and, for the case at hand, an additional term has to be added to  \eqref{sugra11_3form_eom} in the presence of M-theory membranes. Clearly, such a term can affect the integrability of \eqref{sugra11_3form_eom} and should be taken into account. 

We start with the bosonic part of the membrane action
\be 
 S_3=-T_3\int_{{\cal M}_3}\left\{d^3\sigma\,\sqrt{-\hat{g}}+\hat{A}\right\}\; , \eqlabel{membraneaction}
\ee
where $\hat{g}$ and $\hat{A}$ are the pullbacks of the $11$-dimensional space-time metric $g$ and three-form $A$ under the embedding $X^M=X^M(\underline{\sigma} )$ of the membrane world-volume ${\cal M}_3$ into space-time ${\cal M}$. Here, $\underline{\sigma}=(\sigma^0,\sigma^1,\sigma^2)$ are coordinates on the membrane world volume. The membrane tension $T_3$ is given by
\be 
 T_3=\frac{1}{2\pi\sqrt{\beta}}\; . \eqlabel{T3def}
\ee
Adding this action to the bosonic one for $11$-dimensional supergravity, \eqref{sugra11}, and re-computing the equation of motion for $A$ leads to
\begin{equation}
  d \ast G =- \oneon{2} G\wedge G - (2\pi)^4 \beta X_8  -2\kappa_{11}^2 T_3\,\delta({\cal M}_3)\; . \eqlabel{Geom_corr}
\end{equation}
Here, $\delta ({\cal M}_3)$ is an eight-form current associated with the membrane world-volume. It is characterised by the property
\be
 \int_{{\cal M}_3}w=\int_{\cal M}w\wedge\delta ({\cal M}_3)
\ee
for any three-form $w$.

\section{Calabi-Yau five-folds}\label{sec:5-folds}

Our M-theory reduction depends on a range of results on Calabi-Yau five-folds, including results about their topology, their differential geometry and moduli spaces. Perhaps most importantly, to be sure we are not dealing with an empty set, we require some explicit examples of Calabi-Yau five-folds on which consistent M-theory reductions can be carried out. In this chapter, we provide a non-technical summary of the main facts and results for the reader's convenience. For the details we refer to Appendix~\ref{app:cy5}.

We begin by defining what we mean by a Calabi-Yau five-fold $X$. As usual, we require that $X$ be a compact, complex K\"ahler manifold with vanishing first Chern class, $c_1(X)=0$. In addition, $X$ should break supersymmetry by a factor of $1/16$. This means that the holonomy group ${\rm Hol}(X)\subset{\rm SU}(5)$ is a sufficiently large subgroup of ${\rm SU}(5)$ such that in the decomposition
\begin{equation}
 {\bf 16}_{{\rm Spin}(10)}\rightarrow\left[{\bf 10}+\bar{\bf 5}+{\bf 1}\right]_{{\rm SU}(5)}
\end{equation}
of (chiral) spinors  on $X$ under ${\rm SU}(5)$ only the ${\rm SU}(5)$ singlet is invariant under the holonomy group. An immediate consequence is that the Hodge numbers $h^{p,0}(X)=h^{0,p}(X)$ for $p=1,2,3,4$ vanish and that $h^{0,0}(X)=h^{5,0}(X)=h^{0,5}(X)=h^{5,5}(X)=1$. The reason for this additional condition on supersymmetry breaking is to avoid ``non-generic'' cases which lead to a larger number of preserved supersymmetries and additional zero modes (due to $h^{p,0}(X)\neq 0$ for some $p\in\{1,2,3,4\}$), such as $10$-tori, products of lower-dimensional Calabi-Yau manifolds (for example, a product of a three-fold with K3) or products of lower-dimensional Calabi-Yau manifolds with tori (for example, a four-fold times a two-torus). 

Given the restrictions on Hodge numbers discussed above, we remain with six, a priori independent Hodge numbers, namely $h^{1,1}(X)$, $h^{1,2}(X)$, $h^{1,3}(X)$, $h^{2,2}(X)$, $h^{1,4}(X)$ and $h^{2,3}(X)$. For Calabi-Yau four-folds it is known~\cite{Sethi:1996es} that one additional relation between the Hodge numbers can be derived using the index theorem together with the Calabi-Yau condition $c_1(X)=0$. In Appendix~\ref{app:cy5}, we show that the same is true for Calabi-Yau five-folds and we derive the relation
\begin{equation}
 11\, h^{1,1}(X)-10\,h^{1,2}(X)-h^{2,2}(X)+h^{2,3}(X)+10\,h^{1,3}(X)-11\,h^{1,4}(X)=0\; . \eqlabel{Hodgecons0}
\end{equation}
Hence, we are left with five apparently independent Hodge numbers. The precise r\^{o}le of the cohomology groups in the reduction of M-theory will be explained in the following section. Here, we summarise the relation between cohomology groups, M-theory zero modes and flux (see \tabref{tab:modes}).
\TABLE{
\begin{tabular}{|c|c|c|c|}
\hline
\emph{cohomology}&\emph{bosonic zero modes}&\emph{fermionic zero modes}&\emph{flux}\\\hline\hline
$H^{1,1}(X)$&$h^{1,1}(X)$ real, K\"ahler moduli&$\begin{array}{l}\text{$h^{1,1}(X)$ complex,}\\\text{from gravitino}\end{array}$&$-$\\\hline
$H^{1,2}(X)$&$h^{1,2}(X)$ complex, from three-form&$\begin{array}{l}\text{$h^{1,2}(X)$ complex,}\\\text{from gravitino}\end{array}$&$-$\\\hline
$H^{1,3}(X)$&$-$&$\begin{array}{l}\text{$h^{1,3}(X)$ complex,}\\\text{from gravitino}\end{array}$&$G$-flux\\\hline
$H^{2,2}(X)$&$-$&$-$&$G$-flux\\\hline
$H^{1,4}(X)$&$h^{1,4}(X)$ complex structure moduli&$\begin{array}{l}\text{$h^{1,4}(X)$ complex,}\\\text{from gravitino}\end{array}$&$-$\\\hline
$H^{2,3}(X)$&$-$&$-$&$-$\\\hline
\end{tabular}
\caption{Cohomology groups of a Calabi-Yau five-fold $X$ and their relation to zero modes and flux in an M-theory reduction.}
\tablabel{tab:modes}}
As usual, the moduli space of Ricci-flat metrics consists of K\"ahler and complex structure deformations. For Calabi-Yau five-folds they are associated with harmonic $(1,1)$ and $(1,4)$ forms, respectively. Another set of bosonic zero modes arises from the M-theory three-form $A$ and is related to the cohomology $H^{1,2}(X)$. As \tabref{tab:modes} shows, for all these bosonic modes, we have fermionic zero modes counted by the same Hodge number. This suggests an obvious way of arranging modes into one-dimensional super-multiplets. However, the $(1,3)$ sector is somewhat puzzling in that it gives rise to a set of fermionic but not bosonic zero modes. We will come back to this later and show how this apparent mismatch of bosonic and fermionic degrees of freedom can be reconciled with supersymmetry.

\vskip 0.4cm

As discussed before, the equation of motion for the M-theory three-form $A$ leads to an important integrability condition which amounts to the right-hand side of \eqref{Geom_corr} being cohomologically trivial. Let us now consider this condition for the case of an $11$-dimensional space-time of the form ${\cal M}=\mathbb{R}\times X$, with a Calabi-Yau five-fold $X$. The total Pontrjagin class of such a space-time is $p({\cal M})=p(X)$. In general, for a complex manifold $Z$, the Pontrjagin and Chern classes are related by $p_1(Z)=c_1(Z)^2-c_2(Z)$ and $p_2(Z)=c_2(Z)^2-2c_1(Z)c_3(Z)+2c_4(Z)$. Given that $c_1(X)=0$ for a Calabi-Yau five-fold we have $p_1(X)=-2c_2(X)$ and $p_2(X)=c_2(X)^2-2c_4(X)$. Inserting this into the definition~\eqref*{def_x8} of $X_8$, we find
\begin{equation}
 X_8=-\frac{1}{24}c_4(X)\; . \eqlabel{X8res}
\end{equation} 
In general, we also allow four-form flux $G_{\rm flux}$ on $X$ and it is convenient to introduce the re-scaled version
\begin{equation}
 g=\left[\frac{T_3}{2\pi}G_{\rm flux}\right]\; , \eqlabel{gdef}
\end{equation}
where we recall that $T_3$ is the membrane tension defined in \eqref{T3def} and the square brackets indicate the cohomology class. As is clear from the Wess-Zumino term in the membrane action~\eqref*{membraneaction} this re-scaled flux is quantised in integer units, that is, it should be an element of the fourth integer cohomology of $X$. More accurately, taking into account the subtlety explained in Ref.~\cite{Witten:1996md}, the quantisation law reads
\begin{equation}
 g+\frac{1}{2}c_2(X)\in H^4(X,\mathbb{Z})\; . \eqlabel{gquant}
\end{equation} 
Finally, we should allow for membranes which wrap a holomorphic cycle $C\subset X$ of the five-fold, that is membranes with world volume ${\cal M}_3=\mathbb{R}\times C$. The membrane current $\delta ({\cal M}_3)$ then takes the form $\delta ({\cal M}_3)=\delta (C)$. Inserting this current, together with eqs.~\eqref*{X8res} and \eqref*{gdef} into the right-hand side of the $G$ equation of motion~\eqref*{Geom_corr} and taking the cohomology class of the resulting expression, one finds
\begin{equation}
 c_4(X)-12\, g\wedge g = 24\, W\; . \eqlabel{intcond}
\end{equation}
Here, $W=[C]\in H_2(X,\mathbb{Z})$ is the second homology class of the curve $C$, wrapped by the membranes. Eq.~\eqref*{intcond} is a crucial condition which is clearly necessary for consistent M-theory backgrounds based on Calabi-Yau five-folds. When solving this condition, it must be kept in mind that the homology class $W$, having a holomorphic curve representative $C$, must be an effective class in $H_2(X,\mathbb{Z})$, that is, it must be an element in the Mori cone of $X$. 

\vskip 0.4cm

Our task is now to establish the existence of Calabi-Yau five-fold backgrounds which satisfy the above consistency condition. Formally, this amounts to finding Calabi-Yau five-folds $X$, an element $g$ in the fourth cohomology of $X$ and an effective class $W\in H_2(X,\mathbb{Z})$ such that eqs.~\eqref*{gquant} and \eqref*{intcond} are satisfied. In Appendix~\ref{app:cyexamples} we analyse this problem in detail for a number of explicit examples. In particular, we consider torus quotients and complete intersection Calabi-Yau five-folds (CICY five-folds)~\cite{Candelas:1987kf}. 

Let us briefly review some basic properties of CICY five-folds. CICY five-folds are embedded in an ambient space ${\cal A}=\bigotimes_{r=1}^m\mathbb{P}^{n_r}$, given by a product of $m$ projective spaces with dimensions $n_r$. They are defined by the common zero locus of $K=\sum_{r=1}^mn_r-5$ homogeneous polynomials $p_\alpha$ in ${\cal A}$. The polynomials $p_\alpha$ are characterised by their degrees $q^r_\alpha$ in the coordinates of the $r^{\rm th}$ projective factor of the ambient space. A short-hand notation for CICY manifolds is provided by the {\em configuration matrix}
\begin{equation}
 [{\bf n}|{\bf q}]=\left[\begin{array}{l}n_1\\\vdots\\n_m\end{array}\right|\left.\begin{array}{lll}q^1_1&\dots&q^1_K\\
 \vdots&&\vdots\\
 q^m_1&\hdots&q^m_K\\\end{array}\right] \eqlabel{conf1}
 \end{equation} 
which encodes the dimensions of the ambient projective spaces and the (multi)-degrees of the defining polynomials. Such configuration matrices are constrained by the Calabi-Yau condition, $c_1(X)=0$, which for CICY manifolds reads
\begin{equation}
\sum_{\alpha=1}^Kq^r_\alpha=n_r+1
\end{equation}
for all $r$. The simplest CICY five-fold is given by the zero locus of a septic polynomial in $\mathbb{P}^6$ and is represented by the  configuration matrix $[6|7]$. The septic in $\mathbb{P}^6$ is the direct analogue of the best-known example of a Calabi-Yau three-fold, the quintic hypersurface in $\mathbb{P}^4$. In total, there are $11$ CICY five-folds which can be defined in a single projective space and these manifolds are listed in \tabref{tab:cicy1}.

The main results of Appendix~\ref{app:cyexamples} can be summarised as follows. The simplest way of satisfying the integrability condition~\eqref*{intcond} is to turn off flux, $g=0$, and have no membranes so that $W=0$. In this case, a Calabi-Yau five-fold $X$ with vanishing fourth Chern class, $c_4(X)=0$, is required. It can be shown in general that CICY configurations with all $q^r_a\geq 2$ (which includes  configurations with $m=1$ or $K=1$) always have $c_4(X)\neq 0$. In addition, we have verified that no configuration matrix with $m\leq 4$ and $K\leq 4$ leads to $c_4(X)=0$.  For larger configurations with $m>4$ or $K>4$ and at least one $q^r_a<2$ cases with $c_4(X)=0$ might still exist although we have been unable to find an explicit example. It is conceivable that $c_4(X)\neq 0$ for all CICY five-folds. Given the lack of a viable CICY example, we have turned to torus quotients of the form $X=T^{10}/\mathbb{Z}_2^4$. We have shown that, for an appropriate choice of shifts in the $\mathbb{Z}_2$ symmetries, $\mathbb{Z}_2^4$ is freely acting and, hence, $X$ is a manifold. Each $\mathbb{Z}_2$ reduces supersymmetry by $1/2$, so in total it is reduced by a factor of $1/16$. This means that $X$, although its holonomy is merely $\mathbb{Z}_2^4$, is a Calabi-Yau manifold in the sense defined earlier. Clearly, as $X$ admits a flat metric, we have $c_4(X)=0$. It remains an open question whether a Calabi-Yau five-fold with full ${\rm SU}(5)$ holonomy and $c_4(X)=0$ exists. We are not aware of a general mathematical reason which excluded this and it would be interesting to search for such a manifold, for example among toric five-folds. In the present paper, we will not pursue this explicitly.

Next, we should consider the possibility of satisfying the integrability condition~\eqref*{intcond} in the presence of non-vanishing flux but without membranes. The CICY manifolds defined in a single projective space, given in \tabref{tab:cicy1}, all have $b^4(X)=1$ and, hence, there is only a single flux parameter. Eq.~\eqref*{intcond} then turns into a quadratic equation for this parameter. Unfortunately, there is no rational solution to this equation for any of the $11$ cases. This means that the quantisation condition~\eqref*{gquant} cannot be satisfied and, hence, that flux is not sufficient to obtain viable examples for CICYs in a single projective space. Essentially, the reason is that there is only one flux parameter available which is too restrictive. For simple CICYs defined in a product of two projective space, where $b^4(X)=2$ or $b^4(X)=3$ depending on the case, we run into a similar problem. The simplest viable example we have found involves the space
\begin{equation}
X\sim\left[\begin{array}{l}1\\2\\3\end{array}\right|\left.\begin{array}{l}2\\3\\4\end{array}\right]\; ,
\end{equation}
defined in the ambient space ${\cal A}=\mathbb{P}^1\times\mathbb{P}^2\times\mathbb{P}^3$. In this case, we have $b^4(X)=5$ and flux can be parameterized as $g=k_{1,2}J_1J_2+k_{1,3}J_1J_3+k_{2,2}J_2^2+k_{2,3}J_2J_3+k_{3,3}J_3^2$, where $J_r$ are the three K\"ahler forms of the ambient projective spaces, normalised as in \eqref{Pnorm }.  It turns out that both conditions~\eqref*{gquant} and~\eqref*{intcond} can be satisfied for the choice $(k_{1,2},k_{1,3},k_{2,2},k_{2,3},k_{3,3})=(1,3,7/2,0,6)$. 

What about the opposite case of including membranes but setting the flux to zero? A simple viable example is given by the CICY $[7|6\, 2]$ with a membrane wrapping a holomorphic curve with class $W=227\tilde{J}$, where $\tilde{J}$ is the eight-form dual to the ambient space K\"ahler form $J$.

Finally, by combining flux and membranes, the anomaly condition can frequently be satisfied. For example, for the septic, $[6|7]$, with flux $g=kJ^2$ we find the conditions~\eqref*{gquant} and~\eqref*{intcond} are solved for $k=15/2$ and a membrane class $W=6J^4$.

\vskip 0.4cm

In summary, we have demonstrated that the quantisation and integrability conditions~\eqref*{gquant} and~\eqref*{intcond}, necessary for a consistent compactification of M-theory on Calabi-Yau five-fold backgrounds, can be satisfied for a range of simple examples. Flux and membranes are usually necessary and even the septic in $\mathbb{P}^6$ leads to a viable background for appropriate non-zero choices of both flux and membranes. We have also given an example, based on a torus quotient, with $c_4(X)=0$ which is consistent without flux and membranes. We have not been able to find a Calabi-Yau manifold with $c_4(X)=0$ and full ${\rm SU}(5)$ holonomy and it might be interesting to search for such a case, for example among toric Calabi-Yau manifolds.


\section{Compactification on Calabi-Yau five-folds}\label{sec:reduction}

In this section, we consider the compactification of 11-dimensional supergravity on a space-time of the form ${\cal M}=\mathbb{R}\times X$, where $X$ is a Calabi-Yau five-fold. At zeroth order in $\beta$, we start with the background configuration
\be
 ds^2=-dt^2+g_{mn}dx^mdx^n\; ,\qquad G=0\; , \eqlabel{bg}
\ee
where $g_{mn}=g_{mn}(x^p)$ is the Ricci-flat metric on $X$ and $m,n\ldots = 1,\dots ,10$. Clearly, this background solves the leading order bosonic equations of motion~\eqref*{geom} and \eqref*{Geom_class}. At order $\beta$, additional higher-derivative terms appear in the 11-dimensional equations of motion and corrections of the same order will have to be added to the above background. It is not a priori clear that suitable corrections to the background exist in order for it to remain a solution at order $\beta$.
We have seen that the integrability condition~\eqref*{intcond} is a necessary condition for this to be the case. In the absence of flux and membranes, the integrability condition is solved by Calabi-Yau five-folds with $c_4(X)=0$ and, in the previous section, we have given an explicit example of such a five-fold. In Ref.~\cite{Lu:2004ng}, it has been shown that a full solution at order $\beta$ does indeed exist in this case. For five-folds with $c_4(X)\neq 0$ flux and/or membranes need to be included in order to satisfy the integrability condition and we have seen that this can be achieved for a number of simple examples.
In this case, the existence of a full 11-dimensional solution at order $\beta$ has not been analysed in detail. In the presence of flux, one expects a scalar potential in the effective one-dimensional theory. Flux potentials frequently lead to some runaway direction in moduli space and in such cases, one would not expect a {\em static} 11-dimensional solution to exist. The study of 11-dimensional solutions based on five-folds at order $\beta$, generalising the results of Ref.~\cite{Lu:2004ng}, is an interesting subject to which we intend to return in a future publication. In the present paper, we focus on deriving the one-dimensional effective theory for backgrounds where the integrability condition~\eqref*{intcond} is satisfied and under the assumption that a full order $\beta$ background can be found. For now we will focus on the effective action at zeroth order in $\beta$ for which the above simple background is sufficient. Higher order corrections to the effective action and, in particular, the scalar potential due to flux will be discussed later. 

We start with the reduction of the bosonic part of the action before we move on to the fermionic terms in the second part of this section.

\subsection{Performing the dimensional reduction: the bosonic part}

Our starting point is the bosonic part of 11-dimensional supergravity~\eqref*{sugra11} which corresponds to the leading, zeroth order terms in a $\beta$ expansion together with the background~\eqref*{bg}. The order $\beta$ Green-Schwarz term~\eqref*{sugra11_gs} will also play a certain r\^{o}le. We now need to identify the moduli of this background. As discussed in detail in \appref{modspace}, the formalism to deal with Calabi-Yau five-fold moduli spaces is largely similar to the one developed for Calabi-Yau three-folds~\cite{Candelas:1990pi}. Here, we only summarise the essential information needed for the dimensional reduction. As for Calabi-Yau three-folds, the moduli space of Ricci-flat metrics on Calabi-Yau five-folds is (locally) a product of a K\"ahler and a complex structure moduli space which are associated to $(1,1)$ and $(2,0)$ deformations of the metric. They can be described in terms of harmonic $(1,1)$ forms for the K\"ahler moduli space and harmonic $(1,4)$ forms for the complex structure moduli space. We begin with the K\"ahler moduli which we denote by $t^i=t^i(\tau )$, where $i,j,\ldots = 1,\dots ,h^{1,1}(X)$ and $\tau$ is time (for a summary of our index conventions see \appref{conv}). They are real scalar fields and can be defined by expanding the K\"ahler form $J$ on $X$ in terms of a basis $\{\omega_i \}$ of $H^2(X)$ as
\begin{equation}
 J=t^i\omega_i\; .
\end{equation}
The complex structure moduli are denoted by $z^a=z^a(\tau)$, where $a,b,\ldots = 1,\dots , h^{1,4}(X)$, and these are, of course, complex scalar fields.

\vskip 0.4cm

After this preparation, the ansatz for the 11-dimensional metric including moduli can be written as
\begin{equation}
 ds^2=-\frac{1}{4}N(\tau )^2d\tau^2+g_{mn}(t^i,z^a,\bar{z}^{\bar{a}})dx^mdx^n \eqlabel{gansatz}
\end{equation}
where $N=N(\tau )$ is the einbein or lapse function. The lapse function can, of course, be removed by a time reparameterization. However, its equation of motion in the one-dimensional effective theory is the usual zero-energy constraint (the equivalent of the Friedman equation in four-dimensional cosmology). In order not to miss this constraint, we will keep $N$ explicitly in our metric ansatz.

The zero modes of the M-theory three-form $A$ are obtained by an expansion in harmonic forms, as usual. From the Hodge diamond~\eqref*{hodgediamond} of Calabi-Yau five-folds, it is clear that only the harmonic two- and three-forms on $X$ are relevant in this context. For the latter we also introduce a basis $\{\nu_p\}$, where $p,q,\ldots = 1,\dots ,h^{2,1}(X)$. The zero mode expansion for $A$ can then be written as
\be
 A=(\xi^p(\tau )\nu_p +{\rm c.c.})+ N \mu^i(\tau )\,\omega_i\wedge d\tau\; , \eqlabel{Aansatz}
\ee
with $h^{2,1}(X)$ complex scalar fields $\xi^p$ and $h^{1,1}(X)$ real scalars $\mu^i$.  It is clear that the latter correspond to gauge degrees of freedom since $N \mu^i(\tau )\,\omega_i\wedge d\tau = d(f^i(\tau )\,\omega_i)$ with the function $f^i$ being integrals of $N \mu^i$. Note that $N$ enters here merely to ensure worldline reparameterization covariance. Since the fields $\mu^i$ do not represent physical degrees of freedom, the one-dimensional effective action should not depend on these modes. It is, therefore, safe to ignore them in the above ansatz for $A$. Nevertheless, we will find it instructive to keep these modes for now to see explicitly how they drop out of the effective action.

Further zero-modes can arise from membranes if they are included in the compactification, such as moduli of the complex curve $C$ which they wrap and their superpartners. Presently, we will not include these additional modes but rather focus on the modes from pure 11-dimensional supergravity.

\vskip 0.4cm

While the way we have parametrised the zero modes of $A$ in \eqref{Aansatz} appears to be the most natural one, it is not actually the most well-suited ansatz for performing the dimensional reduction. This is due to the fact that we have split a three form into $(2,1)$- and $(1,2)$-pieces (ignoring the gauge part for the moment) the choice of which implicitly depends on the complex structure moduli. If carried through, this leads to an unfavourable and complicated intertwining of kinetic terms of the $(2,1)$- and $(1,4)$-fields in the one-dimensional effective action (that is, terms involving products of the like $\dot{\xi}^p \dot{z}^a$ etc.), which would in turn force us into attempting lengthy field re-definitions in order to diagonalise the kinetic terms.

It would, on the other hand, be much more economic to start out with a formulation in which no such mixing of kinetic terms arises in the first place. Indeed, it is possible to circumvent, yet fully capture, this complication by using real harmonic 3-forms instead of complex $(2,1)$- and $(1,2)$-forms to parametrise the three-form zero modes. Real harmonic 3-forms can be naturally locked to 3-cycles and thus represent topological invariants. In order to employ them in the ansatz for $A$, we first need to introduce a basis $\{ N_{\cal{P}} \}_{\mathcal{P}=1,\ldots,b^3 (X)}$ of real harmonic 3-forms on $X$. Instead of \eqref{Aansatz}, we can then write
\be\eqlabel{Aansatz_3forms}
	A = X^{\cal{P}} (\tau) N_{\cal{P}} + N \mu^i(\tau )\,\omega_i\wedge d\tau \; ,
\ee
with $b^{3}(X) = 2 h^{2,1} (X)$ real scalar fields $X^{\cal{P}}$ and $h^{1,1}(X)$ real scalars $\mu^i$. The two ans\"{a}tze for $A$ are readily related by devising linear maps, denoted $\mathfrak{A}$ and $\mathfrak{B}$, translating back and forth between real harmonic 3-forms and complex harmonic $(2,1)$- or $(1,2)$-forms. These maps implicitly depend on the complex structure moduli, $\mathfrak{A}=\mathfrak{A}(\underline{z},\bar{\underline{z}})$, $\mathfrak{B}=\mathfrak{B}(\underline{z},\bar{\underline{z}})$. In fixed bases, they possess a matrix representation:
\begin{align}
	\nu_p &= \mathfrak{A}_p {}^{\cal{Q}} N_{\cal{Q}} \qquad \qquad \qquad 
	\text{(and: $\bar{\nu}_{\bar{p}} = \bar{\mathfrak{A}}_{\bar{p}} {}^{\cal{Q}} N_{\cal{Q}}$)} \; ,\eqlabel{21_3_rel1_body}\\
	N_{\cal{P}} &= \mathfrak{B}_{\cal{P}} {}^q \nu_q + \bar{\mathfrak{B}}_{\cal{P}} {}^{\bar{q}} \bar{\nu}_{\bar{q}} \; .\eqlabel{21_3_rel2_body}
\end{align}
Inserting \eqrangeref{21_3_rel1_body}{21_3_rel2_body} into \eqrangeref{Aansatz}{Aansatz_3forms}, we learn how the two formulations are related at the level of zero mode fields
\begin{align}
	\xi^p &= X^{\cal{Q}} \mathfrak{B}_{\cal{Q}} {}^p \qquad\qquad \text{(and $\cc$)} , \eqlabel{21_3_rel_xi_X_1} \\
	X^{\cal{P}} &= \xi^q \mathfrak{A}_q {}^{\cal{P}} + \bar{\xi}^{\bar{q}} \bar{\mathfrak{A}}_{\bar{q}} {}^{\cal{P}} \; . \eqlabel{21_3_rel_xi_X_2}
\end{align}
For the reasons outlined above, we henceforth adopt the 3-form formulation. At each step of the calculation, one may, of course, revert if desired to the complex $(2,1)$-form formulation using \eqrangeref{21_3_rel1_body}{21_3_rel_xi_X_2} and the results of \appref{real_form_formalism}, which is devoted to providing a more detailed exposition of the moduli space of real 3-forms on Calabi-Yau five-folds. One particularly important observation is that the moduli space of real 3-forms is itself equipped with a complex structure $\Delta$ inherited from the complex structure of the Calabi-Yau five-fold and explicitly constructed out of $\mathfrak{A}$s and $\mathfrak{B}$s as
\be\eqlabel{3form_cplx_struct_def}
	\Delta_{\cal{P}} {}^{\cal{Q}} := 
	i(\mathfrak{B}_{\cal{P}} {}^q \mathfrak{A}_q {}^{\cal{Q}} - \bar{\mathfrak{B}}_{\cal{P}} {}^{\bar{q}} \bar{\mathfrak{A}}_{\bar{q}} {}^{\cal{Q}}) \; .
\ee
It is readily verified that $\Delta$ satisfies the properties of a complex structure.

The $\mathfrak{A}$ and $\mathfrak{B}$ matrices turn out to be an effective way to parametrize our ignorance of the actual dependence of the $(2,1)$-forms on the complex structure moduli and it would be nice to find explicit expressions instead. However, we are not aware of a method to calculate this dependence explicitly.

\vskip 0.4cm

Returning to the metric ansatz in~\eqref*{gansatz}, we can now compute the eleven dimensional Ricci scalar $R$. As usual, for given values of the complex structure moduli, we introduce local complex coordinates $z^\mu$ and $\bar{z}^{\bar{\mu}}$, where $\mu ,\nu ,\ldots =1,\dots ,5$ and $\bar{\mu} ,\bar{\nu} ,\ldots =\bar{1},\dots ,\bar{5}$, so that  the metric is purely $(1,1)$, that is the components $g_{\mu\bar{\nu}}$ are the only non-vanishing ones. This leads to
\begin{multline}
	\oneon{2}N^2 R	=	   4 N^2 \frac{d}{d \tau} \left(N^{-2} g^{\mu\bar{\nu}}\dot{g}_{\mu\bar{\nu}}\right)
						+    g^{\mu\bar\rho} g^{\nu\bar\sigma} \dot{g}_{\mu\bar\sigma} \dot{g}_{\nu\bar\rho}
						+    g^{\mu\bar\rho} g^{\sigma\bar\nu} \dot{g}_{\mu\sigma} \dot{g}_{\bar\nu\bar\rho}
						\\ + 2 g^{\mu\bar\nu} \dot{g}_{\mu\bar\nu} g^{\sigma\bar\rho} \dot{g}_{\sigma\bar\rho}
						+ 4 N^{-1} \dot{N} g^{\mu\bar\nu}\dot{g}_{\mu\bar\nu} .
\end{multline}
where here and in the following the dot denotes the derivative with respect to $\tau$. Into this expression, we have to insert the expansion of the metric~\eqref*{deltag} which can also be written as
\be\eqlabel{metric_mode_expansion}
	\dot{g}_{\mu\bar{\nu}} = -i \omega_{i,\mu\bar{\nu}} \dot{t}^i\; ,\quad
	\dot{g}_{\mu\nu} = - \oneon{12||\Omega||^2} \Omega_\mu {}^{\bar{\mu}_1\ldots\bar{\mu}_4} 
	  \chi_{a,\bar{\mu}_1\ldots\bar{\mu}_4\nu} \dot{z}^a\; ,\quad
	  \dot{g}_{\bar{\mu}\bar{\nu}}=\left(\dot{g}_{\mu\nu}\right)^\ast \; .
\ee
Here $\{\chi_a\}$, where $a,b,\ldots = 1,\dots ,h^{1,4}(X)$, is a basis of harmonic $(1,4)$ forms.  Further we need the field strength $G=dA$ for the three-form ansatz~\eqref*{Aansatz_3forms} and its Hodge dual which are given by
\be
	G = \dot{X}^{\cal{P}} d\tau \wedge N_{\cal{P}} \; ,\qquad \ast G=-N^{-1} \dot{X}^{\cal{P}} \Delta_{\cal{P}} {}^{\cal{Q}} N_{\cal{Q}}\wedge J^2 \; .
\ee
To derive the second equation we have used the result~\eqref*{3form_hodgestar} for the dual of a real 3-form on a Calabi-Yau five-fold. The $\Delta$ appearing here has been defined in \eqref{3form_cplx_struct_def} and is discussed further in \appref{real_form_formalism}. 

Inserting the ansatz~\eqref*{gansatz}, \eqref*{Aansatz_3forms} together with the last three equations into the bosonic action~\eqref*{sugra11} and integrating over the Calabi-Yau five-fold, one finds the bosonic part of the one-dimensional effective action
\be
\eqlabel{S1B}
	S_{\rm B,kin} = \frac{l}{2} \int d\tau N^{-1} \left\{ \oneon{4} G_{ij}^{(1,1)}(\underline{t}) \dot{t}^i \dot{t}^j  
		+ \oneon{2} G_{\mathcal{PQ}}^{(3)}(\underline{t},\underline{z},\underline{\bar{z}}) \dot{X}^{\cal{P}} \dot{X}^{\cal{Q}} 
		+ 4 V(\underline{t}) G_{a\bar{b}}^{(1,4)} (\underline{z},\underline{\bar{z}}) \dot{z}^a \dot{\bar{z}}^{\bar{b}}  \right\}
\ee
at order zero in the $\beta$ expansion. Here $l = v/\kappa_{11}^{2}$ and $v$ is an arbitrary reference volume of the Calabi-Yau five-fold~\footnote{Related factors of $1/v$ should be included in the definition of the moduli space metrics~\eqref*{G11def}--\eqref*{G3def} but will be suppressed in order to avoid cluttering the notation. These factors can easily be reconstructed from dimensional arguments.}. The moduli space metrics in the $(1,1)$, $(1,4)$ and 3-form sectors are given by
\begin{align}
 G_{ij}^{(1,1)}(\underline{t}) &= 4\int_X\omega_i\wedge\ast\omega_j+8V\tilde{w}_i\tilde{w}_j \; , \eqlabel{G11def}\\
 G_{a\bar{b}}^{(1,4)}(\underline{z},\underline{\bar{z}}) &= \frac{\int_X\chi_a\wedge\bar{\chi}_{\bar{b}}}{\int_X\Omega\wedge\bar{\Omega}} \; , \\
 G_{\mathcal{PQ}}^{(3)}(\underline{t},\underline{z},\underline{\bar{z}}) &= \int_X N_{\cal{P}}\wedge\ast N_{\cal{Q}} \; , \eqlabel{G3def}
\end{align}
where $\tilde{\omega}_i= g^{\mu\bar{\nu}}\omega_{i,\mu\bar{\nu}}$. Since $h^{1,1}(X)$ need not be even, $G_{ij}^{(1,1)}$ is a genuinely real metric that cannot be complexified in general. This is compatible with the anticipated $\susyno=2$ supersymmetry in one dimension, which only demands target spaces of sigma models to be Riemannian manifolds~\cite{Coles:1990hr}. Using the results of \appref{modspace}, these metrics can be computed as functions of the moduli. In the $(1,1)$ sector we have
\begin{equation}
 G_{ij}^{(1,1)}(\underline{t})=8V\left[{\cal G}_{ij}^{(1,1)}(t)-25\frac{\kappa_i\kappa_j}{\kappa^2}\right]=-\frac{2}{3}\kappa_{ij}-\frac{5}{6}\frac{\kappa_i\kappa_j}{\kappa}\; , \eqlabel{G11}
\end{equation}
where $\kappa$ is a quintic polynomial in the K\"ahler moduli given by
\begin{equation}
 \kappa = 5!\, V=d_{i_1\dots i_5}t^{i_1}\dots t^{i_5}\; ,\quad d_{i_1\dots i_5}=\int_X\omega_{i_1}\wedge\dots\wedge\omega_{i_5}\; ,
\end{equation}
$d_{i_1\dots i_5}$ are intersection numbers and $\kappa_i=d_{ii_2\dots i_5}t^{i_2}\dots t^{i_5}$, $\kappa_{ij}=d_{iji_3i_4i_5}t^{i_3}t^{i_4}t^{i_5}$. The standard moduli space metric ${\cal G}_{ij}^{(1,1)}$, as defined in \appref{modspace}, can be obtained from the K\"ahler potential $K^{(1,1)}=-\frac{1}{2}\ln\kappa$ as ${\cal G}_{ij}^{(1,1)}=\partial_i\partial_jK^{(1,1)}$. We note that the physical sigma model metric~\eqref*{G11} differs from the standard moduli space metric ${\cal G}^{(1,1)}_{ij}$ by a term proportional to $\kappa_i\kappa_j$ and a rescaling by the volume. The latter is not really required at this stage and can be removed by a redefinition of time~$\tau$ but it will turn out to be a useful convention in the full supersymmetric version of the effective action. The additional term, however, cannot be removed, for example by a re-scaling of the fields~$t^i$. As a consequence, unlike the standard moduli space metric, the physical metric is not positive definite. In the direction $u^i\sim t^i$ we have $G_{ij}^{(1,1)}u^iu^j<0$ while for all perpendicular directions $u^i$, defined by ${\cal G}^{(1,1)}_{ij}t^iu^j=0$, we have $G_{ij}^{(1,1)}u^iu^j>0$. This means $G^{(1,1)}$ has a Minkowski signature $(-1,+1,\ldots ,+1)$. This is in contrast to, for example, M-theory compactifications on Calabi-Yau three-folds~\cite{Candelas:1990pi,Cadavid:1995bk} where the sigma model metric in the $(1,1)$ sector is identical to the standard moduli space metric and, in particular, is positive definite. In the present case, the appearance of a single negative direction is, of course, not a surprise. Our sigma model metric in the gravity sector can be though of as a ``mini-superspace'' version of the de-Witt metric which is well-known to have precisely one negative eigenvalue~\cite{DeWitt:1967yk}. Here, we see that this negative direction lies in the $(1,1)$ sector. Another difference to the Calabi-Yau three-fold case is the degree of the function $\kappa$. For three-folds $\kappa$ is a cubic while, in the present case, it is a quintic polynomial.

We now turn to the $(1,4)$ moduli space metric $G^{(1,4)}_{a\bar{b}}$ which is, in fact, equal to the standard moduli space metric in this sector and can, hence, be expressed as
\begin{equation}
	G^{(1,4)}_{a\bar{b}}=\partial_a\partial_{\bar{b}}K^{(1,4)}\; ,\quad 
	K^{(1,4)}=\ln\left[i \int_X \Omega \wedge \bar\Omega \right]\eqlabel{G41}
\end{equation}
in terms of the K\"ahler potential $K^{(1,4)}$. This is very similar to the three-fold case. In particular, $G^{(1,4)}_{a\bar{b}}$ is positive definite as it should be, given that the single negative direction arises in the $(1,1)$ sector.

Finally, in the 3-form sector one finds from the results in \appref{real_form_formalism} that the metric can be written as
\begin{equation}
 	G_{\mathcal{PQ}}^{(3)}(\underline{t},\underline{z},\underline{\bar{z}}) = 
		\oneon{2} \Delta_{\cal{\left(P\right.}} {}^{\cal{R}} d_{\left.\mathcal{Q}\right)\mathcal{R}ij} t^i t^j \; ,\quad 
 	d_{\mathcal{P}\mathcal{Q}ij} = \int_X N_{\cal{P}} \wedge N_{\cal{Q}} \wedge\omega_i\wedge\omega_j\; ,\eqlabel{G3}
\end{equation} 
where we have introduced the intersection numbers $d_{\mathcal{P}\mathcal{Q}ij} = - d_{\mathcal{Q}\mathcal{P}ij}$, which are purely topological. The metric~\eqref*{G3} is Hermitian with respect to the complex structure $\Delta$ (see \eqref{G3_hermitian}). 

This completes the definition of all objects which appear in the action~\eqref*{S1B}.

\vskip 0.4cm

We see that this action does not depend on the gauge degrees of freedom $\mu^i$ which appear in the ansatz~\eqref*{Aansatz} for the three-form $A$, as should be the case. This demonstrates $\mu^i$ independence at zeroth order in $\beta$ but what happens at first order in $\beta$? At this order, there are three terms in the 11-dimensional theory, all of them topological, which contribute to $\mu^i$ dependent terms in one dimension. These are the Chern-Simons term $A\wedge G\wedge G$ in \eqref{sugra11}, the Green-Schwarz term~\eqref*{sugra11_gs} and the Wess-Zumino term in the membrane action~\eqref{membraneaction}. Evaluating these three terms leads to the one-dimensional contribution
\be
 S_{\rm B,gauge}=-\frac{l\beta_1}{2}\int d\tau \, N\, \left[12\,g\wedge g+24\,W-c_4(X)\right]_i\mu^i\; , \eqlabel{Sgauge}
\ee
where $\beta_1=(2\pi)^4\beta/v^{4/5}$ is the one-dimensional version of the expansion parameter $\beta$. The notation $[\ldots ]_i$ indicates the components of the eight-form in brackets with respect to a basis $\{\tilde{\omega}^i\}$ of harmonic eight-forms dual to the harmonic two-forms $\{\omega_i\}$. Hence, at order $\beta$ the $\mu^i$ dependent terms do not automatically vanish. However, the bracket in \eqref{Sgauge} vanishes once the integrability condition~\eqref*{intcond} is imposed. Put in a different way, the equation of motion for $\mu^i$ from \eqref{Sgauge} is simply the integrability condition~\eqref*{intcond}
\be
 12\,g\wedge g+24\,W-c_4(X)=0\; .
\ee 
Hence, the r\^{o}le of the gauge modes $\mu^i$ is to enforce the integrability condition at the level of the equations of motion and, once the condition is imposed, the gauge modes disappear from the action as they should. The condition~\eqref*{intcond} can, therefore, also be interpreted as an anomaly cancellation condition which has to be satisfied in order to prevent a gauge anomaly of the M-theory three-form $A$ along the Calabi-Yau $(1,1)$ directions.

\subsection{Performing the dimensional reduction: the fermionic part}

One may ask if an explicit dimensional reduction of the fermionic part of the 11-dimensional action~\eqref*{action11} is really necessary, for in many other cases, once the bosonic terms in the effective action are known the fermionic ones can be inferred from supersymmetry. In the present case, there are a number of reasons why reducing at least some of the fermionic terms might be useful. First of all, the structure of the bosonic action~\eqref*{S1B} points to some features of one-dimensional ${\cal N}=2$ supersymmetry which have not been well-developed in the literature. For example, the bosonic action~\eqref*{S1B} indicates a coupling between the two main types of ${\cal N}=2$ supermultiplets, the $2a$ and $2b$ multiplets, which, to our knowledge, has not been worked out in the literature. Also, in the last section, we have seen that it is important to keep the lapse function as a degree of freedom in the one-dimensional theory, as it generates an important constraint. In the context of supersymmetry, the lapse is part of the  one-dimensional supergravity multiplet which one expects to generate a multiplet of constraints. Therefore, even though gravity is not dynamical in one dimension, we need to consider local one-dimensional ${\cal N}=2$ supersymmetry. Again, it appears this has not been developed in the literature to the extend required for our purposes. We will deal with these problems in \appref{superspace} where we systematically develop one-dimensional ${\cal N}=2$ supersymmetry and supergravity both in component and superspace formalism. At any rate, given that the relevant supersymmetry is not as well established as in some other cases, it seems appropriate to back up our results by reducing some of the 11-dimensional fermionic terms as well. Finally, the list of M-theory zero modes on Calabi-Yau five-folds in \tabref{tab:modes} contains $(1,3)$ fermionic zero modes but no matching bosons. This feature is somewhat puzzling from the point of view of supersymmetry and can certainly not be clarified from the bosonic effective action alone. 

In this section, we will, therefore, reduce the terms in the 11-dimensional action quadratic in fermions. These results together with the bosonic action are sufficient to fix the one-dimensional action in superspace form uniquely and, in addition, provide us with a number of independent checks. Four-fermi terms in the one-dimensional theory are then obtained from the superspace action and we will not derive them by reduction from 11 dimensions. 

\vskip 0.4cm

We should start by writing down a zero mode expansion of the 11-dimensional gravitino $\Psi_M$ on the space-time ${\cal M}=\mathbb{R}\times X$. The covariantly constant, positive chirality spinor on $X$ is denoted by $\eta$ and its negative chirality counterpart by $\eta^\star$ (for a summary of our spinor conventions see \appref{conv}). The spinor $\eta$ is characterised by the annihilation conditions $\gamma^{\bar{\mu}}\eta =0$. Further, by $\omega^{(p,q)}_i$ we denote the harmonic $(p,q)$ forms on $X$. Then, following the known rules for writing down a fermionic zero mode ansatz (see for example Refs.~\cite{Pope:www} \S 1.7.1,~\cite{Roest:2004pk} \S 3.2.5), we have
\begin{align}
	\Psi_0 &= \psi_0(\tau) \otimes \eta^\star + \bar{\psi}_0 (\tau) \otimes \eta \eqlabel{ansatz_gravitino_0_00} , \\
	\Psi_{\bar{\mu}} &= \sum_{p, q} \zeta^{(i)}_{(p,q)} (\tau) \otimes ( \omega^{(p,q)}_{(i),\alpha_1 \ldots \alpha_{(p)} \bar{\beta}_1 \ldots \bar{\beta}_{(q-1)} \bar{\mu}} 
			\gamma^{\alpha_1 \ldots \alpha_{(p)} \bar{\beta}_1 \ldots \bar{\beta}_{(q-1)}} \eta )
	\nonumber \\&+ \sum_{p, q} \zeta^{\prime (i)}_{(p,q)} (\tau) \otimes ( \omega^{(p,q)}_{(i),\alpha_1 \ldots \alpha_{(p)} \bar{\beta}_1 \ldots \bar{\beta}_{(q-1)} \bar{\mu}} \gamma^{\alpha_1 \ldots \alpha_{(p)} \bar{\beta}_1 \ldots \bar{\beta}_{(q-1)}} \eta^\star ), 
	\eqlabel{ansatz_gravitino_internal_00} \\
	\Psi_{\mu} &= (\Psi_{\bar{\mu}})^\ast \; . 
\end{align}
Here, $\zeta_{(p,q)}^{(i)}$ and $\zeta^{\prime (i)}_{(p,q)}$ are one-dimensional complex fermions which represent the zero-modes in the $(p,q)$ sector of the Calabi-Yau five-fold and $\psi_0$ is the one-dimensional gravitino. The sums over $(p,q)$ in~\eqref*{ansatz_gravitino_internal_00} run over all non-trivial cohomology groups of the five-fold. Let us discuss the various $(p,q)$ sectors in the first sum in~\eqref*{ansatz_gravitino_internal_00} in detail. For $(p,q)=(1,4)$ the number of annihilating gamma matrices, $\gamma^{\bar{\mu}}$ exceeds the number of creating ones, $\gamma^\mu$, and, as a result, this term vanishes. Further, for all cases with $q=p+1$ the number of creation and annihilation gamma matrices is identical. Anti-commuting all $\gamma^{\bar{\mu}}$ to the right until they annihilate $\eta$ one picks up inverse metrics $g^{\mu\bar{\nu}}$ which ultimately contract the harmonic $(p,p+1)$ forms $\omega_i^{(p,p+1)}$ to harmonic $(0,1)$ forms. Since the latter do not exist on Calabi-Yau five-folds all terms with $q=p+1$ vanish. This leaves us with the cases where $p\geq q$. Among those, only the terms with $(p,q)=(2,2),(3,2)$ contain both creation and annihilation matrices. For $(p,q)=(2,2)$, anti-commuting leads to a single inverse metric which converts the harmonic $(2,2)$ forms into harmonic $(1,1)$ forms. Therefore, the $(2,2)$ part can effectively be absorbed into the $(1,1)$ term and does not need to be written down independently. The same argument applies to the $(3,2)$ part which can be absorbed into the $(2,1)$ contribution. By the gamma matrix structure and the annihilation property of $\eta^\star$ all but the $(5,0)$ term in the second sum in~\eqref*{ansatz_gravitino_internal_00} vanish. Using the Fierz identity (see \eqref{cy5_fierz_id}) the $(5,0)$ term in the second sum can be converted into a term with the $(1,1)$ structure of the first sum and can, hence, be absorbed by the $(1,1)$ contribution. In summary, all we need to write down explicitly are the $(p,q)$ terms with $q=1$ and $p=1,2,3,4$.

For the same reason as explained in the previous subsection on the bosonic reduction, it is advantageous to use the real 3-form formulation developed in \appref{real_form_formalism} to capture the dynamics of the $(2,1)$-sector while avoiding off-diagonal kinetic terms mixing in time derivatives of $(4,1)$-fields. 
Similarly, we will use the real $\hat{4}$-form formulation, also described in \appref{real_form_formalism}, in the $(1,3)$-sector. A general 4-form, which is always purely topological, can be decomposed into $(1,3)$, $(3,1)$ and $(2,2)$ pieces using the complex structure of the Calabi-Yau five-fold $X$. Henceforth, we will restrict our attention to Calabi-Yau five-folds whose $(2,2)$-forms are completely generated by the product of two $(1,1)$-forms. All the concrete examples of Calabi-Yau five-folds considered in this paper are of this type (see \appref{cyexamples}). In this case, the $(2,2)$-piece of a real 4-form can be split off from the rest in a complex structure independent way and the fate of the $(2,2)$-part of the gravitino ansatz is as described in the previous paragraph. As a shorthand, we will refer to a 4-form that only comprises a $(1,3)$ and a $(3,1)$ piece as a $\hat{4}$-form and given the restriction on $h^{2,2}(X)$, this restriction is also purely topological. The $\hat{4}$-forms are thus well-suited to describe the $(1,3)+(3,1)$-sector of the reduction in a way independent of the complex structure moduli. To this end, it is convenient to choose a particular basis of real 4-forms, $\{ O_{\cal{X}} \}_{\mathcal{X}=1,\ldots,b^4(X)}$, such that the first $2\, h^{1,3}(X)$ 4-forms, denoted $\{ O_{\cal{\hat{X}}} \}_{\mathcal{\hat{X}}=1,\ldots,2h^{1,3}(X)} $, only contain $(1,3)$ and $(3,1)$-pieces and the remaining $h^{2,2}(X)$ 4-forms, denoted $\{ O_{\cal{\tilde{X}}} \}_{\mathcal{\tilde{X}}=1,\ldots,h^{2,2}(X)}$, only contain $(2,2)$-parts. This basis choice is complex structure independent for the class of manifolds under consideration. The $\hat{4}$-forms then lie in the sub-vector space spanned by $\{ O_{\cal{\hat{X}}} \}$. For a general Calabi-Yau five-fold, a more complicated intertwining of the K\"{a}hler and complex structure moduli with the $(1,3)$-fields arises leading to additional interaction terms in the one-dimensional effective action. It should be appreciated that this is a relatively mild restriction as it only affects the $(1,3)$-sector's couplings to fields of other sectors. Our analysis of all other sectors by themselves does not rest on this restriction.

\vskip 0.4cm

After some relabeling, adopting the notation in \appref{modspace} for the harmonic forms and introducing numerical factors for later convenience, the gravitino ansatz now reads
\begin{align}
	\Psi_0 &= \psi_0(\tau) \otimes \eta^\star + \bar{\psi}_0 (\tau) \otimes \eta \eqlabel{ansatz_gravitino_0} , \\
	\Psi_{\bar{\mu}} &= 
			\psi^i(\tau) \otimes (\omega_{i,\alpha_1\bar{\mu}} \gamma^{\alpha_1} \eta) + \frac{i}{4}
			\Lambda^{\cal{P}} (\tau) \otimes (N_{\mathcal{P},\alpha_1\alpha_2\bar{\mu}} \gamma^{\alpha_1\alpha_2} \eta)\nonumber\\
				& + \oneon{4}
			\bar{\Upsilon}^{\hat{\cal{X}}}(\tau) \otimes (O_{{\hat{\mathcal{X}}},\alpha_1\ldots\alpha_3\bar{\mu}} 
				\gamma^{\alpha_1\ldots\alpha_3} \eta) - \oneon{4!}
			\bar{\kappa}^{\bar{a}}(\tau) \otimes (||\Omega ||^{-1} \bar{\chi}_{\bar{a},\alpha_1\ldots\alpha_4\bar{\mu}} 
				\gamma^{\alpha_1\ldots\alpha_4} \eta), 
       \eqlabel{ansatz_gravitino_internal} \\
	\Psi_{\mu} &= (\Psi_{\bar{\mu}})^\ast ,\eqlabel{ansatz_gravitino_1}
\end{align}
The four terms in \eqref{ansatz_gravitino_internal} correspond to the $(1,1)$, $(2,1)$, $(3,1)$ and $(4,1)$ sectors, respectively. The harmonic $(1,1)$ forms are denoted by $\omega_i$, where $i,j,\ldots =1,\dots ,h^{1,1}(X)$, the real 3-forms are denoted by $N_{\cal{P}}$, where $\mathcal{P,Q},\ldots =1,\dots ,b^{3}(X)$, the real $\hat{4}$-forms by $O_{\hat{\cal{X}}}$, where $\hat{\cal{X}},\hat{\cal{Y}},\ldots = 1,\dots , 2h^{1,3}(X)$ and the $(1,4)$ forms by $\chi_a$, where $a,b,\ldots =1,\dots ,h^{1,4}(X)$. In the same order, the associated zero modes, which are complex one-dimensional fermions, are denoted by $\psi^i$, $\Lambda^{\cal{P}}$, $\Upsilon^{\hat{\cal{X}}}$ and $\kappa^a$. 
It is clear that the number of zero modes cannot be reduced any further and that these four types of modes are independent. Three of them, the $(1,1)$, 3-form and $(1,4)$ modes pair up with corresponding bosonic zero modes in the same sectors. The $\hat{4}$-form modes, however, have no bosonic zero mode partners, as mentioned earlier and one of our tasks will be to understand how they can be incorporated into a supersymmetric one-dimensional effective theory.

Had we written the second term in \eqref{ansatz_gravitino_internal} in $(2,1)$-language $\Psi_{\bar{\mu}} = \ldots - 1/4 \lambda^p (\tau) \otimes (\nu_{p,\alpha_1\alpha_2\bar{\mu}} \gamma^{\alpha_1\alpha_2} \eta) + \ldots$, we would have identified a set of $h^{2,1}(X)$ complex one-dimensional fermions in this sector. From \eqref{ansatz_gravitino_internal} however, there appear to be $b^3 (X) = 2 h^{2,1}(X)$ complex one-dimensional fermions. This apparent factor of two discrepancy in the number of degrees of freedom is resolved by observing that a successive insertion of \eqrangeref{21_3_rel1}{21_3_cplx_coords} into the second term in \eqref{ansatz_gravitino_internal} leads to a constraint in the form of a projection condition on the 3-form fermions $\Lambda^{\cal{P}}$
\be\eqlabel{3ferm_constraint}
	P_+ {}_{\cal{P}} {}^{\cal{Q}} \Lambda^{\cal{P}} = \Lambda^{\cal{Q}} \; , \qquad\qquad (\text{and:} \;\;
	P_- {}_{\cal{P}} {}^{\cal{Q}} \bar{\Lambda}^{\cal{P}} = \bar{\Lambda}^{\cal{Q}} ) \; ,
\ee
where $P_\pm {}_{\cal{P}} {}^{\cal{Q}}$ were defined in \eqref{3form_projectors_def}. This condition, which is equivalent to $P_- {}_{\cal{P}} {}^{\cal{Q}} \Lambda^{\cal{P}}=0$, precisely halves the number of degrees of freedom so as to match the counting in $(2,1)$-language. In other words, there are $1/2 \, b^3 (X) = h^{2,1}(X)$ complex one-dimensional fermions in this sector, as claimed in \tabref{tab:modes}. It can be shown that this constraint also applies to the time derivative and supersymmetry transformation of $\Lambda^{\cal{P}}$
\be\eqlabel{3ferm_constraint_dot_susy}
	P_+ {}_{\cal{P}} {}^{\cal{Q}} \dot{\Lambda}^{\cal{P}} = \dot{\Lambda}^{\cal{Q}} \; , \qquad\qquad\qquad
	P_+ {}_{\cal{P}} {}^{\cal{Q}} (\delta_\epsilon \Lambda^{\cal{P}}) = \delta_\epsilon \Lambda^{\cal{Q}} \; ,
\ee
implying in particular that the projection operators commute with both supersymmetry and time translation when acting on $\Lambda^{\cal{P}}$
\be\eqlabel{3ferm_projector_susy_commutator}
	\left[ P_\pm {}_{\cal{P}} {}^{\cal{Q}}, \partial_0 \right] \Lambda^{\cal{P}} = 0 \; , \qquad\qquad
	\left[ P_\pm {}_{\cal{P}} {}^{\cal{Q}}, \delta_\epsilon \right] \Lambda^{\cal{P}} = 0 \; .
\ee
The projection condition is thus preserved under both operations as is required by consistency. \Eqrangeref{3ferm_constraint}{3ferm_projector_susy_commutator} will play important r\^{o}les in finding the correct superspace formulation for this sector later in section~\ref{sec:N=2}.

By complete analogy, we learn that the $\hat{4}$-form sector really only contains $h^{1,3}(X)$ complex one-dimensional fermions (cf. \tabref{tab:modes}) and not $2\, h^{1,3}(X)$ as is suggested by the third term in \eqref{ansatz_gravitino_internal}. By using \eqref{31_4_new_rel1,31_4_new_rel3} and the third term in \eqref{ansatz_gravitino_internal}, we infer
\be\eqlabel{4ferm_constraint}
	P_+ {}_{\hat{\cal{Y}}} {}^{\hat{\cal{X}}} \Upsilon^{\hat{\cal{Y}}} = \Upsilon ^{\hat{\cal{X}}} \; , \qquad\qquad (\text{and:} \;\;
	P_- {}_{\hat{\cal{Y}}} {}^{\hat{\cal{X}}} \bar{\Upsilon}^{\hat{\cal{Y}}} = \bar{\Upsilon}^{\hat{\cal{X}}} ) \; ,
\ee
thereby halving the number of degrees of freedom. The projection operators $P_\pm {}_{\hat{\cal{Y}}} {}^{\hat{\cal{X}}}$ were defined in \eqref{4form_projectors_def}. \Eqref{4ferm_constraint} implies
\begin{align}\eqlabel{4ferm_constraint_dot_susy}
	&P_+ {}_{\hat{\cal{Y}}} {}^{\hat{\cal{X}}} \dot{\Upsilon}^{\hat{\cal{Y}}} = \dot{\Upsilon}^{\hat{\cal{X}}} \; ,
	&&\left[ P_\pm {}_{\hat{\cal{Y}}} {}^{\hat{\cal{X}}}, \partial_0 \right] \Upsilon^{\hat{\cal{Y}}} = 0 \; , \\
	&P_+ {}_{\hat{\cal{Y}}} {}^{\hat{\cal{X}}} (\delta_\epsilon \Upsilon^{\hat{\cal{Y}}}) = \delta_\epsilon \Upsilon^{\hat{\cal{X}}} \; , 
	&&\left[ P_\pm {}_{\hat{\cal{Y}}} {}^{\hat{\cal{X}}}, \delta_\epsilon \right] \Upsilon^{\hat{\cal{Y}}} = 0 
		\eqlabel{4ferm_projector_susy_commutator}
\end{align}
guaranteeing the preservation of the projection condition under time translation and supersymmetry. The compatibility conditions~\eqrangeref*{4ferm_constraint_dot_susy}{4ferm_projector_susy_commutator} are, of course, required for consistency.

\vskip 0.4cm

In order to reduce the fermion terms, we also need explicit expressions for the vielbein, its time derivative and the spin connection. In particular, it should be kept in mind that the gravitino ansatz~\eqref*{ansatz_gravitino_0}--\eqref*{ansatz_gravitino_1} implicitly depends on the vielbein since the curved index gamma matrices $\gamma_\mu$ that appear have to be replaced by flat index gamma matrices $\gamma_{\underline{\mu}}$ via $\gamma_\mu=e_\mu {}^{\underline{\nu}}\gamma_{\underline{\nu}}$. We begin with the vielbein. From the metric ansatz~\eqref*{gansatz} with the $10$-dimensional metric taken to be purely $(1,1)$ its non-zero components are $e_0 {}^\UZ = -N/2$, $e_\mu {}^{\underline\nu} $ and $e_{\bar\mu} {}^{\underline{\bar\nu}}$, so that $g_{\mu\bar{\nu}}=e_\mu {}^{\underline{\rho}} e_{\bar{\nu}} {}^{\underline{\bar{\sigma}}}\eta_{\underline{\rho}\underline{\bar{\sigma}}}$ is the Ricci-flat metric on the Calabi-Yau five-fold. Of course, the $10$-dimensional part of the vielbein depends on the Calabi-Yau K\"ahler moduli $t^i=t^i(\tau)$ and the complex structure moduli $z^a=z^a(\tau )$ and, hence, its time-derivative is non-zero. From the time derivative~\eqref*{metric_mode_expansion} for the metric one finds
\begin{align}
	\dot{e}_\mu {}^{\underline\nu} &= - \frac{i}{2} \omega_{i,\mu} {}^\rho e_{\rho} {}^{\underline\nu} \dot{t}^i , \\
	\dot{e}_\mu {}^{\underline{\bar\nu}} &= - \oneon{12||\Omega||^2} \Omega_{\mu} {}^{\bar{\mu}_1\ldots\bar{\mu}_4} 
	  \chi_{a,\bar{\mu}_1\ldots\bar{\mu}_4\rho} e^{\rho\underline{\bar\nu}} \dot{z}^a ,
\end{align}
and similarly for the complex conjugates. From the equations above and the covariant constancy of the vielbein, we find expressions for the 11-dimensional spin-connection $\omega_N {}^{\underline{Q}\underline{R}}$. Its only non-zero components are given by
\begin{align}
	\omega_\mu {}^{\underline{\nu} \UZ} &= -i N^{-1} \omega_{i,\mu} {}^\rho e_\rho {}^{\underline\nu} \dot{t}^i , \eqlabel{spinconn1} \\
	\omega_\mu {}^{\underline{\bar\nu} \UZ} &= -\oneon{6||\Omega||^2} N^{-1} \Omega_{\mu} {}^{\bar{\mu}_1\ldots\bar{\mu}_4} 
	  \chi_{a,\bar{\mu}_1\ldots\bar{\mu}_4\rho} e^{\rho\underline{\bar\nu}} \dot{z}^a , \eqlabel{spinconn2}
\end{align}
plus their complex conjugates and the components $\omega_m{}^{\underline{n}\underline{p}}$ of the Calabi-Yau spin connection, computed from the $10$-dimensional vielbein $e_m {}^{\underline{n}}$.  The complex conjugates of the components listed above are, of course, also present. The components of the eleven dimensional covariant derivative, defined in \eqref{covder}, then become
\begin{align}
	D_0 &= \partial_0 , \\
	D_\mu &= \tilde{D}_\mu  + \frac{i}{2} N^{-1} \omega_{i,\mu\bar\nu} \dot{t}^i \gamma^{\bar\nu} \Gamma^\UZ
		+ \oneon{12||\Omega||^2} N^{-1} \Omega_{\mu} {}^{\bar{\mu}_1\ldots\bar{\mu}_4} 
	 		\chi_{a,\bar{\mu}_1\ldots\bar{\mu}_4\nu} \dot{z}^a \gamma^\nu \Gamma^\UZ , \\
	D_{\bar\mu} &= (D_\mu)^\ast ,
\end{align}
where $\tilde{D}_\mu$ is the covariant derivative on the Calabi-Yau five-fold.

\vskip 0.4cm

We are now ready to perform the reduction. Inserting the gravitino ansatz \eqrangeref*{ansatz_gravitino_0}{ansatz_gravitino_1} into the fermionic action~\eqref*{sugra11_fermions} produces a vast number of terms -- even when restricting to terms quadratic in fermions. Each of these terms contains a product of a certain number of gamma matrices sandwiched between two spinors $\eta$ or $\eta^\star$. Luckily, on a Calabi-Yau five-fold there only exist a very limited number of non-vanishing such spinor bilinears, namely $\eta^\dagger\eta$, $J_{\mu\bar{\nu}}$, $\Omega_{\mu_1\dots\mu_5}$ and their complex conjugates (see Appendix~\appref*{cydg} for details). As a result, many terms in the reduction vanish immediately, due to their gamma matrix structure. The remaining terms can be split into two types. The first type leads to one-dimensional fermion kinetic terms and such terms originate from the 11-dimensional Rarita-Schwinger term in the action~\eqref*{sugra11_fermions}. The second type leads to one-dimensional Pauli terms, that is couplings between two fermions and the time derivative of a boson, which descend from all the remaining terms in the action~\eqref*{sugra11_fermions}, quadratic in fermions. 

After inserting the gravitino ansatz and integrating over the Calabi-Yau manifold, the Rarita-Schwinger term gives rise to the following fermion kinetic terms
\begin{multline}
	S_{\rm F,kin} = - \frac{l}{2} \int d\tau \frac{i}{2} \left\{ G^{(1,1)}_{ij} (\underline{t}) ( \psi^i \dot{\bar{\psi}}^j - \dot{\psi}^i \bar\psi^j ) 
			+ G_{\mathcal{PQ}}^{(3)}(\underline{t},\underline{z},\underline{\bar{z}}) 
			   (\Lambda^{\cal{P}} \dot{\bar{\Lambda}}^{\cal{Q}} - \dot{\Lambda}^{\cal{P}} \bar{\Lambda}^{\cal{Q}})
		\right. \\ \left. 
			+  3 G^{(\hat{4})}_{\hat{\cal{X}}\hat{\cal{Y}}}(\underline{t}) 
			    (\Upsilon^{\hat{\cal{X}}} \dot{\bar{\Upsilon}}^{\hat{\cal{Y}}} - \dot{\Upsilon}^{\hat{\cal{X}}} \bar{\Upsilon}^{\hat{\cal{Y}}})
			+ 4 V(t) G^{(1,4)}_{a\bar{b}}(\underline{z},\underline{\bar{z}}) (\kappa^a \dot{\bar{\kappa}}^{\bar{b}} - \dot{\kappa}^a \bar{\kappa}^{\bar{b}})
		\right\}\; .
\eqlabel{s11f}\end{multline}
Here, $G_{ij}^{(1,1)}$, $G_{\mathcal{PQ}}^{(3)}$ and $G_{a\bar{b}}^{(1,4)}$ are the moduli space metrics for the $(1,1)$, 3-form and $(1,4)$ bosons exactly as defined in the previous sub-section (see \eqrangeref{G11def}{G3def}). Since there are no $\hat{4}$-form bosons, we have not yet encountered the metric $G^{(\hat{4})}_{\hat{\cal{X}}\hat{\cal{Y}}}$. It is given by
\be
	G^{(\hat{4})}_{\hat{\cal{X}}\hat{\cal{Y}}}(\underline{t}) = \int_X O_{\hat{\cal{X}}} \wedge\ast O_{\hat{\cal{Y}}} 
		= - d_{\hat{\mathcal{X}}\hat{\mathcal{Y}}i} t^i\; ,\quad
	d_{\hat{\mathcal{X}}\hat{\mathcal{Y}}i} = \int_X O_{\hat{\cal{X}}} \wedge O_{\hat{\cal{Y}}} \wedge \omega_i
\ee
in terms of the intersection numbers $d_{\hat{\mathcal{X}}\hat{\mathcal{Y}}i}=d_{\hat{\mathcal{Y}}\hat{\mathcal{X}}i}$, which are purely topological for the class of five-folds we are considering. To evaluate $\ast O_{\hat{\cal{Y}}}$ in the above integral we have used the result for the Hodge dual of $\hat{4}$-forms from \eqref{4form_hodgestar}.

Reducing the other fermion bilinear terms in the 11-dimensional action~\eqref*{sugra11_fermions} we find for the Pauli terms
\bea\eqlabel{action1_f_pauli}
	S_{\rm F,Pauli} = \frac{l}{2} \int d\tau \left\{
		  \frac{i}{2} N^{-1} G^{(1,1)}_{ij}(\underline{t}) (\psi^i\psi_0 + \bar{\psi}^i\bar\psi_0) \dot{t}^j
		+ \frac{i}{2} G^{(1,1)}_{ij,k}(\underline{t}) (\psi^k\bar{\psi}^i + \bar{\psi}^k\psi^i) \dot{t}^j   \right. \\ \left. 
			+ i N^{-1} G^{(3)}_{\mathcal{PQ}}(\underline{t},\underline{z},\bar{\underline{z}}) 
				(\Lambda^{\cal{P}}\psi_0 + \bar{\Lambda}^{\cal{P}}\bar\psi_0) \dot{X}^{\cal{Q}}
			+ i G^{(3)}_{\mathcal{PQ},i}(\underline{t},\underline{z},\bar{\underline{z}}) 
				(\psi^i\bar{\Lambda}^{\cal{P}} + \bar{\psi}^i\Lambda^{\cal{P}}) \dot{X}^{\cal{Q}} \right. \\ \left. 
			- \frac{i}{2} G^{(3)}_{\mathcal{PQ},a}(\underline{t},\underline{z},\bar{\underline{z}}) 
				\Lambda^{\cal{P}} \bar\Lambda^{\cal{Q}} \dot{z}^a
			+ G^{(3)}_{\mathcal{PQ},a}(\underline{t},\underline{z},\bar{\underline{z}}) 
				\kappa^a \bar\Lambda^{\cal{P}} \dot{X}^{\cal{Q}}	\right. \\ \left. 
			+ \frac{i}{2} G^{(3)}_{\mathcal{PQ},\bar{a}}(\underline{t},\underline{z},\bar{\underline{z}}) 
				\Lambda^{\cal{P}} \bar\Lambda^{\cal{Q}} \dot{\bar{z}}^{\bar{a}} 
			- G^{(3)}_{\mathcal{PQ},\bar{a}}(\underline{t},\underline{z},\bar{\underline{z}}) 
				\bar\kappa^{\bar{a}} \Lambda^{\cal{P}} \dot{X}^{\cal{Q}}  \right. \\ \left.  
		+ 2i V G^{(1,4)}_{a\bar{b}, c}(\underline{z}, \underline{\bar{z}}) \kappa^a \bar{\kappa}^{\bar{b}} \dot{z}^c 
		- 2i V G^{(1,4)}_{a\bar{b}, \bar{c}}(\underline{z}, \underline{\bar{z}}) \kappa^a \bar{\kappa}^{\bar{b}} \dot{\bar{z}}^{\bar{c}}     \right. \\ \left. 
		- 4 N^{-1} V G^{(1,4)}_{a\bar{b}}(\underline{z}, \underline{\bar{z}}) 
			(\psi_0 \kappa^a \dot{\bar{z}}^{\bar{b}} - \bar\psi_0 \bar{\kappa}^{\bar{b}} \dot{z}^a)  \right. \\ \left. 
		- \oneon{3!} K_i G^{(1,4)}_{a\bar{b}}(\underline{z}, \underline{\bar{z}}) 
			( \psi^i \bar{\kappa}^{\bar{b}} \dot{z}^a - \bar{\psi}^i \kappa^a \dot{\bar{z}}^{\bar{b}} )
	\right\} .
\eea
This completes the dimensional reduction of the fermionic part of the 11-dimensional action at the level of terms quadratic in fermions. Our complete result for the one-dimensional effective action in components, four-fermi terms not included, is given by the sum of the bosonic action \eqref*{S1B} and the two fermionic parts \eqref*{s11f} and \eqref*{action1_f_pauli}. 
Next, we have to verify that this action is indeed invariant under one-dimensional ${\cal N}=2$ local supersymmetry, as it should be. In the following section, we will do this by writing down a superspace action whose associated component action coincides with our reduction result. This superspace action then also determines the four-fermion terms, which we have not explicitly computed from the dimensional reduction.

\section{Supersymmetry and Calabi-Yau five-folds}\label{sec:N=2}

Compactification on Calabi-Yau five-folds reduces the number of supersymmetries by a factor of $16$, so the effective theory derived in the previous section should, in fact, have one-dimensional ${\cal N}=2$ supersymmetry. We will now show that this is indeed the case. Our first step is to identify how the five-fold zero modes have to be arranged in one-dimensional ${\cal N}=2$ supermultiplets. This is done by reducing the 11-dimensional supersymmetry transformations to one dimension and comparing the result with the known supersymmetry transformations of the various types of one-dimensional multiplets. Then, we write down a superspace action and show that its associated component action, after integrating out auxiliary fields and neglecting four-fermi terms, is identical to the component action derived from reduction. As we have already mentioned, the required one-dimensional ${\cal N}=2$ theories have not been worked out in sufficient detail and generality for our purposes. We have, therefore, included a systematic exposition of both globally  and locally supersymmetric one-dimensional ${\cal N}=2$ theories, tailored to our needs, in \appref{superspace}. Here, we will briefly summarise the main results of this appendix, focusing on the structure of the multiplets and other information necessary to relate ${\cal N}=2$ superspace and component actions. 

\vskip 0.4cm

One-dimensional, ${\cal N}=2$ superspace (``supertime'') is labelled by coordinates $(\tau ,\theta ,\bar{\theta})$ where $\theta$ is a complex one-dimensional spinor\footnote{For our spinor conventions see \appref{conv}.} and $\bar{\theta}$ its complex conjugate. General superfields are functions of these coordinates and can, as usual, be expanded in powers of $\theta$ and $\bar{\theta}$ to obtain their component fields. Since $\theta^2=\bar{\theta}^2=0$, only four terms arise in such an expansion, namely the theta-independent term and the ones proportional to $\theta$, $\bar{\theta}$ and $\theta\bar{\theta}$. In order to develop the geometry of supertime one needs to introduce a supervielbein, a superconnection and supertorsion and solve the Bianchi identities subject to certain constraints on the torsion tensor. This is explicitly carried out in \appref{superspace} and here we simply cite the main results. The field content of the supergravity multiplied can most easily be read off from the component expansion of the super-determinant $\mathcal{E}$ of the supervielbein. It is given by
\be
 \mathcal{E}= - N - \frac{i}{2}\theta\bar{\psi}_0 - \frac{i}{2}\bar{\theta}\psi_0\; ,
\ee
where $N$ is a real scalar, the einbein or lapse function and $\psi_0$ is a complex fermion, the one-dimensional gravitino or lapsino. 

A $2a$ multiplet is a real supermultiplet, that is a supermultiplet $\phi$ satisfying $\phi^\dagger =\phi$. Its component expansion is given by
\be
 \phi =\varphi+i\theta\psi +i\bar{\theta}\bar{\psi}+\frac{1}{2}\theta\bar{\theta}f\; ,
\ee 
and contains the real scalars $\varphi$ and $f$ and the complex fermion $\psi$. The highest component $f$ turns out to be an auxiliary field so we remain with a real scalar and a complex fermion as the physical degrees of freedom.

A $2b$ multiplet, $Z$, is defined by the constraint $\bar{\cal D}Z=0$, where ${\cal D}$ is the super-covariant derivative
\be
	\sderiv = \left( 1 - \frac{i}{2}N^{-1}\bar\theta\psi_0 - \oneon{4}N^{-2}\theta\bar\theta\psi_0\bar\psi_0 \right) \partial_\theta
		+ \left( \frac{i}{2}N^{-1}\bar\theta - \oneon{4}N^{-2}\theta\bar\theta\bar\psi_0 \right) \partial_0 
		- \frac{i}{2}N^{-1}\bar\theta\bar\psi_0 \partial_{\bar\theta} 
\ee
and $\bar{\cal D}$ its conjugate. For the component expansion of a $2b$ multiplet one finds
\be
	Z = z + \theta\kappa + \frac{i}{2}N^{-1}\theta\bar{\theta}(\dot{z}-\psi_0\kappa)\; ,
\ee
with a complex scalar $z$ and a complex fermion $\kappa$. Unlike a $2a$ multiplet, a $2b$ multiplet does not contain an auxiliary field so that its physical field content consists of a complex scalar and a complex fermion. This distinction in physical field content between $2a$ and $2b$ multiplets will be useful in identifying the supermultiplet structure of the five-fold zero modes. 

For both $2a$ and $2b$ multiplets fermionic versions exist, that is multiplets satisfying the same constraint as their bosonic counterparts but with a fermion as the lowest component. Here, we only need the fermionic $2b$ multiplet, $R$, defined by the constraint $\bar{\cal D}R=0$. Its component expansion
\be
 R=\rho +\theta h + \frac{i}{2}N^{-1}\theta\bar{\theta}(\dot{\rho}-\psi_0 h)
\ee
is analogous to that of an ordinary $2b$ multiplet except that the lowest component, $\rho$, is now a (complex) fermion, while $h$ is a complex scalar. As we will see, for a suitable chosen action, the scalar $h$ is an auxiliary field so that the fermion $\rho$ is the only physical degree of freedom. 

A superfield action can now be written as an integral $\int d\tau\,d^2\theta\, \mathcal{E}$ over some function of the above fields and their super-covariant derivatives, where $d^2\theta = d\theta\, d\bar{\theta}$. Explicit superfield actions and their component expansions as required for our purposes are given in \appref{superspace}.

\subsection{$\susyno=2$ supersymmetry transformations and multiplets}

We should now identify how the zero modes of M-theory on Calabi-Yau five-folds fall into super-multiplets of one-dimensional ${\cal N}=2$ supersymmetry. It is a plausible assumption that bosonic and fermionic zero modes that arise from the same sector of harmonic $(p,q)$ forms on the five-fold pair up into supermultiplets. For example, the $h^{1,1}(X)$ K\"ahler moduli $t^i$ should combine with the same number of $(1,1)$ fermions $\psi^i$. Since the K\"ahler moduli $t^i$ are real scalars the resulting $h^{1,1}(X)$ supermultiplets must be of type $2a$. In the $(1,4)$ sector, on the other hand, we have $h^{1,4}(X)$ complex scalars $z^a$ (the complex structure moduli) and the same number of complex fermions $\kappa^a$ so one expects $h^{1,4}(X)$ supermultiplets of type $2b$. The 3-form sector is somewhat more peculiar. There are $b^3(X)$ real scalars $X^{\cal{P}}$ and the same number of complex fermions $\Lambda^{\cal{P}}$ fitting nicely into $b^3(X)$ $2a$ multiplets. However, we also need to take into account the constraint \eqref*{3ferm_constraint} on the fermions, which halves their number. The result is a set of constrained $2a$ multiplets with the same number of degrees of freedom as $1/2 \, b^3(X)$ $2b$ multiplets, reminding us of their original nature. This leaves us with the $\hat{4}$-form fermions $\Upsilon^{\hat{\cal{X}}}$. They have no bosonic zero mode partners so cannot be part of either the standard $2a$ or $2b$ multiplets. The natural guess is for them to form $2\, h^{1,3}(X)$ fermionic $2b$ multiplets. As for the 3-form fermions, there is the constraint~\eqref*{4ferm_constraint}, which reduces their number to by a factor of two. That is, we have $h^{1,3}(X)$ complex one-dimensional fermions in this sector. Finally, the lapse function $N$ and the component $\psi_0$ of the 11-dimensional gravitino should form the one-dimensional gravity multiplet. We now verify this assignment of supermultiplets by a reduction of the 11-dimensional supersymmetry transformations.

\vskip 0.4cm

Our task is to reduce the 11-dimensional supersymmetry transformations~\eqref*{sugra11_susy_transf} for the metric ansatz~\eqref*{gansatz}, the associated spin connection~\eqrangeref*{spinconn1}{spinconn2}, the three-form ansatz~\eqref*{Aansatz} and the gravitino ansatz~\eqref*{ansatz_gravitino_0}--\eqref*{ansatz_gravitino_1}. We denote the spinor parameterising 11-dimensional supersymmetry transformations by $\epsilon^{(11)}$ in order to distinguish it from its one-dimensional counterpart $\epsilon$. The 11-dimensional spinor can then be decomposed as
\be\eqlabel{susyparam}
	\epsilon^{(11)} = \frac{i}{2} \epsilon \otimes \eta^\star - \frac{i}{2} \bar\epsilon \otimes \eta\; ,
\ee
where $\eta$ is the covariantly constant spinor on the Calabi-Yau five-fold. Inserting all this into the 11-dimensional supersymmetry transformations and collecting terms proportional to the same harmonic Calabi-Yau forms we find the supersymmetry transformations of the various zero modes. For the lapse function $N$ and the time component $\psi_0$ of the gravitino they are
\bea
	\delta_\epsilon N &= - \epsilon\bar\psi_0, &\qquad \delta_\epsilon\psi_0 &= i \dot{\epsilon}, &\qquad \delta_\epsilon \bar\psi_0 &= 0 \\
	\delta_{\bar\epsilon} N &= \bar\epsilon\psi_0, &\qquad \delta_{\bar\epsilon} \psi_0 &= 0, &\qquad \delta_{\bar\epsilon} \bar\psi_0 &= - i\dot{\bar{\epsilon}} \; .
\eqlabel{susytrans_1d_sugra}
\eea
These transformations are identical to the one for a one-dimensional ${\cal N}=2$ supergravity multiplet as can be seen by comparing with \appref{superspace}. 

For the other zero modes we find the supersymmetry transformations
\begin{align}
	(1,1): \;\;
	&\delta_\epsilon t^i = - \epsilon \psi^i , && 
	&&\delta_\epsilon \psi^i = 0, && \delta_\epsilon \bar{\psi}^i = \frac{i}{2} N^{-1} \epsilon \dot{t}^i + \ldots  
	\eqlabel{susytrans_1d_11} , \\
	\text{3-form}: \;\;
	&\delta_\epsilon X^{\cal{P}} = - \epsilon \Lambda^{\cal{P}} , && 
	&&\delta_\epsilon \Lambda^{\cal{P}} = 0, && \delta_\epsilon \bar{\Lambda}^{\cal{P}} = i N^{-1} \epsilon P_- {}_{\cal{Q}} {}^{\cal{P}} 
		\dot{X}^{\cal{Q}} + \ldots 
	\eqlabel{susytrans_1d_3} , \\
	\text{$\hat{4}$-form}: \;\;
	&\delta_\epsilon \Upsilon^{\hat{\cal{X}}} = 0 + \ldots,  && \delta_\epsilon \bar\Upsilon^{\hat{\cal{X}}} = 0, && \, && \, \eqlabel{susytrans_1d_4} \\
	(1,4): \;\;
	&\delta_\epsilon z^a = i \epsilon \kappa^a, && \delta_\epsilon \bar{z}^{\bar{a}} = 0, 
	&&\delta_\epsilon \kappa^a = 0, &&\delta_\epsilon \bar{\kappa}^{\bar{a}} = N^{-1} \epsilon \dot{\bar{z}}^{\bar{a}} + \ldots
	\eqlabel{susytrans_1d_41},
\end{align}
and similarly for the $\bar\epsilon$-variation. The dots indicate terms cubic in fermions which we have omitted\footnote{It may be a bit surprising that the transformations above do not seem to mix fields of different types (that is $(1,1)$, $(1,4)$, etc.) despite the plethora of cross-sector interaction terms in the action. However, this is merely an artifact due to the omission of $(\text{fermi})^3$ terms. That is, the sector-mixing terms in the transformations are all of order $(\text{fermi})^3$, which can be seen by taking the full, off-shell supersymmetry transformations of \appref{superspace} and eliminating the auxiliary fields.}. To arrive at the last equation in~\eqref*{susytrans_1d_3}, we have performed a compensating transformation, making use of a local fermionic symmetry. Namely, the action \eqref*{s11f,action1_f_pauli} is invariant under
\be\eqlabel{3form_symm}
	\delta \Lambda^{\cal{P}} = P_- {}_{\cal{Q}} {}^{\cal{P}} l^{\cal{Q}} \; , \qquad\qquad\qquad (\text{and: } 
	\delta \bar\Lambda^{\cal{P}} = P_+ {}_{\cal{Q}} {}^{\cal{P}} \bar{l}^{\cal{Q}} ) \; ,
\ee
for a set of local complex fermionic parameters $l^{\cal{Q}}$, while all other fields do not transform. The constraint~\eqref*{3ferm_constraint} on $\Lambda^{\cal{P}}$ may be viewed as a gauge choice with respect to this symmetry. The form of the last equation in~\eqref*{susytrans_1d_3} then guarantees the preservation of this gauge choice under a supersymmetry transformation as required by \eqref{3ferm_constraint_dot_susy}. Even though the $\hat{4}$-form fermions $\Upsilon^{\hat{\cal{X}}}$ are subject to the same kind of constraint (cf. \eqref{4ferm_constraint}), there is no associated local symmetry. This is because the proof that~\eqref*{3form_symm} is a symmetry crucially hinges on the Hermiticity of the 3-form metric (cf. \eqref{G3_hermitian}), but the $\hat{4}$-form metric is not Hermitian.

Again, comparing with the results for the supersymmetry transformations of the various one-dimensional ${\cal N}=2$ multiplets given in \appref{superspace}, we confirm the assignment of zero modes into supermultiplets discussed above. In particular, the transformation of the $\hat{4}$-form fermions $\Upsilon^{\hat{\cal{X}}}$ indicates that they should indeed be part of fermionic $2b$ supermultiplets. 

To summarise these results, we write down the explicit off-shell component expansion for all superfields in terms of the Calabi-Yau five-fold zero modes and appropriate auxiliary fields. Taking into account the component structure of the various supermultiplets derived in \appref{superspace}, we have
\begin{align}
	\text{SUGRA} && (2a): \quad		
		&\mathcal{E} = - N - \frac{i}{2} \theta\bar\psi_0 - \frac{i}{2} \bar\theta\psi_0 , \eqlabel{superfield_E} \\
	(1,1) && (2a): \quad
		&T^i = t^i + i \theta \psi^i + i \bar\theta \bar\psi^i + \oneon{2} \theta\bar\theta f^i , \eqlabel{superfield_T} \\
	\text{3-form} && (2a): \quad
		&\mathcal{X}^{\cal{P}} = X^{\cal{P}} + i \theta \Lambda^{\cal{P}} + i \bar\theta \bar{\Lambda}^{\cal{P}} 
			+ \oneon{2} \theta\bar\theta g^{\cal{P}} , \eqlabel{superfield_X} \\
	\text{$\hat{4}$-form} && (2b)-{\rm fermionic}: \quad
		&\mathcal{R}^{\hat{\cal{X}}} = \Upsilon^{\hat{\cal{X}}} + \theta H^{\hat{\cal{X}}} 
			+ \frac{i}{2} N^{-1} \theta\bar\theta (\dot{\Upsilon}^{\hat{\cal{X}}} - \psi_0 H^{\hat{\cal{X}}}) , \eqlabel{superfield_R} \\
	(1,4) && (2b): \quad
		&Z^a = z^a + \theta \kappa^a + \frac{i}{2} N^{-1} \theta\bar\theta (\dot{z}^a - \psi_0 \kappa^a ) , \eqlabel{superfield_Z}
\end{align}
where $f^i$, $g^{\cal{P}}$ and $H^{\hat{\cal{X}}}$ are bosonic auxiliary fields. These auxiliary fields can, of course, not be obtained from the reduction (since 11-dimensional supersymmetry is realised on-shell) and have to be filled in ``by hand''. Full, off-shell supersymmetry transformations for all the above components are given in \appref{superspace}.

\subsection{The one-dimensional effective action in superspace}

Having identified the relevant supermultiplets and their components our next step is to write down an ${\cal N}=2$ superspace version of the one-dimensional effective theory. For the most part, an appropriate form for the superspace action can be guessed based on the bosonic action~\eqref*{S1B}. Basically, all one has to do is to promote the bosonic fields in this action to their associated superfields, replace time derivatives by super-covariant derivatives ${\cal D}$ or $\bar{\cal D}$ and integrate over superspace. 
In addition, we need to implement the constraint~\eqref*{3ferm_constraint} on the 3-form fermions $\Lambda^{\cal{P}}$ at the superspace level. The superpartner of the constraint \eqref*{3ferm_constraint} turns out to be
\be\eqlabel{3ferm_constraint_spartner}
	g^{\cal{P}} = N^{-1} \Delta_{\cal{Q}} {}^{\cal{P}} \dot{X}^{\cal{Q}} + N^{-1} (\psi_0 \Lambda^{\cal{P}} - \bar\psi_0 \bar{\Lambda}^{\cal{P}}) \; .
\ee
Note that since the only object in this equation depending on the complex structure moduli is $\Delta_{\cal{Q}} {}^{\cal{P}}$, it follows that $\Delta_{\cal{Q}} {}^{\cal{P}} {}_{,a} \dot{X}^{\cal{Q}} = 0$.
Constraints~\eqref*{3ferm_constraint,3ferm_constraint_spartner} form a constraint multiplet and can hence be obtained from a single complex superspace equation
\be\eqlabel{3ferm_constraint_sspace}
	P_- {}_{\cal{P}} {}^{\cal{Q}} (\underline{Z},\bar{\underline{Z}}) \sderiv \mathcal{X}^{\cal{P}} = 0 \; , \qquad\qquad\qquad (\text{and $\cc$}) \; ,
\ee
where $P_- {}_{\cal{P}} {}^{\cal{Q}} (\underline{Z},\bar{\underline{Z}})$ is the superspace version of the projection operator $P_- {}_{\cal{P}} {}^{\cal{Q}}$ defined in \eqref{3form_projectors_def}. The superspace constraint \eqref*{3ferm_constraint_sspace} follows from a superspace action by introducing a set of $b^3 (X)$ complex fermionic Lagrange multiplier superfields $L_{\cal{P}}$
\be\eqlabel{3ferm_lagr_mult_sfield}
	L_{\cal{P}} = L^{(0)}_{\cal{P}} + \theta L^{(1)}_{\cal{P}} + \bar\theta L^{(2)}_{\cal{P}} + \oneon{2} \theta\bar\theta L^{(3)}_{\cal{P}} \; .
\ee
The action for the fermionic Lagrange multiplier superfields is then given by
\be
	-\frac{l}{2}\int d\tau d^2 \theta\, \mathcal{E} 
		\left( L_{\cal{Q}} P_- {}_{\cal{P}} {}^{\cal{Q}} (\underline{Z},\bar{\underline{Z}}) \sderiv \mathcal{X}^{\cal{P}} -  
		\bar{L}_{\cal{Q}} P_+ {}_{\cal{P}} {}^{\cal{Q}} (\underline{Z},\bar{\underline{Z}}) \bar\sderiv \bar{\mathcal{X}}^{\cal{P}}\right) \; .
\ee
This takes care of all but the fermionic multiplets in the $\hat{4}$-form sector whose superfield action has to be inferred from the fermionic component action~\eqref*{s11f}, \eqref*{action1_f_pauli}. In particular, the $\hat{4}$-form part of the superspace action should be such that the bosons $H^{\hat{\cal{X}}}$ in the fermionic multiplets are non-dynamical. As for the 3-form case, we need to implement the constraint~\eqref*{4ferm_constraint} on the $\hat{4}$-form fermions $\Upsilon^{\hat{\cal{X}}}$ at the superspace level. The superpartner of the constraint~\eqref*{4ferm_constraint} is simply
\be\eqlabel{4ferm_constraint_spartner}
	P_+ {}_{\hat{\cal{Y}}} {}^{\hat{\cal{X}}} H^{\hat{\cal{Y}}} = H^{\hat{\cal{X}}} \; , \qquad\qquad (\text{and $\cc$}) \; .
\ee
\Eqref{4ferm_constraint,4ferm_constraint_spartner} are part of a single superspace equation
\be\eqlabel{4ferm_constraint_sspace}
	P_- {}_{\hat{\cal{Y}}} {}^{\hat{\cal{X}}}  (\underline{Z},\bar{\underline{Z}}) \mathcal{R}^{\hat{\cal{Y}}} = 0 \; , \qquad\qquad (\text{and $\cc$}) \; ,
\ee
which can be obtained from a superspace action principle
\be
	-\frac{l}{2}\int d\tau d^2 \theta\, \mathcal{E} 
		\left( L_{\hat{\cal{X}}} P_- {}_{\hat{\cal{Y}}} {}^{\hat{\cal{X}}} (\underline{Z},\bar{\underline{Z}}) \mathcal{R}^{\hat{\cal{Y}}} -  
		\bar{L}_{\hat{\cal{X}}} P_+ {}_{\hat{\cal{Y}}} {}^{\hat{\cal{X}}} (\underline{Z},\bar{\underline{Z}}) \bar{\mathcal{R}}^{\hat{\cal{X}}}\right)
\ee
by means of a set of $2\, h^{1,3}(X)$ complex fermionic Lagrange multiplier superfields $L_{\hat{\cal{X}}}$, which have the same component expansion as in \eqref{3ferm_lagr_mult_sfield}. $P_\pm {}_{\hat{\cal{Y}}} {}^{\hat{\cal{X}}}  (\underline{Z},\bar{\underline{Z}})$ are the superspace versions of the projection operators $P_\pm {}_{\hat{\cal{Y}}} {}^{\hat{\cal{X}}}$ defined in \eqref{4form_projectors_def}.

Combining all this, the suggested superspace action is
\begin{multline}
	S_1  = - \frac{l}{2} \int d\tau\, d^2 \theta\, \mathcal{E} \left\{ 
	 	   G^{(1,1)}_{ij} (\underline{T}) \sderiv T^i \bar\sderiv T^j 
		+ G_{\mathcal{PQ}}^{(3)}(\underline{T},\underline{Z},\underline{\bar{Z}}) \sderiv \mathcal{X}^{\cal{P}} \bar\sderiv \mathcal{X}^{\cal{Q}}
		- 3 G^{(\hat{4})}_{\hat{\cal{X}}\hat{\cal{Y}}} (\underline{T}) \mathcal{R}^{\hat{\cal{X}}} \bar{\mathcal{R}}^{\hat{\cal{Y}}}
		\right. \\ \left.		
		+ 4 V(\underline{T}) G^{(1,4)}_{a\bar{b}} 
			(\underline{Z},\bar{\underline{Z}}) \sderiv Z^a \bar\sderiv \bar{Z}^{\bar{b}} 
		+ \left( L_{\cal{Q}} P_- {}_{\cal{P}} {}^{\cal{Q}} (\underline{Z},\bar{\underline{Z}}) \sderiv \mathcal{X}^{\cal{P}} 
		+ L_{\hat{\cal{X}}} P_- {}_{\hat{\cal{Y}}} {}^{\hat{\cal{X}}} (\underline{Z},\bar{\underline{Z}}) \mathcal{R}^{\hat{\cal{Y}}} + \cc \right) \right\}\; .
\eqlabel{superspace_action_1d}
\end{multline}
This action can be expanded out in components using the formul\ae~presented earlier and systematically developed in \appref{superspace}. The result can be split into $(1,1)$, 3-form, $\hat{4}$-form and $(1,4)$ parts by writing
\be\eqlabel{superspace_action_1d_comp}
	S_1 = \frac{l}{2} \int d\tau\left\{ \mathcal{L}^{(1,1)} + \mathcal{L}^{(3)} + \mathcal{L}^{(\hat{4})} + \mathcal{L}^{(1,4)}\right\}\; .
\ee
For these four parts of the Lagrangian in~\eqref*{superspace_action_1d_comp} we find, after taking into account the constraints~\eqref*{3ferm_constraint,3ferm_constraint_spartner} and using the formul\ae~provided in \appref{real_form_formalism}
\begin{align}
	&\begin{aligned}
	\mathcal{L}^{(1,1)} &= \oneon{4} N^{-1} G^{(1,1)}_{ij}(\underline{t}) \dot{t}^i \dot{t}^j
			- \frac{i}{2} G^{(1,1)}_{ij}(\underline{t}) (\psi^i \dot{\bar\psi}^j - \dot{\psi}^i\bar{\psi}^j)
			+ \oneon{4} N G^{(1,1)}_{ij}(\underline{t}) f^i f^j \\
			&+ \frac{i}{2} N^{-1} G^{(1,1)}_{ij}(\underline{t}) (\psi^i\psi_0 + \bar{\psi}^i\bar\psi_0) \dot{t}^j
			+ \oneon{2} N^{-1} G^{(1,1)}_{ij}(\underline{t}) \psi_0\bar\psi_0 \psi^i \bar{\psi}^j \\
			&- \oneon{2} N G^{(1,1)}_{ij,k}(\underline{t}) (\psi^i\bar{\psi}^j f^k - \psi^k\bar{\psi}^j f^i - \psi^i\bar{\psi}^k f^j)
			+ \frac{i}{2} G^{(1,1)}_{ij,k}(\underline{t}) (\psi^k\bar{\psi}^i + \bar{\psi}^k\psi^i) \dot{t}^j \\
			&- N G^{(1,1)}_{ij,kl}(\underline{t}) \psi^i\bar{\psi}^j\psi^k\bar{\psi}^l , \eqlabel{L11}
	\end{aligned} \\
	&\begin{aligned}
	\mathcal{L}^{(3)} &= \oneon{2} N^{-1} G^{(3)}_{\mathcal{PQ}}(\underline{t},\underline{z},\bar{\underline{z}}) \dot{X}^{\cal{P}} \dot{X}^{\cal{Q}}
			- \frac{i}{2} G^{(3)}_{\mathcal{PQ}}(\underline{t},\underline{z},\bar{\underline{z}}) 
				(\Lambda^{\cal{P}} \dot{\bar\Lambda}^{\cal{Q}} - \dot{\Lambda}^{\cal{P}} \bar{\Lambda}^{\cal{Q}}) \\
			&+ i N^{-1} G^{(3)}_{\mathcal{PQ}}(\underline{t},\underline{z},\bar{\underline{z}}) 
				(\Lambda^{\cal{P}}\psi_0 + \bar{\Lambda}^{\cal{P}}\bar\psi_0) \dot{X}^{\cal{Q}}
			+ N^{-1} G^{(3)}_{\mathcal{PQ}}(\underline{t},\underline{z},\bar{\underline{z}}) 
				\psi_0\bar\psi_0 \Lambda^{\cal{P}} \bar{\Lambda}^{\cal{Q}} \\
			&- \oneon{2} N G^{(3)}_{\mathcal{PQ},i}(\underline{t},\underline{z},\bar{\underline{z}}) \Lambda^{\cal{P}}\bar{\Lambda}^{\cal{Q}} f^i
			+ i G^{(3)}_{\mathcal{PQ},i}(\underline{t},\underline{z},\bar{\underline{z}}) 
				(\psi^i\bar{\Lambda}^{\cal{P}} + \bar{\psi}^i\Lambda^{\cal{P}}) \dot{X}^{\cal{Q}} \\
			&- \oneon{2} G^{(3)}_{\mathcal{PQ},i}(\underline{t},\underline{z},\bar{\underline{z}}) \Lambda^{\cal{P}} \bar{\Lambda}^{\cal{Q}} 
				(\psi_0 \psi^i - \bar\psi_0 \bar\psi^i) 
			- N G^{(3)}_{\mathcal{PQ},ij}(\underline{t},\underline{z},\bar{\underline{z}}) \Lambda^{\cal{P}}\bar{\Lambda}^{\cal{Q}}\psi^i\bar{\psi}^j \\
			&- \frac{i}{2} G^{(3)}_{\mathcal{PQ},a}(\underline{t},\underline{z},\bar{\underline{z}}) 
				\Lambda^{\cal{P}} \bar\Lambda^{\cal{Q}} (\dot{z}^a - 2 \psi_0 \kappa^a) 
			+ \frac{i}{2} G^{(3)}_{\mathcal{PQ},\bar{a}}(\underline{t},\underline{z},\bar{\underline{z}}) 
				\Lambda^{\cal{P}} \bar\Lambda^{\cal{Q}} (\dot{\bar{z}}^{\bar{a}} + 2 \bar\psi_0 \bar\kappa^{\bar{a}}) \\
			&+ G^{(3)}_{\mathcal{PQ},a}(\underline{t},\underline{z},\bar{\underline{z}}) 
				\kappa^a \bar\Lambda^{\cal{P}} \dot{X}^{\cal{Q}}
			- G^{(3)}_{\mathcal{PQ},\bar{a}}(\underline{t},\underline{z},\bar{\underline{z}}) 
				\bar\kappa^{\bar{a}} \Lambda^{\cal{P}} \dot{X}^{\cal{Q}}  
			- N G^{(3)}_{\mathcal{PQ},a\bar{b}}(\underline{t},\underline{z},\bar{\underline{z}}) 
				\Lambda^{\cal{P}} \bar\Lambda^{\cal{Q}} \kappa^a \bar\kappa^{\bar{b}} \\
			&- i N G^{(3)}_{\mathcal{PQ},ia}(\underline{t},\underline{z},\bar{\underline{z}}) 
				\Lambda^{\cal{P}} \bar\Lambda^{\cal{Q}} \bar\psi^i \kappa^a 
			- i N G^{(3)}_{\mathcal{PQ},i\bar{a}}(\underline{t},\underline{z},\bar{\underline{z}}) 
				\Lambda^{\cal{P}} \bar\Lambda^{\cal{Q}} \psi^i \bar\kappa^{\bar{a}}  ,
			\eqlabel{L3}
	\end{aligned} \\
	&\begin{aligned}
	\mathcal{L}^{(\hat{4})} &= - \frac{3i}{2} G^{(\hat{4})}_{\hat{\cal{X}}\hat{\cal{Y}}}(\underline{t}) 
			( \Upsilon^{\hat{\cal{X}}} \dot{\bar{\Upsilon}}^{\hat{\cal{Y}}} - \dot{\Upsilon}^{\hat{\cal{X}}} \bar{\Upsilon}^{\hat{\cal{Y}}} ) 
		+ 3 N G^{(\hat{4})}_{\hat{\cal{X}}\hat{\cal{Y}}}(\underline{t}) H^{\hat{\cal{X}}} \bar{H}^{\hat{\cal{Y}}} \\ &
		+ 3 i N G^{(\hat{4})}_{\hat{\cal{X}}\hat{\cal{Y}},i}(\underline{t}) 
			(\psi^i \Upsilon^{\hat{\cal{X}}} \bar{H}^{\hat{\cal{Y}}} + \bar{\psi}^i \bar{\Upsilon}^{\hat{\cal{Y}}} H^{\hat{\cal{X}}} ) 
		+ \frac{3}{2} N G^{(\hat{4})}_{\hat{\cal{X}}\hat{\cal{Y}},i}(\underline{t}) \Upsilon^{\hat{\cal{X}}} \bar{\Upsilon}^{\hat{\cal{Y}}} f^i \\ &
		+ 3 N G^{(\hat{4})}_{\hat{\cal{X}}\hat{\cal{Y}},ij}(\underline{t}) \Upsilon^{\hat{\cal{X}}} \bar{\Upsilon}^{\hat{\cal{Y}}} \psi^i \bar{\psi}^j
		- \frac{3}{2} G^{(\hat{4})}_{\hat{\cal{X}}\hat{\cal{Y}},i}(\underline{t}) 
			\Upsilon^{\hat{\cal{X}}} \bar{\Upsilon}^{\hat{\cal{Y}}} ( \psi_0 \psi^i - \bar{\psi}_0 \bar{\psi}^i ) , 
			\eqlabel{L4}
	\end{aligned} \\
	&\begin{aligned}\eqlabel{L14}
	\mathcal{L}^{(1,4)} &= 4 N^{-1} V G^{(1,4)}_{a\bar{b}}(\underline{z}, \underline{\bar{z}}) \dot{z}^a \dot{\bar{z}}^{\bar{b}}
		- 2i V G^{(1,4)}_{a\bar{b}}(\underline{z}, \underline{\bar{z}}) 
			(\kappa^a \dot{\bar{\kappa}}^{\bar{b}} - \dot{\kappa}^a \bar{\kappa}^{\bar{b}}) \\ &
		- 4 N^{-1} V G^{(1,4)}_{a\bar{b}}(\underline{z}, \underline{\bar{z}}) 
			(\psi_0 \kappa^a \dot{\bar{z}}^{\bar{b}} - \bar\psi_0 \bar{\kappa}^{\bar{b}} \dot{z}^a)
		+ 4 N^{-1} V G^{(1,4)}_{a\bar{b}}(\underline{z}, \underline{\bar{z}}) \psi_0 \bar\psi_0 \kappa^a \bar{\kappa}^{\bar{b}} \\ &
		+ 2i V G^{(1,4)}_{a\bar{b}, c}(\underline{z}, \underline{\bar{z}}) \kappa^a \bar{\kappa}^{\bar{b}} \dot{z}^c 
		- 2i V G^{(1,4)}_{a\bar{b}, \bar{c}}(\underline{z}, \underline{\bar{z}}) \kappa^a \bar{\kappa}^{\bar{b}} \dot{\bar{z}}^{\bar{c}} \\ &
		- \oneon{12} N K_i G^{(1,4)}_{a\bar{b}}(\underline{z}, \underline{\bar{z}}) \kappa^a \bar{\kappa}^{\bar{b}} f^i
		- \frac{2}{3} N K_{ij} G^{(1,4)}_{a\bar{b}}(\underline{z}, \underline{\bar{z}}) \kappa^a \bar{\kappa}^{\bar{b}} \psi^i \bar{\psi}^j \\ &
		- \oneon{3!} K_i G^{(1,4)}_{a\bar{b}}(\underline{z}, \underline{\bar{z}}) 
			\psi^i \bar{\kappa}^{\bar{b}} (\dot{z}^a - \oneon{2} \psi_0 \kappa^a)
		+ \oneon{3!} K_i G^{(1,4)}_{a\bar{b}}(\underline{z}, \underline{\bar{z}}) 
			\bar{\psi}^i \kappa^a (\dot{\bar{z}}^{\bar{b}} + \oneon{2} \bar{\psi}_0 \bar{\kappa}^{\bar{b}})\; .
	\end{aligned} 
\end{align}
We should now compare this Lagrangian with our result obtained from dimensional reduction in the previous section. To do this, we first have to integrate out the auxiliary fields $f^i$ and $H^{\hat{\cal{X}}}$. A quick inspection of their equations of motion derived from eqs.~\eqref*{L11}--\eqref*{L14} shows that they are given by fermion bilinears. Hence, integrating them out only leads to additional four-fermi terms. Since we have not computed four-fermi terms in our reduction from 11 dimensions they are, in fact, irrelevant for our comparison. All other terms, that is purely bosonic terms and terms bilinear in fermions, coincide with our reduction result~\eqref*{S1B}, \eqref*{s11f} and \eqref*{action1_f_pauli}. This shows that \eqref{superspace_action_1d} is indeed the correct superspace action. 

Both the lapse function $N$ and the gravitino $\psi_0$ are non-dynamical and their equations of motion lead to constraints. For the lapse, this constraint implies the vanishing of the Hamiltonian associated with the Lagrangian~\eqref*{L11}--\eqref*{L14} and it reads (after integrating out the $(1,1)$ and $\hat{4}$-form auxiliary fields $f^i$ and $H^{\hat{\cal{X}}}$)
\begin{multline}
	   \oneon{4}  G^{(1,1)}_{ij}(\underline{t}) ( \dot{t}^i + 2i \psi^i\psi_0 + 2i \bar{\psi}^i\bar\psi_0) \dot{t}^j 
	+ \oneon{2} G^{(3)}_{\mathcal{PQ}}(\underline{t},\underline{z},\bar{\underline{z}}) (\dot{X}^{\cal{P}}
			+ 2 i \Lambda^{\cal{P}}\psi_0 + 2 i \bar{\Lambda}^{\cal{P}}\bar\psi_0)) \dot{X}^{\cal{Q}} \\
	+ 4 V G^{(1,4)}_{a\bar{b}}(\underline{z}, \underline{\bar{z}}) ( \dot{z}^a \dot{\bar{z}}^{\bar{b}}
			- \psi_0 \kappa^a \dot{\bar{z}}^{\bar{b}} + \bar\psi_0 \bar{\kappa}^{\bar{b}} \dot{z}^a) + (\text{fermi})^4 = 0\;  .
\end{multline}
The equation of motion for $\psi_0$ generates the superpartner of this Hamiltonian constraint and implies the vanishing of the supercurrent.

\vskip 0.4cm

Let us now discuss some of the symmetries of the above one-dimensional action. The action~\eqref*{superspace_action_1d} is manifestly invariant under super-worldline reparametrizations $\{\tau,\theta,\bar\theta\}$ $\rightarrow$ $\{\tau^\prime (\tau,\theta,\bar\theta)$, $\theta^\prime (\tau,\theta,\bar\theta)$, $\bar{\theta}^\prime (\tau,\theta,\bar\theta)\}$, which, in particular, includes worldline reparametrizations $\tau \rightarrow \tau^\prime (\tau)$ (that is, one-dimensional diffeomorphisms) and local $\susyno=2$ supersymmetry. Note that the super-determinant of the supervielbein $\mathcal{E}$, which transforms as a super-density, is precisely what is needed to cancel off the super-jacobian from the change of $d\tau\, d^2 \theta$, so that $d\tau\, d^2 \theta\,\mathcal{E}$ is an invariant measure.

In particular, the theory is invariant under worldline reparametrizations, $\tau\rightarrow\tau '(\tau)$ which can be seen as a remnant of the diffeomorphism invariance of the eleven dimensional action~\eqref*{action11}. Here, the lapse function, $N$, 
plays the same r\^{o}le as the ``vielbein'' and it transforms as a co-vector under worldline reparametrizations. The transformation properties of the different types of component fields under worldline reparametrizations are summarized in \tabref{WR}.
\TABLE[h]{
\hspace{15cm}\vspace{-0.2cm}\vspace{0.2cm} 
\newline
\begin{tabular}{|c|l|} \hline
	\emph{Name} & \emph{WR transformation} $\tau \rightarrow \tau^\prime (\tau)$ \\ \hline\hline
	scalar & $z^a \rightarrow z^a {}^\prime (\tau^\prime) = z^a(\tau)$ \\
	co-vector & $N \rightarrow N^\prime (\tau^\prime) = \frac{\partial \tau}{\partial \tau^\prime} N (\tau)$ \\
	spin-$1/2$ & $\kappa^a \rightarrow \kappa^a {}^\prime (\tau^\prime) = \kappa^a (\tau)$ \\
	spin-$3/2$ & $\psi_0 \rightarrow \psi_0^\prime (\tau^\prime) = \frac{\partial \tau}{\partial \tau^\prime} \psi_0 (\tau)$ \\ \hline
\end{tabular}\caption{Worldline reparametrization (WR) covariance.}\tablabel{WR}}
The bosonic matter fields $t^i$, $X^{\cal{P}}$ and $z^a$ and the bosonic auxiliary fields $f^i$, $g^{\cal{P}}$ and $H^{\hat{\cal{X}}}$ transform as scalars, whereas the fer\-mi\-onic matter fields $\psi^i$, $\Lambda^{\cal{P}}$, $\Upsilon^{\hat{\cal{X}}}$ and $\kappa^a$ transform as spin-$1/2$ fields. Finally, the gravitino $\psi_0$ transforms as a spin-$3/2$ field. 

The 3-form scalars $X^{\cal{P}}$ arise as zero-modes of the M-theory three-form $A$ and, hence, they are axion-like scalars with associated shift transformations acting as
\be
	X^{\cal{P}} (\tau) \rightarrow {X^{\cal{P}}}^\prime (\tau) = X^{\cal{P}} (\tau) + c^{\cal{P}} \; ,
\ee
where the $c^{\cal{P}}$ are a set of complex constants. It is easy to see that the component action~\eqref*{L11}--\eqref*{L14} only depends on $\dot{X}^{\cal{P}}$ but not on $X^{\cal{P}}$ and that, hence, the action is invariant under the above shifts. Also in the 3-form sector, there is a local fermionic symmetry of the form $\delta \Lambda^{\cal{P}} = P_- {}_{\cal{Q}} {}^{\cal{P}} l^{\cal{Q}}$ as discussed around \eqref{3form_symm}.

\section{Flux and the one-dimensional scalar potential}\label{sec:flux}

We have seen that, unless one works with a Calabi-Yau five-fold $X$ satisfying $c_4(X)=0$, flux and/or membranes are required in order to satisfy the anomaly condition~\eqref*{intcond}. At order $\beta$, both flux and membranes are expected to contribute to a scalar potential in the one-dimensional effective theory. So far, we have worked at zeroth order in $\beta$ but, in this section, we will calculate the leading order $\beta$ contributions to the scalar potential. Given the need for flux and/or membranes in many five-fold compactifications this potential is clearly of great physical significance.

\subsection{Calculating the scalar potential from 11 dimensions}

There are three terms in the 11-dimensional theory which can contribute at order $\beta$ to a scalar potential in the one-dimensional effective theory: The non-topological $R^4$ terms~\eqref*{R4} evaluated on the five-fold background, the kinetic terms $G\wedge\ast G$ for the four-form field strength if flux is non-zero and the volume term in the membrane action~\eqref*{membraneaction} provided wrapped membranes are present. We will now discuss these terms in turn starting with the $R^4$ one. 

Due to its complicated structure, the reduction of this term on a Calabi-Yau five-fold background is not straightforward. Also, this term depends on the unknown four-curvature of the five-fold and the only hope of arriving at an explicit result is that it becomes topological when evaluated on a five-fold background. A fairly tedious, although in principle straightforward calculation shows that this is indeed the case and that it can be expressed in terms of the fourth Chern class, $c_4(X)$, of the five-fold. Explicitly, we find that \eqref{R4} reduces to
\be
 \frac{l\beta_1}{4}\int d\tau\, N\, \frac{1}{24}\,c_{4i}(X)t^i\; ,
\ee
where $\beta_1=(2\pi )^4\beta /v^{4/5}$ and we have expanded the fourth Chern class as $c_4(X)=c_{4i}(X)\tilde{\omega}^i$ into a basis of harmonic $(4,4)$-forms $\tilde{\omega}^i$ dual to the harmonic $(1,1)$-forms $\omega_i$.

Next, we consider the contribution of a membrane wrapping a holomorphic curve $C$ in $X$ with second homology class $W$. Using the explicit parametrisation $X^0=\sigma^0$, $X^\mu =X^\mu (\sigma )$, $X^{\bar{\mu}}=X^{\bar{\mu}}(\bar{\sigma})$, where $\sigma =(\sigma^1+i\sigma^2)/\sqrt{2}$ for the curve $C$, the first term in the membrane action~\eqref*{membraneaction} reduces to
\be
 - \frac{l\beta_1}{4}\int d\tau\, N\, W_it^i\; .
\ee
Here, we have expanded the membrane class as $W=W_i\tilde{\omega}^i$ into our basis of harmonic $(4,4)$-forms.

Finally, we need to consider four-form flux. In terms of the rescaled four-form $g$ (see~\eqref{gdef}) the ansatz for flux can be written as
\be\eqlabel{4form_flux_ansatz}
	g = \oneon{2} n^{\cal{X}} O_{\cal{X}} = n^e\sigma_e+(m^x\varpi_x +{\rm c.c.})\; ,
\ee
where $\{ O_{\cal{X}} \}$ with $\mathcal{X,Y},\ldots=1,\ldots,b^4(X)$ is a basis of real harmonic 4-forms, $\{\sigma_e\}$ with $e,f,\ldots = 1,\dots ,h^{2,2}(X)$ is a basis of real harmonic $(2,2)$-forms, $\{\varpi_x\}$ with $x,y,\ldots = 1,\dots ,h^{1,3}(X)$ is a basis of harmonic $(1,3)$-forms and we used the Hodge decomposition to split a real 4-form into $(1,3)$, $(3,1)$ and $(2,2)$ parts. The factor of $1/2$ has been introduced for convenience in view of the flux quantisation condition~\eqref*{gquant}, which demands that $n^{\cal{X}}$ be an even (odd) integer depending on whether $c_2 (X)$ is even (odd). An essential ingredient in the reduction is the 10-dimensional Hodge dual of $g$. From the results in~\eqref{dualforms} we see that this is given by
\be
 \ast g 	= n^e\left(J\wedge\sigma_e-\frac{i}{2}J^2\wedge\tilde{\sigma}_e+\frac{1}{12}\tilde{\tilde{\sigma}}_eJ^3\right) 
 		-  \left(m^xJ\wedge\varpi_x+{\rm c.c.}\right) \; .
\ee
We recall from \appref{cydg} that $\tilde{\sigma}_e$ is a harmonic $(1,1)$-form which is obtained from $\sigma_e$ by a contraction with the inverse metric $g^{\mu\bar{\nu}}$.  Likewise, $\tilde{\tilde{\sigma}}_e$ is a scalar on $X$, obtained from $\sigma_e$ by contraction with two inverse metrics. Following the discussion in \appref{modspace} these objects can be written as
\be
 \tilde{\sigma}_e=ik_e^i\omega_i\; ,\qquad \tilde{\tilde{\sigma}}_e=-\frac{5}{\kappa}k_e^i\kappa_i\; ,
\ee 
where $k_e^i$ is a set of (moduli-dependent) coefficients.
Combining these results the four-form kinetic term $\frac{1}{2\kappa_{11}^2}\int_{\cal M} (-\oneon{2})G\wedge\ast G$ reduces to
\be
 -\frac{l\beta_1}{4}\int d\tau \frac{N}{2}\left[ n^en^f\left(d_{efi}t^i+\frac{1}{2}k_f^id_{eijk}t^jt^k-\frac{5}{12\kappa}k_e^i\kappa_id_{ejkl}t^jt^kt^l\right)-\left(m^x\bar{m}^{\bar{y}}d_{x\bar{y}i}t^i +{\rm c.c.}\right)\right],
\ee 
where we have used some of the intersection numbers introduced in \appref{modspace}.

We introduce a one-dimensional scalar potential ${\cal U}$ by
\be
 S_{\rm B,pot}=-\frac{l}{4}\int d\tau\, N\, {\cal U}\; . \eqlabel{SBpot}
\ee
This expression should be added to the bosonic action~\eqref*{S1B}. Then, by combining the three contributions above, we find that
\begin{align}
 {\cal U} &= \beta_1\left[\left(\frac{1}{2}(g\wedge g)_{(2,2)i}-\frac{1}{2}(g\wedge g)_{(1,3)i}+W_i-\frac{1}{24}c_{4i}(X)\right)t^i\right. \eqlabel{Vl1}\\
 &\qquad\left.+\frac{1}{4}n^en^fk_f^i\left({\cal G}^{(1,1)}_{ik}-\frac{25}{12}\frac{\kappa_i\kappa_k}{\kappa^2}\right){\cal G}^{(1,1)kj}d_{ejlm}t^lt^m\right]\eqlabel{Vl2}\; .
\end{align} 
Let us pause to discuss this result. The first line is linear in the K\"ahler moduli $t^i$ with coefficients which are almost identical to the components of the anomaly condition~\eqref*{intcond}. In fact, only the sign of $(g\wedge g)_{(1,3)}$, the contribution from the $(1,3)$ part of the flux, is opposite to what it is in the anomaly condition~\eqref*{intcond}. The sign difference between the $(2,2)$ and $(1,3)$ flux parts in \eqref{Vl1} can be traced back to a sign difference in the formul\ae~\eqref*{dualforms} for the Hodge duals which read $\ast\sigma = J\wedge\sigma +\dots$ for $(2,2)$ forms and $\ast\varpi = -J\wedge\varpi$ for $(1,3)$ forms. We have checked this sign difference carefully. We are, therefore, led to conclude that, after using the anomaly condition~\eqref*{intcond}, the first part~\eqref*{Vl1} of the scalar potential reduced to a linear term which depends only on $(1,3)$ flux. As will become clear in the following such a term is not consistent with one-dimensional ${\cal N}=2$ supersymmetry. On the other hand, the second part~\eqref*{Vl2} of the potential which only depends on $(2,2)$ flux can be written in a supersymmetric form, as we will see. Hence, $(2,2)$ flux is consistent with one-dimensional ${\cal N}=2$ supersymmetry while $(1,3)$ flux breaks it explicitly. This conclusion is also supported by analysing the eleven-dimensional Killing spinor equations and the conditions for ${\cal N}=2$ supersymmetry in the presence of fluxes~\cite{Gillard:2004xq}. While there may not be anything wrong with this explicit breaking, we have set out to study M-theory compactifications which preserve one-dimensional ${\cal N}=2$ supersymmetry. We will, therefore, focus on $(2,2)$ flux and set the $(1,3)$ flux to zero in the subsequent discussion.

The decomposition in \eqref{4form_flux_ansatz} of four-form flux into $(1,3)$, $(3,1)$ and $(2,2)$ pieces depends on the complex structure and therefore the condition for unbroken ${\cal N}=2$ supersymmetry, namely that the $(1,3)$ and $(3,1)$ parts of the four-form flux vanish, $g_{(1,3)}=g_{(3,1)}=0$, \emph{a priori} leads to a potential for the complex structure moduli. In other words, the complex structure moduli are only allowed to fluctuate in such a way as to keep the four-form flux purely of $(2,2)$ type. With the decomposition~\eqref*{31_4_rel3} inserted into \eqref{4form_flux_ansatz}, the condition $g_{(1,3)}=0$ becomes
\be\eqlabel{g13_zero_implicit_14_potential}
	m^x (\underline{z},\underline{\bar{z}}) = n^{\cal{X}} \mathfrak{D}_{\cal{X}} {}^x (\underline{z},\underline{\bar{z}}) = 0 \; , 
	\qquad\qquad (\text{and $\cc$}) \; .
\ee
However, it is not known whether the $\mathfrak{D}_{\cal{X}} {}^x$ and hence the resulting potential for the $z^a$ can be calculated explicitly. It is important to recall that in our analysis of bosonic and fermionic 4-form fields we are restricting to Calabi-Yau five-folds that satisfy \eqref{22_vertical} and, in this case, the potential vanishes, that is the complex structure moduli are restored as flat directions in the moduli space, because for such manifolds the split of a four-form into a $(2,2)$-piece and a $(1,3)+(3,1)$-piece is complex structure independent. This can also be seen by noting that in this case the condition~\eqref*{g13_zero_implicit_14_potential} turns into the complex structure independent equation $n^{\cal{\hat{X}}} = 0$, with the help of the decomposition~\eqref*{31_4_new_rel3}. Moreover, the $(2,2)$ flux itself, $g_{(2,2)} = \oneon{2} n^{\cal{\tilde{X}}} O_{\cal{\tilde{X}}} = n^e \sigma_e$, becomes a complex structure independent quantity.

This leaves us with the second part~\eqref*{Vl2} of the scalar potential and, in order to write this into a more explicit form, we need to compute the coefficients $k_e^i$. This has, in fact, been done in \eqref{kei}. Inserting these results and using eqs.~\eqref*{GtGti} and \eqref*{Gti} we finally find for the scalar potential
\be 
 {\cal U}=\frac{1}{2}G^{(1,1)ij}{\cal W}_i{\cal W}_j\; ,\qquad {\cal W}_i=\frac{\partial{\cal W}}{\partial t^i}\; , \eqlabel{UW0}
\ee
where the ``superpotential'' ${\cal W}$ is given by
\be
 {\cal W}(\underline{t})=\frac{\sqrt{\beta_1}}{3}d_{eijk}n^et^it^jt^k\eqlabel{superpot}
\ee
and $G^{(1,1)ij}$ is the inverse of the physical $(1,1)$ moduli space metric \eqref*{G11}.
The fact that the scalar potential can be written in terms of a superpotential in the usual way suggests it can be obtained as the bosonic part of a superfield expression. This is indeed the case and the term we have to add to the superspace action~\eqref*{superspace_action_1d} is simply
\be
	S_{\rm pot} = - \frac{l}{2} \int d\tau \,d^2\theta\,\mathcal{E}\,{\cal W}(\underline{T})\; . \eqlabel{Spot}
\ee 
Indeed, combining this term with the $(1,1)$ kinetic term in the superspace action~ \eqref*{superspace_action_1d} and working out the bosonic component action using \eqref{L11} one finds the terms
\be
	\frac{l}{2}\int d\tau\, \frac{N}{4} \left( G^{(1,1)}_{ij}f^if^j - 2 f^i{\cal W}_i \right) \; ,
\ee
which, after integrating out the $(1,1)$ auxiliary fields
\begin{equation}
f^i=G^{(1,1)ij}\,{\cal W}_i\; , \eqlabel{fW}
\end{equation}
reproduce the correct scalar potential. 

It is, perhaps, at first surprising that the formula~\eqref*{UW0} for the scalar potential in terms of the superpotential looks exactly like the one in global supersymmetry and does not seem to have the usual supergravity corrections such as the analogue of the famous $-3|{\cal W}|^2$ term in four-dimensional $\susyno = 1$ supergravity. However, we have to keep in mind that the physical moduli space metric $G^{(1,1)}_{ij}$ differs from the standard moduli space metric ${\cal G}^{(1,1)}_{ij}$ and it this difference which encodes the supergravity corrections to the scalar potential. Specifically, let us formally introduce a ``K\"ahler covariant derivative'' $D_i{\cal W}={\cal W}_i+\frac{\partial K^{(1,1)}}{\partial t^i}{\cal W}={\cal W}_i-t_i{\cal W}$, where we recall that $K^{(1,1)}=-\frac{1}{2}\ln\kappa$ and we have used \eqref{tli} in the second equality. Moreover, we note that, from \eqref{Gti}, the inverse $G^{(1,1)ij}$ can be written as
\be
 G^{(1,1)ij}=\frac{1}{8V}\left({\cal G}^{(1,1)ij}-\frac{4}{9}t^it^j\right)\; .
\ee
Combining these results and using $\kappa = 5!\, V$ we can also write the scalar potential~\eqref*{UW0} as
\be
 {\cal U}=\frac{15}{2}e^{2K^{(1,1)}}\left({\cal G}^{(1,1)ij}D_i{\cal W}D_j{\cal W}-\frac{1}{2}{\cal W}^2\right)\; ,
\ee  
which resembles the expression for the four-dimensional $\susyno=1$ supergravity potential quite closely.

\vskip 0.4cm

Finally, we should point out that the superpotential~\eqref*{superpot} can be obtained from a Gukov-type formula
\be
	{\cal W} (\underline{t}) = \oneon{3} \int_X G_{\rm flux}\wedge J^3\; .
\ee
This integral is, in fact, the only topological integral, linear in flux,  one can build using the two characteristic forms $J$ and $\Omega$ of the five-fold and $G_{\rm flux}$. In this sense, it is the natural expression for the superpotential. Here, we have explicitly verified by a reduction form $11$ dimensions that it gives the correct answer.

\vskip 0.4cm

When $(2,2)$ flux is non-vanishing, another set of bosonic terms arises from the Chern-Simons term $A\wedge G\wedge G$ in the 11-dimensional action~\eqref*{sugra11}. Writing the complete ansatz for the four-form field strength $G$, including flux and zero modes, one has
\be\eqlabel{Gcompl}
 G = G^{(2,2)}_{\rm flux} + \dot{X}^{\cal{P}} d\tau \wedge N_{\cal{P}} = \frac{2\pi}{T_3}n^e\sigma_e + \dot{X}^{\cal{P}} d\tau \wedge N_{\cal{P}}\; .
\ee
Here, we recall that $\{N_{\cal{P}}\}$, where $\mathcal{P,Q},\ldots =1,\dots ,b^3(X)$, are a basis of real harmonic 3-forms and $X^{\cal{P}}$ are the associated 3-form zero modes. Inserting this ansatz into the 11-dimensional Chern-Simons term one finds
\be\eqlabel{SBCS}
	S_{\rm B,CS} = - \frac{l}{2} \int d\tau\frac{\sqrt{\beta_1}}{3} d_{\mathcal{PQ}e}n^e\dot{X}^{\cal{P}} X^{\cal{Q}} \; ,
\ee
where $d_{\mathcal{PQ}e} = \int_X N_{\cal{P}} \wedge N_{\cal{Q}} \wedge \sigma_e$. Note that~\eqref*{SBCS} is linear in flux and, hence, appears at order $\sqrt{\beta}$. It represents a one-dimensional Chern-Simons term.

\subsection{A closer look at the bosonic action and the scalar potential}

We would now like to discuss some features of the bosonic effective action. To begin with, we summarise our result for the complete bosonic action up to and including order $\beta$. The bosonic action depends on three sets of fields, the real $(1,1)$ moduli $t^i$, the real 3-form moduli $X^{\cal{P}}$ and the complex $(1,4)$ moduli $z^a$. It can be written as a sum of three parts
\be
 S_{\rm B}=S_{\rm B,kin}+S_{\rm B,pot}+S_{\rm B,CS}
\ee
which, from eqs.~\eqref*{S1B}, \eqref*{SBpot}, \eqref*{UW0} and \eqref*{SBCS}, are given by
\begin{align}
	S_{\rm B,kin} &= \frac{l}{2} \int d\tau N^{-1} \left\{ \oneon{4} G_{ij}^{(1,1)}(\underline{t}) \dot{t}^i \dot{t}^j  
		+ \oneon{2} G_{\mathcal{PQ}}^{(3)}(\underline{t},\underline{z},\underline{\bar{z}}) \dot{X}^{\cal{P}} \dot{X}^{\cal{Q}}
		+ 4 V(\underline{t}) G_{a\bar{b}}^{(1,4)} (\underline{z},\underline{\bar{z}}) \dot{z}^a \dot{\bar{z}}^{\bar{b}}  \right\} \; , \\
	S_{\rm B,pot} &= - \frac{l}{4}\int d\tau\, N\, {\cal U} \; , \\
	S_{\rm B,CS} &= - \frac{l}{2} \int d\tau\frac{\sqrt{\beta_1}}{3} d_{\mathcal{PQ}e}n^e\dot{X}^{\cal{P}} X^{\cal{Q}} \; ,
\end{align}		
with the scalar potential ${\cal U}$ and superpotential ${\cal W}$
\be
 {\cal U}=\frac{1}{2}G^{(1,1)ij}{\cal W}_i{\cal W}_j\; ,\qquad {\cal W}(\underline{t})=\frac{\sqrt{\beta_1}}{3}d_{eijk}n^et^it^jt^k\; .
 \eqlabel{UW}
\ee 
The $(1,1)$ metric $G^{(1,1)}_{ij}$ has been defined in \eqref{G11}, the 3-form metric $G_{\mathcal{PQ}}^{(3)}$ in \eqref{G3} and the $(1,4)$ metric $G^{(1,4)}_{a\bar{b}}$ in \eqref{G41}. The first two parts of this action can be schematically written as
\be
 S_{\rm B,kin}+S_{\rm B,pot}= \frac{l}{2}\int d\tau\left\{N^{-1}\,G_{IJ}(\underline{\phi})\dot{\phi}^I\dot{\phi}^J - \frac{N}{2} \,{\cal U}(\underline{\phi})\right\}\; ,
\ee 
where we have collectively denoted the various types fields of fields by $(\phi^I)$$=$$(t^i, X^{\cal{P}}, z^a, \bar{z}^{\bar{b}})$ and $G_{IJ}$ is a block-diagonal metric which contains the above moduli space metrics in the appropriate way. The associated equations of motion then have the general form
\be
	\frac{1}{N}\frac{d}{d\tau}\left(\frac{\dot{\phi}^I}{N}\right) + \Gamma^I_{JK} \frac{\dot{\phi}^J}{N} \frac{\dot{\phi}^K}{N} + 
	\oneon{4} G^{IJ}\frac{\partial{\cal U}}{\partial\phi^J}+C^I=0\; , \eqlabel{eom1d}
\ee
where $\Gamma^I_{JK}$ is the Christoffel connection associated to $G_{IJ}$ and $C^I$ is the contribution from the Chern-Simons term. Since the Chern-Simons term only depends on $X^{\cal{P}}$, we have $C^i=C^a=C^{\bar{b}}=0$.

\vskip 0.4cm

Are there any static solutions, that is, solutions with all $\phi^I={\rm const}$ in the presence of a flux potential? Since the potential ${\cal U}$ only depends on the $(1,1)$ moduli, it is certainly consistent with the equations of motion~\eqref*{eom1d} to set all other fields to constants. For vacua without $(2,2)$ flux (but possibly with membranes) this can also be done for the $(1,1)$ moduli $t^i$. In this case the scalar potential vanishes identically and the moduli space is completely degenerate.

In the presence of $(2,2)$ flux the situation is more complicated. First, one should look for vacua with constant scalars which preserve the ${\cal N}=2$ supersymmetry of the one-dimensional theory. Finding such vacua amounts to setting the supersymmetry variations of all fermions to zero and solving the resulting Killing spinor equations, as usual. For the various $2b$ multiplets the supersymmetry variations of their fermion components vanishes automatically for constant scalar fields and vanishing fermions, as can be seen directly from the results in \appref{localsusy}. On the other hand, the supersymmetry variations of the fermions residing in $2a$ multiplets require a bit more care. For the 3-form fermions $\Lambda^{\cal{P}}$ one has from \eqref{sspace_curved_transf_2a,sspace_curved_transf_2a_bar}
\be
	\delta_\epsilon \Lambda^{\cal{P}} = 0 \; , \qquad
	\delta_\epsilon \bar\Lambda^{\cal{P}} = -\oneon{2}\epsilon g^{\cal{P}} = 0 \; ,
\ee
after inserting the constraint in \eqref{3ferm_constraint_spartner} determining $g^{\cal{P}}$. For the $(1,1)$ fermions $\psi^i$ the transformations lead to
\begin{equation}
  \delta_\epsilon\psi^i = 0\; ,\qquad \delta_\epsilon\bar{\psi}^i = -\oneon{2}\epsilon f^i=-\oneon{2}\epsilon G^{(1,1)ij}\, {\cal W}_j\; ,
\end{equation}
again assuming vanishing fermions and constant scalars. Hence, constant scalar field vacua which preserve ${\cal N}=2$ supersymmetry are characterised by the ``F-term'' equations
\begin{equation}
 {\cal W}_i=0\; .
\end{equation}
\Eqref{UW} shows that solutions to these F-term equations are stationary points of the scalar potential, although, unlike in four-dimensional ${\cal N}=1$ supergravity, they need not be minima since the $(1,1)$ metric $G^{(1,1)}$ is not positive definite. Another interesting difference to four-dimensional supergravity is that the scalar potential always vanishes for solutions of the F-term equations. 

Let us now consider explicit examples to see whether the F-term equations have interesting solutions for our examples. From the general form of ${\cal W}$ in \eqref{UW}, it is clear that for a single $(1,1)$ modulus, that is, $h^{1,1}(X)=1$, the only solution to the F-term equations is $t^1=0$. This corresponds to vanishing Calabi-Yau volume so we should certainly not trust our one-dimensional effective theory at this point. Moving on to Calabi-Yau manifolds with $h^{1,1}=2$ we start with the second example in \tabref{tab:cicy2}, a co-dimension one CICY in the ambient space ${\cal A}=\mathbb{P}^3\times\mathbb{P}^3$ with configuration matrix
\begin{equation}
X\sim\left[\begin{array}{l}3\\3\end{array}\right|\left.\begin{array}{l}4\\4\end{array}\right]\; .
\end{equation}
The discussion in \appref{cyexamples} shows that the anomaly condition for this CICY can be satisfied for a range of fluxes and an appropriate number of membranes. Since $h^{2,2}(X)=3$, we have three flux parameters $n_1$, $n_2$, $n_3$ and flux can explicitly be written as $g=n_1J_1^2+n_2J_1J_2+n_3J_2^2$. Then, one finds for the K\"ahler potential and superpotential
\begin{equation}
\kappa = 40 t_1^3t_2^2+40t_1^2t_2^3\; ,\qquad {\cal W}=\frac{4}{3}n_1t_2^3+4(n_1+n_2)t_1t_2^2+\frac{4}{3}n_3t_1^3+4(n_2+n_3)t_1^2t_2\; .
\end{equation} 
It is easy to see that setting, for example, $n_1=n_3=3$ and $n_2=-4$ the F-term equations are satisfied along the flat direction $t_1=t_2$. Moreover, this flat direction consists of minima of the potential with zero cosmological constant. The existence of such minima is of considerable importance for our M-theory compactifications. A general problem of compactifications with flux is the tendency of producing large potential energies above the compactification scale due to the quantised nature of flux. Such high scales of potential energy are of course problematic as they invalidate the low-energy effective theory. We have just seen an example where this problem can be avoided due to a flat direction with vanishing vacuum energy in the two-dimensional K\"ahler moduli space. This means, at least to first order in our $\beta$ expansion, self-consistent five-fold compactifications of M-theory with $(2,2)$ flux exist.

As the next example shows this is by no means automatic. Consider the first example in \tabref{tab:cicy2},  a co-dimension one CICY in the ambient space ${\cal A}=\mathbb{P}^1\times\mathbb{P}^5$ with configuration matrix
\begin{equation}
X\sim\left[\begin{array}{l}1\\5\end{array}\right|\left.\begin{array}{l}2\\6\end{array}\right]\; . \eqlabel{cicy260}
\end{equation}
As in the previous example, the anomaly condition can be satisfied for a range of fluxes and with the appropriate number of membranes (see the discussion around \eqref{cicy26}). This manifold has $h^{2,2}(X)=2$ and the flux can be written as $g=n_1J_1J_2+n_2J_2^2$ with two flux parameters $n_1$ and $n_2$. The K\"ahler potential and superpotential for the model are given by
\be 
 \kappa = 30 t_1t_2^4+2t_2^5\; ,\qquad W=6n_2t_1t_2^2+\left(2n_1+\frac{2}{3}n_2\right)t_2^3\; . \eqlabel{kw26}
\ee 
In this case, the F-term equations imply that $t_2=0$ and the above expression for the K\"ahler potential shows that the Calabi-Yau volume vanishes for this value. Hence, there is no viable supersymmetric minimum in this case.

\vskip 0.4cm

We should now discuss the scalar potential in some of the cases where solutions to the F-term equations cannot be found.
In general, we note that under a rescaling $t^i\rightarrow \lambda t^i$ the $(1,1)$ metric scales as $G^{(1,1)}_{ij}(\lambda\underline{t})=\lambda^3G^{(1,1)}_{ij}(\underline{t})$ and the superpotential as ${\cal W}(\lambda\underline{t})=\lambda^3{\cal W}(\underline{t})$. This means that the scalar potential scales as ${\cal U}(\lambda\underline{t})=\lambda{\cal U}(\underline{t})$, so is homogeneous with degree one. When discussing the implications of this scaling behaviour it has to be kept in mind that the metric $G^{(1,1)}_{ij}$ has signature $(-1,+1,\ldots  ,+1)$ with the negative direction $u^i$ given by $u^i\sim t^i$. Whether this negative direction is ``probed'' by the scalar potential depends on the structure of the superpotential and its derivatives. If it is, the potential will be of the form ${\cal U}=-c\lambda$, where $c$ is a positive constant. This indicates an instability which leads to a rapid growth of the Calabi-Yau volume and decompactification. Clearly, this is  always the case for examples  with $h^{1,1}(X)=1$ where the metric is just a negative number. For $h^{1,1}(X)>1$ the picture is less clear and what happens depends on the choice of Calabi-Yau manifold and flux.

Let us consider two explicit examples. At the end of section~\ref{sec:5-folds} we have discussed how to satisfy the anomaly condition for the septic, $[6|7]$, by a combination of flux and membranes. For this case we have
\be
 \kappa = 7t^5\; ,\qquad {\cal W}=\frac{35}{2}\sqrt{\beta_1}t^3\; ,
\ee 
where $t$ is the single $(1,1)$ modulus. After a short computation, using eqs.~\eqref*{G11} and \eqref*{UW} this leads to the scalar potential
\be
 {\cal U}=-\frac{525}{4}\beta_1t\; . \eqlabel{Uquintic}
\ee
As expected, the potential is negative and results in a rapid growth of the volume. Our compactification can only be trusted for large Calabi-Yau volume, that is $t \gg 1$. In this case the scale of the scalar potential~\eqref*{Uquintic} is quite large and it is questionable if we can trust our low-energy theory.  

For an example with $h^{1,1}(X)=2$ we return to the manifold in \eqref{cicy260} with K\"ahler and superpotential as in \eqref{kw26} which did not exhibit F-flat directions. We find for the scalar potential
\be
 {\cal U}=\frac{\beta_1(15n_1+2n_2)}{6(15t_1+t_2)}\left((3n_1-2n_2)t_2^2-36n_2t_1t_2\right)\; .
\ee 
We recall that the K\"ahler cone of this CICY is given by $t_1>0$ and $t_2>0$. Now choose the fluxes to be $n_1=0$ and $n_2=1/2$. Then the above potential is strictly negative in the K\"ahler cone of $X$ and such that both $t_1$ and $t_2$ will grow. For $n_1=1$ and $n_2=-1/2$, on the other hand, the above potential is strictly positive in the K\"ahler cone. Gradients are such that $t_2$ contracts and, as a result, the total volume goes to zero (while $t_1$ slowly expands). As for the septic, for large volume, $t_1\gg 1$, $t_2\gg 1$, the scalar potential is large and it is not clear that the low-energy theory is valid.

In summary, a first look at the one-dimensional effective theory at order $\beta$ indicates a number of possibilities to obtain self-consistent compactifications with vanishing vacuum energy. First of all, for some Calabi-Yau five-folds the anomaly condition can be satisfied without the inclusion of flux, either if $c_4(X)=0$ or if a non-zero $c_4(X)$ can be compensated for by membranes, and, in this case, the scalar potential vanishes identically. An interesting general feature of the scalar potential is that it vanished for supersymmetric vacua, that is, for solutions to the F-term equations. We have shown that such solutions to the F-term equations do indeed exist for some five-folds and that they correspond to flat directions of the potential. The general structure of the scalar potential means that the vacuum energy vanishes along those flat directions.
If supersymmetric flat directions do not exist, the scalar potential, which is homogeneous of degree one in the K\"ahler moduli, is generally large for large volume and it is questionable whether one can trust the effective theory. Taken at face value, this scalar potential may either lead to a rapid expansion or a rapid contraction of the Calabi-Yau volume, depending on the case. Calabi-Yau five-folds with $h^{1,1}=1$ such as the septic do not have F-flat directions and always contract. For $h^{1,1}>1$, supersymmetric flat directions may or may not exist. If they do not exist, one can have rapid expansion or contraction, depending on the choice of Calabi-Yau manifolds and flux.

\section{Conclusion and Outlook}\label{sec:concl}

In this paper, we have considered compactifications of M-theory on Calabi-Yau five-fold backgrounds, leading to one-dimensional effective theories with ${\cal N}=2$ supersymmetry. In the absence of flux and membranes, such five-fold backgrounds are solutions to M-theory at zeroth order in the $\beta\sim\kappa_{11}^{4/3}$ expansion of the theory but at first order in $\beta$ one encounters a non-trivial consistency condition~\eqref*{intcond}. This condition ensures the absence of a gauge anomaly of the M-theory three-form $A$ on a five-fold background. It requires a cancellation between the fourth Chern class, $c_4(X)$, of the Calabi-Yau five-fold $X$, the square, $G \wedge G$, of the flux $G$ and the charge, $W$, of a membrane wrapping a holomorphic curve in $X$.

We have studied explicit examples of Calabi-Yau five-folds to check whether and how this condition can be satisfied. The simplest possibility is to use a five-fold satisfying $c_4(X)=0$, without any membranes or flux. We have constructed an explicit example of such a five-fold with vanishing fourth Chern class, based on a quotient of a $10$--torus by a freely-acting $\mathbb{Z}_2^4$ symmetry. Although such a torus quotient has merely $\mathbb{Z}_2^4$ rather than ${\rm SU}(5)$ holonomy, it still breaks supersymmetry by a factor of $16$
and, hence, all our subsequent results apply to this example. As another class of examples, we have studied complete intersection Calabi-Yau five-folds (CICY five-folds) which are defined as the common zero locus of homogeneous polynomials in a projective space or a product of projective spaces. The simplest example of such a CICY five-fold is the septic in $\mathbb{P}^6$, the analogue of the famous quintic Calabi-Yau three-fold in $\mathbb{P}^4$. We have shown for a wide range of CICY five-folds that $c_4(X)\neq 0$ and it is conceivable that this holds for all CICY five-folds. It remains an open question as to whether Calabi-Yau five-folds with full ${\rm SU}(5)$ holonomy and $c_4(X)=0$ exist, for example among toric five-folds. 

For CICY five-folds we have shown that the anomaly condition can frequently be solved by a cancellation between $c_4(X)\neq 0$ and appropriate flux and/or membranes. In particular, this can be achieved for the septic in $\mathbb{P}^6$ when both flux and membranes are included. Given the large number of topologically different Calabi-Yau five-folds and the fact that many of the simplest examples can already be made to work we can expect a large and rich class of consistent M-theory five-fold compactifications. It is for such anomaly-free compactifications that we have set out to compute the associated one-dimensional ${\cal N}=2$ effective theory. To this end, we have developed the general properties of  Calabi-Yau five-folds with regards to their topology, differential geometry and moduli spaces. In particular, there are six a priori independent Hodge numbers, $h^{1,1}(X)$, $h^{1,2}(X)$, $h^{1,3}(X)$, $h^{2,2}(X)$, $h^{1,4}(X)$ and $h^{2,3}(X)$.
However, the Calabi-Yau condition $c_1(X)=0$ together with the index theorem leads to one linear relation between those six Hodge numbers, so that only five of them are effectively independent. 

M-theory zero modes on five-folds can be classified according to the sector of harmonic $(p,q)$ forms they are related to. For the bosonic zero modes, we have metric K\"ahler moduli, related to the $(1,1)$ sector and the metric complex structure moduli, related to the $(1,4)$ sector. Further bosonic zero modes in the $(2,1)$ sector arise from the three-form $A$. All these bosonic zero modes have associated fermionic partners which originate from the same $(p,q)$ sector of the five-fold. In addition, we also find $(1,3)$ fermionic zero modes that do not have any bosonic partners, a feature which seems at first puzzling from the viewpoint of supersymmetry.

After identifying these zero modes, we have reduced both the bosonic and fermion bilinear terms in 11 dimensions to obtain the one-dimensional effective action, initially at zeroth order in the $\beta$ expansion. In order to understand the supersymmetry of this effective action, we have systematically developed one-dimensional global and local ${\cal N}=2$ superspace, extending previously known results. Based on these results, it turned out that the $(1,4)$ zero modes reside in $2b$ multiplets while the $(1,1)$ multiplets reside in $2a$ multiplets. The complex $(2,1)$ zero modes are best described collectively as real 3-form fields forming $2a$ multiplets and subject to a constraint halving the number of fermions. This was necessary in order to keep under control the otherwise intricate intertwining with the complex structure moduli. It was found that the fermionic $(1,3)$ zero modes are compatible with supersymmetry. However, the complex structure moduli also intertwine with those modes. For this sector, we restricted our analysis to five-folds whose $(2,2)$-forms are generated by the product of two $(1,1)$-forms. The fermionic $(1,3)$ and $(3,1)$ modes together, or $\hat{4}$-form modes for short, could then be described by constrained fermionic $2b$ multiplets. For all those multiplets, we have then written down a non-linear supersymmetric sigma model in superspace  and we have verified that the component version of this sigma model precisely reproduces our reduction result from $11$ dimensions. Interesting properties of this sigma model are the ``non-standard'' form of the $(1,1)$ sigma model metric which differs from the standard Calabi-Yau moduli space metric and the mixing between $2a$ and $2b$ multiplets. We also stress that {\em local} one-dimensional ${\cal N}=2$ supersymmetry is required in order to properly describe the constraints which are the remnants of (super)gravity in one dimension.

In a next step we have extended our results to order $\beta$ effects and we have computed the one-dimensional scalar potential which arises at this order. After imposing the anomaly condition, it turns out that this potential has two parts, depending on $(1,3)$ and $(2,2)$ flux, respectively. We have not been able to find a supersymmetric description of the $(1,3)$ part of this scalar potential and we conclude that $(1,3)$ flux is not compatible with one-dimensional ${\cal N}=2$ supersymmetry. Since this is a complex structure dependent statement, keeping full ${\cal N}=2$ supersymmetry in the presence of non-zero four-form flux induces restrictions on the complex structure moduli. The explicit form of these restrictions and how they can be implemented, for example in terms of a potential, is not known. In order to nonetheless make concrete statements about flux, we therefore focussed on Calabi-Yau five-folds for which $(1,3)$ flux can be set to zero without imposing restrictions on the complex structure moduli. In particular, we restricted to Calabi-Yau five-folds whose $(2,2)$-forms are completely generated by wedging together two $(1,1)$-forms. All the explicit examples of Calabi-Yau five-folds presented in this paper are of this type. In this case, the $(2,2)$ flux potential is complex structure independent and allows for a fully supersymmetric description. It is associated to the auxiliary fields in the $(1,1)$ $2a$ multiplets and is given in terms of a superpotential which only depends on $(1,1)$ moduli and is cubic in those fields. We find that this superpotential can also be directly obtained from a Gukov-type formula. 

A first look at the properties of the effective theory suggests different possibilities for self-consistent compactifications with small, or rather vanishing vacuum energy. For compactifications without flux (but possibly with membranes) the potential vanishes identically. In the presence of $(2,2)$ flux and depending on the case, there may be supersymmetric flat directions with vanishing vacuum energy. The property of zero vacuum energy for supersymmetric solutions is facilitated by the form of the scalar potential which vanishes for vanishing F-terms. Models with flux but without flat directions have a rather large scalar potential and it is not clear if the effective theory can be trusted. Na\"{\i}vely, such scalar potentials can lead to a rapid expansion or contraction of the Calabi-Yau volume, depending on the Calabi-Yau manifold and the choice of fluxes. We have constructed explicit examples for all these cases.

Our results open up a whole range of applications, particularly in the context of moduli space ``cosmology''. For example, the question as to whether the system can evolve towards a state with three large and seven small spatial dimensions can be studied as a dynamical problem in the five-fold moduli space. The effect of a scalar potential, from flux or other, non-perturbative sources not discussed in the present paper, is of course crucial in such a discussion. Another interesting aspect of such a cosmological analysis might be the study of various types of topological phase transitions for Calabi-Yau five-folds. 
These and related issues are centred around the question of how a one-dimensional theory can evolve to effectively become four-dimensional and thereby become a viable description of the ``late'' universe. Such a question might even be studied in a ``mini-superspace'' quantised version of our one-dimensional theory.

There are also a number of more theoretical issues in relation to our results. It would be interesting to find the ``uplift'' of certain solutions to our one-dimensional theory by studying supersymmetric solutions to the 11-dimensional theory based on Calabi-Yau five-folds. In particular, our results for the flux scalar potential indicate that such solutions should not exist in the presence of $(1,3)$ flux. Another interesting aspect concerns the possibility of mirror symmetry for Calabi-Yau five-folds. One might speculate that M-theory on five-folds is mirror symmetric to F-theory on five-folds (times a circle). Both compactifications lead to one-dimensional ${\cal N}=2$ supersymmetric theories in one dimension and a first test for mirror symmetry would be provided by a comparison of the one-dimensional theories derived in the present paper with the ones obtained from an F-theory reduction on five-folds. 
Several of these problems are currently under investigation.

\acknowledgments

The authors are very grateful to A.~Barrett for collaboration in the early stages of this work. We would also like to thank P.~Candelas, X.~de~la~Ossa and G.~Papadopoulos for discussions and we are grateful to the referee for careful reading and constructive comments. A.S.H.~acknowledges the award of a postgraduate studentship by the Institute for Mathematical Sciences, Imperial College London, and thanks the Albert-Einstein-Institute for hospitality and generous financial support. A.L.~is supported by the EC 6th Framework Programme MRTN-CT-2004-503369 and would like to thank the Albert-Einstein-Institute for hospitality. The research of K.S.S.~was supported in part by the EU under contract MRTN-CT-2004-005104, by the STFC under rolling grant PP/D0744X/1 and by the Alexander von Humboldt Foundation through the award of a Research Prize. K.S.S.~would like to thank the Albert-Einstein-Institute and CERN for hospitality during the course of the work.

\section*{Appendix}
\appendix

\section{Index conventions and spinors}\applabel{conv}

In this section, we summarise notations and conventions used throughout the paper. Indices for space-time or superspace in the various relevant dimensions are listed in \tabref{spaceindices}.
\TABLE[ht]{
\begin{tabular}{|p{2cm}|p{2cm}|p{7.5cm}|} \hline
	\emph{symbols}							& \emph{range}			& \emph{meaning} \\ \hline\hline
	$A,B,C,\ldots$								& $0,\theta,\bar\theta$		& one-dimensional $\susyno=2$ superspace indices \\
	$M,N,P,\ldots$								& $0,\ldots,10$				& $D=11$ space-time indices \\
	$m,n,p,\ldots$							 	& $1,\ldots,10$				& $D=10$ Euclidean indices \\
	$\mu,\nu,\ldots$							& $1,\ldots,5$				& $D=10$ Euclidean holomorphic indices \\
	$\bar{\mu},\bar{\nu},\ldots$					& $\bar{1},\ldots,\bar{5}$		& $D=10$ Euclidean anti-holomorphic indices \\
	\hline
\end{tabular}
\caption{Curved space-time indices and superspace indices. Tangent space indices are denoted by the same letters but are underlined.}
\tablabel{spaceindices}}
Indices in this table are curved indices and we refer to their corresponding tangent space indices by underlining the same set of letters. Multiple indices are always symmetrized or anti-symmetrized with weight one. In addition, we need a range of index types for the various cohomology groups of Calabi-Yau five-folds. They are listed in \tabref{modindices}.
\TABLE[ht]{
\begin{tabular}{|p{2cm}|p{2cm}|p{7.5cm}|} \hline
	\emph{symbols}		& \emph{range}		& \emph{meaning} \\ \hline\hline
	$i,j,\ldots$				& $1,\ldots,h^{1,1}$ 		& $(1,1)$-moduli \\
	$p,q,\ldots$			& $1,\ldots,h^{2,1}$		& $(2,1)$-moduli \\
	$x,y,\ldots$			& $1,\ldots,h^{1,3}$		& $(1,3)$-moduli \\
	$e,f,\ldots$			& $1,\ldots,h^{2,2}$		& $(2,2)$-flux \\
	$a,b,\ldots$			& $1,\ldots,h^{1,4}$		& $(1,4)$-moduli \\
	$\mathcal{P,Q},\ldots$	& $1,\ldots,b^3$		& 3-form moduli \\
	$\mathcal{X,Y},\ldots$	& $1,\ldots,b^4$		& 4-form moduli \\
	$\mathcal{\hat{X},\hat{Y}},\ldots$ & $1,\ldots,2h^{1,3}$ & $\hat{4}$-form moduli ($\hat{4}=(1,3)+(3,1)$)\\
	$\mathcal{A,B},\ldots$	& $1,\ldots,b^5$		& 5-form periods \\	
	\hline
\end{tabular}
\caption{Indices for Calabi-Yau five-fold cohomology.}
\tablabel{modindices}}
For the index types $p,q,\ldots$, $x,y,\ldots$ and $a,b,\ldots$, the barred versions are also present and are used to label the complex conjugate of the respective moduli fields.

\vskip 0.4cm
We now turn to our spinor conventions and start in 11 dimensions. We denote the 11-dimensional coordinates by $x^M$ and choose the 11-dimensional Minkowski metric $\eta_{\underline{M}\underline{N}}$ to have mostly plus signature, so $\eta_{\underline{M}\underline{N}} = \diag (-1,+1,\ldots,+1)$. The eleven dimensional gamma matrices $\Gamma^{\underline{M}}$ satisfy the Clifford algebra
\be\eqlabel{cliff11}
	\{ \Gamma^{\underline{M}}, \Gamma^{\underline{N}} \} = 2 \eta^{\underline{M}\underline{N}} \id_{32\times 32} .
\ee
Dirac spinors $\Psi$ in $11$ dimensions have $32$ complex components and are anti-commuting objects. We are working in the Majorana representation in which the charge conjugation matrix is equal to~$\id$ so that Majorana spinors $\Psi$ are real, that is $\Psi^\ast = \Psi$. The gamma matrices in this representation are also real, $(\Gamma^{\underline{M}})^\ast = \Gamma^{\underline{M}}$, and all spatial gamma matrices are symmetric, $(\Gamma^{\underline{m}})^T = \Gamma^{\underline{m}}$, whereas the timelike gamma matrix is anti-symmetric, $(\Gamma^{\underline{0}})^T = - \Gamma^{\underline{0}}$. These properties combine into the following formul\ae:
\be\eqlabel{gamma_11d_dagger}
	(\Gamma^{\underline{M}})^\dagger = \Gamma^{\underline{0}} \Gamma^{\underline{M}} \Gamma^{\underline{0}} \qquad\text{and}\qquad
	(\Gamma^{\underline{M}})^T = \Gamma^{\underline{0}} \Gamma^{\underline{M}} \Gamma^{\underline{0}} .
\ee
Curved gamma matrices $\Gamma^M$ are constructed by contracting with an inverse vielbein $\Gamma^M = e_{\underline{N}}^M \Gamma^{\underline{N}}$.

In 10 Euclidean dimensions with coordinates $x^m$  we introduce complex coordinates by
\be\eqlabel{real_complex_coords}
 z^\mu =\frac{1}{\sqrt{2}}\left(x^\mu+i\, x^{\mu +5}\right)\; ,\qquad 
 \bar{z}^{\bar{\mu}} =\frac{1}{\sqrt{2}}\left(x^{\bar{\mu}}-i\, x^{\bar{\mu} +5}\right)\; .
\ee
Tensors transform from real to complex coordinates accordingly.

The 10-dimensional gamma matrices $\gamma^{\underline{m}}$ satisfying the Clifford algebra
\be
	\{ \gamma^{\underline{m}}, \gamma^{\underline{n}} \} = 2 \delta^{\underline{m}\underline{n}} \id_{32\times 32}\,  .
	\eqlabel{Cliff10}
\ee
In accordance with our 11-dimensional conventions they are chosen to be real matrices and are, hence, also symmetric.
The ten dimensional chirality operator $\gamma^{(11)}$ is given by
\be
	\gamma^{(11)} = i \gamma^{\underline{1}} \cdots \gamma^{\underline{10}}  \eqlabel{gamma11},
\ee
and it satisfies the relations $(\gamma^{(11)})^2 = \id_{32\times 32}$, $(\gamma^{(11)})^\ast = - \gamma^{(11)}$, $(\gamma^{(11)})^T = - \gamma^{(11)}$ and $\{\gamma^{(11)} , \gamma^{\underline{m}} \}$ $=$ $0$. Ten-dimensional Dirac spinors $\eta$ are $32$-dimensional complex, as in $11$ dimensions, and are taken to be commuting. Positive (negative) chirality spinors $\eta$ are then defined by $\gamma^{(11)}\eta = \eta$ ($\gamma^{(11)}\eta = -\eta$). Written in complex coordinates the anti-commutation relations for the gamma matrices read
\be\eqlabel{cy5_hol_cliff_alg}
	\{ \gamma^{\underline{\mu}}, \gamma^{\bar{\underline{\nu}}} \} = 2\, \delta^{\underline{\mu}\bar{\underline{\nu}}} \id_{32\times 32} , \qquad\qquad 
	\{ \gamma^{\underline{\mu}},\gamma^{\underline{\nu}} \} = \{ \gamma^{\bar{{\underline{\mu}}}}, \gamma^{\bar{\underline{\nu}}} \} = 0\; .
\ee
As usual, the gamma matrices in complex coordinates can be interpreted as creation and annihilation operators. If one defines a ``ground state'' $\eta$ by
\be
 \gamma^{\bar{\mu}}\eta =0
\ee
then $\eta$ has positive and $\eta^\star$ negative chirality. The other spinor states are obtained by acting with up to five creation operators $\gamma^\mu$ on $\eta$.

Finally, in one dimension, there is only one gamma matrix, a $1\times 1$ matrix, which we take to be~$-i$. One-dimensional Dirac spinors $\psi$ are complex one-component anti-commuting objects and we often denote their complex conjugate by $\bar{\psi} := (\psi)^\ast$. Spinorial differentiation and Berezin integration are the same operations and satisfy the relations
\bea
	&\partial_\theta \theta = 1 , \;\; \partial_\theta \bar\theta = 0 , \;\; \partial_{\bar\theta} \theta = 0 , \;\; \partial_{\bar\theta} \bar\theta = 1 , \\
	&\partial_\theta^2 = 0, \;\; \partial_{\bar\theta}^2 = 0, \;\; \{ \partial_\theta, \partial_{\bar\theta} \} = 0 \; ,
\eea
where $\partial_\theta := \partial/\partial\theta$ and $\partial_{\bar\theta} := \partial/\partial\bar\theta = - (\partial_\theta)^\ast$. Complex conjugation of a product of two anti-commuting objects is defined to be $(\psi_1 \psi_2)^\ast = \bar{\psi}_2 \bar{\psi}_1$. Note the change of order on the right hand side. The rules for Berezin integration can be read off by replacing $\partial_\theta \rightarrow \int d\theta$ and $\partial_{\bar\theta} \rightarrow \int d\bar\theta$. We also abbreviate $d^2 \theta = d\theta d\bar\theta$ so that
\be
 \int d^2\theta\, \theta\bar\theta = -1\; .
\ee

\vskip 0.4cm

The relation between $11$-, $10$- and one-dimensional gamma matrices is summarised by the decomposition
\be\eqlabel{gamma_matrix_dim_red_split}
	\Gamma^\UZ = (- i) \otimes \gamma^{(11)} , \qquad \Gamma^{\underline{m}} = \id_{1\times 1} \otimes \gamma^{\underline{m}} ,
\ee
where the tensor product between a complex number and a $32\times 32$ matrix has been introduced solely to make contact with similar formul\ae~for compactifications to more than one dimension. As can be checked quickly, the matrices~\eqref*{gamma_matrix_dim_red_split} indeed satisfy the 11-dimensional anti-commutation relations \eqref*{cliff11,gamma_11d_dagger}, provided the $\gamma^{\underline{m}}$ satisfy the 10-dimensional anti-commutation relations~\eqref*{Cliff10}. Dirac spinors $\Psi$ in 11-dimensions can be written as (linear combinations of) tensor products of the form $\psi\otimes\eta$, where $\psi$ and $\eta$ are one- and $10$-dimensional spinors, respectively.  An 11-dimensional Majorana spinor $\Psi$ can be decomposed as
\begin{equation}
 \Psi=\psi\otimes\eta +\bar{\psi}\otimes\eta^\star\; .
\end{equation}


\section{Calabi-Yau five-folds}\applabel{cy5}
In this appendix, we develop the necessary tools to deal with Calabi-Yau five-folds and present some examples relevant to our discussion in the main text. Of course, much of the formalism will be analogous to Calabi-Yau three-folds and four-folds and we will borrow heavily from the literature, particularly from Refs.~\cite{Candelas:1990pi,Hubsch:1992nu,Candelas:1987is}.

\subsection{Basic topological properties}
For the purpose of this paper, we define a Calabi-Yau five-fold to be a five complex-dimensional compact K\"ahler manifold $X$ with vanishing first Chern class, $c_1(X)=0$, and holonomy ${\rm Hol}(X)\subset{\rm SU}(5)$ sufficiently large to allow only one out of $16$ supersymmetries. By the last condition we mean that in the decomposition
\begin{equation}
 {\bf 16}_{{\rm Spin}(10)}\rightarrow \left[{\bf 10}+\bar{\bf 5}+{\bf 1}\right]_{{\rm SU}(5)}
\end{equation} 
of the ${\bf 16}$ spinor representation under ${\rm SU}(5)$ only the ${\rm SU}(5)$ singlet is invariant under ${\rm Hol}(X)$. Hence, for positive chirality, we have precisely one covariantly constant spinor $\eta$. In particular, this means that $10$-dimensional tori, direct products such as between three-folds and four tori and similar spaces are excluded from our considerations. The correspondence between covariantly constant spinors and harmonic $(0,p)$ forms then implies that the Hodge numbers of $X$ are constrained  by $h^{0,p}(X)=h^{p,0}(X)=0$ for $p=1,2,3,4$ and $h^{0,0}(X)=h^{0,5}(X)=h^{5,0}(X)=h^{5,5}(X)=1$. Consequently, the Hodge diamond of a Calabi-Yau five-fold has the following general form
\begin{equation}
 \begin{array}{ccccccccccc}
           &          &          &          &          &     1   &          &          &          &          &            \\
           &          &          &          &    0    &          &   0     &         &          &           &            \\
           &          &          &    0    &          &h^{1,1}&        &   0    &          &          &             \\ 
           &          &   0    &           &h^{1,2}&          &h^{1,2}&     &  0      &         &              \\     
           &  0      &         &h^{1,3}&          &h^{2,2}&        &h^{1,3}&     &     0    &              \\ 
    1    &          &h^{1,4}&          &h^{2,3}&          &h^{2,3}&        &h^{1,4}&    &    1     \\
          &  0      &         &h^{1,3}&          &h^{2,2}&        &h^{1,3}&     &     0    &              \\ 
           &          &   0    &           &h^{1,2}&          &h^{1,2}&     &  0      &         &              \\     
           &          &          &    0    &          &h^{1,1}&        &   0    &          &          &             \\ 
           &          &          &          &    0    &          &   0     &         &          &           &            \\
           &          &          &          &          &     1   &          &          &          &          &
 \end{array}\eqlabel{hodgediamond}
\end{equation} 
with six, a priori independent Hodge numbers. For the Betti numbers $b^i(X)$ this implies
\begin{equation}
\begin{array}{lllllll}
 b^0(X)&=&1&\quad&b^1(X)&=&0\nonumber\\
 b^2(X)&=&h^{1,1}(X)&\quad&b^3(X)&=&2h^{1,2}(X)\nonumber\\
 b^4(X)&=&2h^{1,3}(X)+h^{2,2}(X)&\quad&b^5(X)&=&2+2h^{1,4}(X)+2h^{2,3}(X)\nonumber
 \end{array}
\end{equation} 
and $b^i(X)=b^{10-i}(X)$ for $i>5$. The Euler number $\eta (X)$ of $X$ can, therefore, be written as
\begin{equation}
 \eta (X)\equiv\sum_{i=0}^{10}(-1)^ib^i(X)=2h^{1,1}(X)-4h^{1,2}(X)+4h^{1,3}(X)+2h^{2,2}(X)-2h^{1,4}(X)-2h^{2,3}(X)\; .\eqlabel{Euler0}
\end{equation} 
For Calabi-Yau four-folds it is known~\cite{Sethi:1996es} that one additional relation between the Hodge numbers can be derived by an  index theorem calculation using the Calabi-Yau condition $c_1(X)=0$.
As we will now see, a similar procedure can be carried out for Calabi-Yau five-folds. First recall the general form of the index theorem
\be
 \chi (X,V)\equiv\sum_{i=0}^{{\rm dim}(X)}(-1)^i{\rm dim}H^i(X,V)=\int_X{\rm ch}(V)\wedge{\rm Td}(TX)\; , \eqlabel{indtheorem}
\ee
for a vector bundle $V$ on $X$. We would now like to apply this theorem to the specific bundles $V=\wedge^qT^\star X$, where $q=0,1,2,3$. The cohomology groups of these bundles can be written as $H^i(X,V)=H^i(X,\wedge^qT^\star X)\simeq H^{i,q}(X)$ and they are, hence, directly related to the Hodge numbers of $X$. 
For the subsequent calculation, it is convenient to use the splitting principle and write the Chern class and character of the tangent bundle as
\begin{equation}
 c(TX)=1+c_1(TX)+c_2(TX)+\dots = \prod_i(1+x_i)\; ,\quad {\rm ch}(TX)=\sum_ie^{x_i}\; . \eqlabel{cch}
\end{equation}
Then we have
\be
 {\rm ch}(\wedge^qT^\star X)\wedge {\rm Td}(TX)=\sum_{i_1>i_2>\dots >i_q}e^{-x_{i_1}}\dots e^{-x_{i_q}}\prod_j\frac{x_j}{1-e^{-x_j}}\; .
 \ee
Expanding this expression and re-writing it in terms of Chern classes using \eqref{cch} we find from the index theorem~\eqref*{indtheorem}
\begin{align*}
 \chi_0&= h^{0,0}-h^{1,0}+h^{2,0}-h^{3,0}+h^{4,0}+h^{5,0}\\&=\frac{1}{1440}\int_X\left[-c_2c_1^3+c_1^2c_3-c_1c_4+3c_1c_2^2\right]\nonumber\\
 \chi_1&= h^{0,1}-h^{1,1}+h^{2,1}-h^{3,1}+h^{4,1}-h^{5,1}\\&=\frac{1}{480}\int_X\left[-c_1^3c_2+c_1^2c_3-21c_1c_4+3c_1c_2^2-20c_5\right]\nonumber\\
 \chi_2&= h^{0,2}-h^{1,2}+h^{2,2}-h^{3,2}+h^{4,2}-h^{5,2}\\&=\frac{1}{720}\int_X\left[-c_1^3c_2+c_1^2c_3-31c_1c_4+3c_1c_2^2+330 c_5\right]\nonumber\\
 \chi_3&= h^{0,3}-h^{1,3}+h^{2,3}-h^{3,3}+h^{4,3}-h^{5,3}\\&=-\frac{1}{720}\int_X\left[-c_1^3c_2+c_1^2c_3-31c_1c_4+3c_1c_2^2+330 c_5\right]\nonumber
\end{align*}
where we have used the short-hand notation $\chi_q=\chi (X,\wedge^qT^\star X)$ and $c_i=c_i(TX)$. Inserting the non-trivial information about Hodge numbers from the Hodge diamond~\eqref*{hodgediamond} together with $c_1(X)=0$ the above equation for $\chi_0$ is trivially satisfied while the one for $\chi_3$ is equivalent to the one for $\chi_2$. The remaining two relations for $\chi_1$ and $\chi_2$ turn into
\bea
 \chi_1&= -h^{1,1}+h^{1,2}-h^{1,3}+h^{1,4}=-\frac{1}{24}\int_Xc_5 \; , \\ 
 \chi_2&= -h^{1,2}+h^{2,2}-h^{2,3}+h^{1,3}=\frac{11}{24}\int_Xc_5 \; .
\eea 
Subtracting these two equations from one another and comparing with \eqref{Euler} results in the standard formula
\be
 \eta (X)=\int_Xc_5(X) \eqlabel{Euler}
\ee
for the Euler number $\eta (X)$ of the five-fold $X$. Eliminating $c_5$, on the other hand, leads to the relation
\begin{equation}
 11 h^{1,1}(X)-10h^{1,2}(X)-h^{2,2}(X)+h^{2,3}(X)+10h^{1,3}(X)-11h^{1,4}(X)=0 \eqlabel{Hodgecons}
\end{equation}
which only depends on Hodge numbers. Hence, five-folds are characterised by five rather than six independent Hodge numbers. 

Other relevant topological invariants of Calabi-Yau five-folds, apart from the Hodge numbers and the Euler number, are the Chern classes $c_2(X)$, $c_3(X)$ and $c_4(X)$, the intersection numbers $d_{{i_1}\dots {i_5}}$ of five eight-cycles and various other intersection numbers which we will introduce later. 

\vskip 0.4cm

As we have seen in the main part of the paper, compactification of M-theory requires a Calabi-Yau five-fold $X$, a fourth cohomology class $g\in H^4(X)$ and an effective second cohomology class $W\in H_2(X,\mathbb{Z})$ satisfying the integrability and quantisation conditions
\begin{equation}
 c_4(X)-12\,g\wedge g=24\,W\; ,\quad g+\frac{1}{2}c_2(X)\in H^4(X,\mathbb{Z})\; . \eqlabel{anomcond}
\end{equation}
Physically, $g$ corresponds to a four-form flux and $W$ is the homology class of a holomorphic curve in $X$ which is wrapped by membranes. Clearly, there are a number of qualitatively different ways one might try to solve these conditions. Probably the simplest possibility is to find a Calabi-Yau five-fold $X$ with $c_4(X)=0$. In this case, one can set the flux $g$ and the membrane class $W$ to zero. For Calabi-Yau five-folds with $c_4(X)\neq 0$ one can ask if the conditions can be satisfied with either flux or membranes individually or by a combination of both. We should now discuss if and how these possibilities can be realised and to do so we need to turn to specific examples of Calabi-Yau five-folds.

\subsection{Examples of Calabi-Yau five-folds}\applabel{cyexamples}

\subsubsection{Complete intersection Calabi-Yau five-folds}

Perhaps the simplest class of Calabi-Yau manifolds is obtained from complete intersections in a projective space or a product of projective spaces (see, for example, Ref.~\cite{Hubsch:1992nu} for a review). In the case of Calabi-Yau three-folds, the best known example of such complete intersection Calabi-Yau manifolds (CICY) is the quintic in $\mathbb{P}^4$, defined as the zero locus of a homogeneous degree five polynomial in $\mathbb{P}^4$. For the case of Calabi-Yau five-folds, the direct analogue of the quintic in $\mathbb{P}^4$ is the septic in $\mathbb{P}^6$, that is the zero locus of a homogeneous degree seven polynomial in $\mathbb{P}^6$. 

In order to define CICY five-folds more generally, we first introduce an ambient space ${\cal A}=\bigotimes_{r=1}^m\mathbb{P}^{n_r}$, as a product of $m$ projective spaces with dimension $n_r$. Each of these projective spaces comes equipped with a K\"ahler form $J_r$ which we normalise such that
\begin{equation}
 \int_{\mathbb{P}^{n_r}}J_r^{n_r}=1\; .\eqlabel{Pnorm}
\end{equation} 
We are interested in the common zero locus of polynomials $p_\alpha$, where $\alpha =1,\ldots , K$, which are homogenous of degree $q^r_\alpha$ in the coordinates of the factor $\mathbb{P}^{n_r}$ in ${\cal A}$. In order for this zero locus to be five-dimensional we need, of course,
\begin{equation}
K=\sum_{r=1}^mn_r-5\; . \eqlabel{dimcons}
\end{equation}
It is useful to summarise the dimensions of the various projective spaces together with the (multi-) degrees of the polynomials in a {\em configuration matrix}
\begin{equation}
 [{\bf n}|{\bf q}]=\left[\begin{array}{l}n_1\\\vdots\\n_m\end{array}\right|\left.\begin{array}{lll}q^1_1&\dots&q^1_K\\
 \vdots&&\vdots\\
 q^m_1&\hdots&q^m_K\\\end{array}\right] \eqlabel{conf2}
 \end{equation} 
We note that every column in the $\mathbf{q}$ part of this matrix corresponds to the multi-degree of one of the defining polynomials. As an example, using this short-hand notation, the septic in $\mathbb{P}^6$ can be written as $[6|7]$. The total Chern class of such a CICY is given be the well-known formula~\cite{Hubsch:1992nu}
\begin{equation}
	c([\mathbf{n}|\mathbf{q}]) = \frac{ \prod_{r=1}^m (1+J_r)^{n_r+1} } { \prod_{\alpha=1}^K (1+\sum_{s=1}^m q_\alpha^s J_s) } . 
	\eqlabel{cherntotal}
\end{equation}
and the various individual Chern classes $c_q([\mathbf{n}|\mathbf{q}])$ can be obtained by expanding the above expression and extracting terms of order $q$ in the K\"ahler forms $J_r$.  For the first Chern class this leads to
\begin{equation}
 c_1([\mathbf{n}|\mathbf{q}])=\sum_{r=1}^m\left(n_r+1-\sum_{\alpha=1}^Kq^r_\alpha\right)J_r\; .
\end{equation} 
Hence, the Calabi-Yau condition $c_1(X)=0$ translates into the simple numerical constraints
\begin{equation}
 \sum_{\al=1}^Kq^r_\al=n_r+1, \qquad \forall r = 1,\ldots , m \eqlabel{cycondition}
\end{equation}
on the multi-degrees of the defining polynomials. This means the rows in the $\mathbf{q}$ part of the configuration matrix always have to sum up to the dimension of the associated projective space plus~1 in order for the complete intersection to be a Calabi-Yau space. 
In our application to M-theory compactifications, higher Chern classes and $c_4(X)$ in particular, play a crucial r\^{o}le. By expanding \eqref{cherntotal} we find for CICYs
\begin{align}
	c_2([\mathbf{n}|\mathbf{q}])&= c_2^{rs} J_r J_s = 
		\oneon{2}\sum_{r,s=1}^m \left[-(n_r+1)\delta^{rs} + \sum_{\al=1}^K q_\al^r q_\al^s \right] J_r J_s \; , \\ 
	c_3([\mathbf{n}|\mathbf{q}])&= c_3^{rst} J_r J_s J_t = 
		\oneon{3}\sum_{r,s,t=1}^m \left[(n_r+1)\delta^{rst} - \sum_{\al=1}^K q_\al^r q_\al^s q_\al^t \right] J_r J_s J_t \; , \\
         c_4 ([\mathbf{n}|\mathbf{q}]) &= c_4^{rstu} J_r J_s J_t J_u = \oneon{4}
	\left[ -(n_r+1)\delta^{rstu} + \sum_{\al=1}^K q_\al^r q_\al^s q_\al^t q_\al^u + 2 c_2^{rs} c_2^{tu} \right] J_r J_s J_t J_u \; , \eqlabel{c4}\\
         c_5 ([\mathbf{n}|\mathbf{q}]) &= c_5^{r_1 \ldots r_5} J_{r_1} \cdots J_{r_5} \nonumber \\ &= \oneon{5}
	\left[ (n_r+1)\delta^{r_1 \ldots r_5} - \sum_{\al=1}^K q_\al^{r_1} \cdots q_\al^{r_5} 
		+ 5 c_3^{\left(r_1 r_2 r_3 \right.} c_2^{\left. r_4 r_5 \right)} \right] J_{r_1} \cdots J_{r_5} \; , \eqlabel{c5}
\end{align}
where $c_1([\mathbf{n}|\mathbf{q}])=0$ has been used to simplify the expressions. The fourth Chern class should be written in terms of a set of harmonic eight-forms $\{\tilde{J}^r\}$ as $c_4(X)=\tilde{c}_{4r}\tilde{J}^r$.  If we choose these forms to be dual to the K\"ahler forms $J_r$, that is,
\begin{equation}
 \int_XJ_r\wedge\tilde{J}^s=\delta_r^s\; ,
\end{equation}
it is easy to see that $c_{4r}$ can be obtained from the coefficients which appear in the formula~\eqref*{c4} by
\begin{equation}
 \tilde{c}_{4r}=d_{rstuv}c_4^{stuv}\; ,\quad  d_{i_1\dots i_5}=\int_XJ_{i_1}\wedge\dots\wedge J_{i_5}\; . \eqlabel{intersec}
\end{equation}
The intersection numbers $d_{i_1\dots i_5}$ can be explicitly computed from the identity 
\begin{equation}
 \int_X w=\int_{\cal A}w\wedge\mu\; ,\quad \mu =\bigwedge_{\al=1}^K\left(\sum_{r=1}^mq^r_\al J_r\right)\; , \eqlabel{mudef}
\end{equation}
which converts integration of a $10$-form $w$ over $X$ into an integration over the ambient space. In carrying out the latter the normalisation~\eqref*{Pnorm} must be taken into account. The calculation of Hodge numbers is straightforward for CICYs with $q^r_\alpha>0$ for all $r$ and $\alpha$. In this case, repeated application of the Lefshetz hyperplane theorem (see, for example, Ref.~\cite{Hubsch:1992nu}) shows that
\begin{equation}
 H^{p,q}(X)\simeq H^{p,q}({\cal A})\;\mbox{  for }\; p+q\neq 5\; .
\end{equation}
Hence, all cohomology groups except the middle ones are isomorphic to the ambient space cohomology groups for such CICYs. The only non-vanishing Hodge numbers of $\mathbb{P}^n$ are $h^{p,p}(\mathbb{P}^n)=1$ and, by applying the K\"{u}nneth formula $H^n(Y\otimes Z)=\bigoplus_{i+j=n}H^i(Y)\otimes H^j(Z)$, one can easily compute the Hodge numbers of the ambient space ${\cal A}$ from this result. Combining these facts, one finds for CICYs with all $q^r_a>0$ that
\begin{align}
 h^{1,1}(X)&= h^{1,1}({\cal A})=m\\
 h^{1,2}(X)&= 0\\
 h^{1,3}(X)&= 0\\
 h^{2,2}(X)&= h^{2,2}({\cal A})=\frac{m(m-1)}{2}+\#\{r|n_r\geq 2\}\; .
\end{align} 
The first of these equations means that the restrictions of the ambient space K\"ahler forms $J_r$ to $X$ form a basis of the second cohomology. The last equation implies that the four-forms $J_r\wedge J_s$ span $H^{2,2}(X)$. Let us define the six-cycles $C_{rs}=[\mathbf{n}|\mathbf{q}\,\mathbf{e}_r\,\mathbf{e}_s]\subset X$, where $\mathbf{e}_r$ is a vector with one in the $r^{\rm th}$ entry and zero elsewhere. The measure $\mu_{rs}$ for these six-cycles is given by $\mu_{rs}=\mu\wedge J_r\wedge J_s$ where $\mu$ is the measure of the CICY as in \eqref{mudef}. It follows that 
$\int_{C_{rs}}w=\int_{\cal A}w\wedge \mu_{rs}=\int_Xw\wedge J_r\wedge J_s$ for all six-forms $w$. Hence, the forms $J_r\wedge J_s$ are Poincar\'e dual to the six-cycles $C_{rs}$ and are, therefore, integral.
Two remaining Hodge numbers need to be determined, namely $h^{1,4}(X)$ and $h^{2,3}(X)$. This can be accomplished by calculating the Euler number from eqs.~\eqref*{Euler}, \eqref*{c5} and then using the two constraints \eqref*{Euler0} and \eqref*{Hodgecons}. 

For CICYs where some $q^r_\al$ vanish a more refined version of the above reasoning can sometimes be applied~\cite{Hubsch:1992nu}. In more complicated cases, the Hodge numbers must be calculated using spectral sequence methods~\cite{Green:1987cr}. For CICY three-folds it is known that $h^{1,1}(X)$ can be larger than $m$ in such cases so that not all $(1,1)$ classes descend from the ambient space. A similar phenomenon can be expected for CICY five-folds.  In general, one can also expect $h^{1,2}(X)$ and $h^{1,3}(X)$ to acquire non-zero values. The detailed analysis of these issues is somewhat outside our main line of investigation and will not be pursued here.

Another useful feature of CICYs whose second cohomology is spanned by the ambient space K\"ahler forms $J_r$ is that the Mori cone, the cone of effective cohomology classes in $H_2(X,\mathbb{Z})\simeq H^8(X,\mathbb{Z})$, is given by positive integer linear combinations $n_r\tilde{J}^r$ of the eight-forms $\tilde{J}^r$ dual to $J_r$.

\vskip 0.4cm

It is useful to have some explicit  examples of CICYs available. The simplest sub-class consists of CICY five-folds which can be defined in a single projective space. In this case, a linear polynomial constraint simply amounts to a reduction of the ambient space dimension by one. In other words, a configuration matrix of the form $[n|q_1\,\dots \,q_{K-1}\, 1]$ in $\mathbb{P}^n$ is equivalent to a configuration matrix $[n-1|q_1\,\dots\, q_{K-1}]$ in $\mathbb{P}^{n-1}$. Hence, we can require that all $q_\al >1$ without restricting generality. It is then a simple combinatorial exercise to write down all configurations in a single projective space, subject, of course, to the dimension constraint~\eqref*{dimcons} and the Calabi-Yau condition~\eqref*{cycondition}. One finds $11$ cases which are listed in \tabref{tab:cicy1}.
\TABLE[t]{
\begin{tabular}{|l|l|l|l|l|l|}
\hline
$[n|q_1\dots q_K]$&$c_2(X)/J^2$&$c_4(X)/J^4$&$\eta (X)$&$h^{1,4}(X)$&$h^{2,3}(X)$\\\hline\hline
$[6|7]$&$21$&$819$&$-39984$&$1667$&$18327$\\\hline
$[7|6\,2]$&$16$&$454$&$-32544$&$1357$&$14917$\\\hline
$[7|5\,3]$&$13$&$259$&$-19440$&$811$&$8911$\\\hline
$[7|4\,4]$&$12$&$198$&$-14208$&$593$&$6513$\\\hline
$[8|5\,2\,2]$&$12$&$234$&$-23280$&$971$&$10671$\\\hline
$[8|4\,3\,2]$&$10$&$136$&$-13392$&$559$&$6139$\\\hline
$[8|3\,3\,3]$&$9$&$99$&$-9720$&$406$&$4456$\\\hline
$[9|4\,2\,2\,2]$&$9$&$114$&$-14592$&$609$&$6689$\\\hline
$[9|3\,3\,2\,2]$&$8$&$78$&$-9648$&$403$&$4423$\\\hline
$[10|3\,2\,2\,2\,2]$&$7$&$58$&$-8832$&$369$&$4049$\\\hline
$[11|2\,2\,2\,2\,2\,2]$&$6$&$39$&$-6912$&$289$&$3169$\\\hline
\end{tabular}
\caption{The $11$ CICY five-folds which can be defined in a single projective space. The Hodge numbers $h^{1,2}(X)=h^{1,3}(X)=0$ and $h^{1,1}(X)=h^{2,2}(X)=1$ for all manifolds.}
\tablabel{tab:cicy1}}
\enlargethispage*{\baselineskip} 
For comparison, in the case of Calabi-Yau three-folds there exist five CICYs which can be defined in a single projective space~\cite{Hubsch:1992nu}. In fact, CICY three-folds in arbitrary products of projective spaces have been classified~\cite{Candelas:1987kf} and about $8000$ manifolds have been found. No similar classification is available for CICY five-folds but it is reasonable to assume that their number is significantly larger than $8000$. Here, we will be content with the three examples of CICY five-folds defined in a product of projective spaces given in \tabref{tab:cicy2}.
\TABLE[t]{
\begin{tabular}{|l|l|l|l|l|l|l|l|}
\hline
$[\mathbf{n}|\mathbf{q}]$&$c_2(X)$&$c_4(X)$&$\eta (X)$&$h^{1,1}(X)$&$h^{2,2}(X)$&$h^{1,4}(X)$&$h^{2,3}(X)$\\\hline\hline
$\left[\begin{array}{l}1\\5\end{array}\right|\left.\begin{array}{l}2\\6\end{array}\right]$&$\begin{array}{l}12J_1J_2+\\15J_2^2\end{array}$&$\begin{array}{l}2610\tilde{J}^1+\\4542\tilde{J}^2\end{array}$&$-32280$&$2$&$2$&$1347$&$14797$\\\hline
$\left[\begin{array}{l}3\\3\end{array}\right|\left.\begin{array}{l}4\\4\end{array}\right]$&$\begin{array}{l}6J_1^2+\\16J_1J_2+\\6J_2^2\end{array}$&$\begin{array}{l}3600\tilde{J}^1+\\3600\tilde{J}^2\end{array}$&$-28608$&$2$&$3$&$1194$&$13115$\\\hline
$\left[\begin{array}{l}1\\2\\3\end{array}\right|\left.\begin{array}{l}2\\3\\4\end{array}\right]$&$\begin{array}{l}3J_2^2+\\12J_2J_3+\\6J_3^2+\\6J_1J_2+\\8J_1J_3\end{array}$&$\begin{array}{l}84\tilde{J}^1+\\114\tilde{J}^2+\\130\tilde{J}^3\end{array}$&$-24480$&$3$&$5$&$1023$&$11225$\\\hline
\end{tabular}
\caption{Examples of CICY five-folds defined in a product of projective spaces. The Hodge numbers $h^{1,2}(X)=h^{1,3}(X)=0$ for all manifolds.}
\tablabel{tab:cicy2}}

\vskip 0.4cm

\begin{sloppypar} 
We should now discuss the various possibilities to satisfy the M-theory conditions~\eqref*{anomcond} for compactifications on CICY five-folds. We will not attempt to address this question in a systematic way but merely analyse a number of examples. Our main goal is to show that CICY five-folds for consistent M-theory compactifications exist and to give a flavour of how restrictive the conditions are. 
As discussed earlier, the simplest option is to compactify on a manifold with $c_4(X)=0$, without flux and membranes. Do CICY five-folds with $c_4(X)=0$ exist? From \eqref{intersec} we see that the intersection numbers are positive, that is, $d_{i_1\dots i_5}\geq 0$. Further, it is clear from \eqref{c4} that all components $c_4^{rstu}\geq 0$. This means the fourth Chern class of CICY five-folds is positive in the sense that $\tilde{c}_{4r}\geq 0$ for all $r$. If the configuration matrix is such that $q^r_\alpha \geq 2$ for all $r$ and $\alpha$, the coefficients  $c_4^{rstu}$ are strictly positive. From the first \eqref{intersec} this shows that $c_4(X)\neq 0$ for all such CICY five-folds. In particular, it follows that all CICY five-folds defined in a single projective space ($m=1$) and all co-dimension one five-folds ($K=1$) have $c_4(X)\neq 0$. (The former fact is, of course, confirmed by \tabref{tab:cicy1}.) So, we are left with CICY five-folds satisfying $m>1$, $K>1$ and $q^r_\al<2$ for at least one component. We have scanned all such configurations for $m\leq 4$ and $K\leq 4$ and the only examples with $c_4(X)=0$ we have found are spaces such as
\begin{equation}
 X\sim\left[\begin{array}{l}4\\2\\2\end{array}\right|\left.\begin{array}{lll}5&0&0\\0&3&0\\0&0&3\end{array}\right] \; ,
\end{equation}
which correspond to the direct product of a Calabi-Yau three-fold $Y$ (the quintic in the above example) with two tori $T^2$. Clearly, $c_4(Y\times T^2\times T^2)=0$ but such a space only has holonomy ${\rm SU}(3)$. It breaks a quarter of the supersymmetry and is, therefore, not a Calabi-Yau five-fold in the sense defined at the beginning of this appendix. In summary, for $m\leq 4$ and $K\leq 4$ we have not found any proper CICY five-folds with holonomy ${\rm SU}(5)$ which satisfy $c_4(X)=0$. We cannot exclude that larger configurations with this property exist although we have not been able to find any explicit examples. 
\end{sloppypar}

Given the lack of CICY five-folds with $c_4(X)=0$, we can ask if the conditions~\eqref*{anomcond} can be satisfied by including flux and membranes. To analyse this question let us start with the five-folds in a single projective space which are listed in \tabref{tab:cicy1}. We write the fourth Chern class as $c_4(X)={\cal C}J^4$, where the numbers ${\cal C}$ can be read off from \tabref{tab:cicy1} and the flux as $g=k J^2$ for some number $k$. In the absence of membranes ($W=0$) the anomaly condition~\eqref*{anomcond} is then solved for flux values
\begin{equation}
 k=\pm\sqrt{\frac{{\cal C}}{12}}\; .
\end{equation} 
For the $11$ cases in \tabref{tab:cicy1}, it can be checked that the resulting values of $k$ are never rational. This means, it is impossible to satisfy the flux quantisation condition~\eqref*{anomcond} for such values of $k$. We conclude that, in the absence of membranes the $11$ CICY five-folds in a single projective space cannot be used for consistent M-theory compactifications. 

Does the inclusion of membranes help? We begin with the septic, $[6|7]$, whose properties are listed in the first row of \tabref{tab:cicy1}.  For the right-hand sides of the anomaly and quantisation conditions~\eqref*{anomcond} we find in this case
\begin{equation}
 c_4(X)-12 g^2=(819-12k^2)J^4\; ,\quad g-\frac{1}{2}c_2(X)=\left(k+\frac{21}{2}\right)J^2\; .
\end{equation}
Setting the flux to $k=15/2$, the anomaly condition can then be satisfied for a membrane wrapping a holomorphic curve with class $W=6J^4$. Recalling that $J^2$ is an integral class, the flux quantisation condition is also satisfied for this value of $k$. Hence, by including flux and membranes the M-theory conditions can be satisfied for the septic. 

While the M-theory conditions for CICY five-folds in a single projective space cannot be satisfied with flux only, can they be satisfied for membranes only? Let us look at the example $[7|6\, 2]$ which corresponds to the second row in \tabref{tab:cicy1}. From \eqref{mudef} we know that the measure for this manifold is given by $\mu=12J^2$. This means
\begin{equation}
\int_XJ\wedge J^4=12\; .
\end{equation}
Comparing this with the definition $\int_XJ\wedge\tilde{J}=1$ of the dual eight-form $\tilde{J}$ we learn that $\tilde{J}=J^4/12$ and that this is an integral class. Given that $c_4(X)=454J^4$, the anomaly condition can then be satisfied by setting the flux to zero and by including a membrane which wraps a holomorphic curve with class $W=227\tilde{J}$.

\vskip 0.4cm

In order to find viable examples with flux only we need to consider CICYs defined in products of projective spaces. Let us start with the first example in \tabref{tab:cicy2}, a co-dimension one CICY five-folds with configuration matrix
\begin{equation}
X\sim\left[\begin{array}{l}1\\5\end{array}\right|\left.\begin{array}{l}2\\6\end{array}\right]\; ,\eqlabel{cicy26}
\end{equation}
defined in the ambient space ${\cal A}=\mathbb{P}^1\otimes\mathbb{P}^5$. Writing the flux as $g=k_{1,2}J_1J_2+k_{2,2}J_2^2$ one finds for the right-hand-sides of the anomaly and quantisation condition~\eqref*{anomcond}
\begin{align}
 c_4(X)-12g^2 &= (2610-72\, k_{2,2}^2)\tilde{J}^1+(4542-144\, k_{1,2}k_{2,2}-24\,k_{2,2}^2)\tilde{J}^2\; ,\eqlabel{anexam}\\
 g+\frac{1}{2}c_2(X) &= (k_{1,2}+6)J_1J_2+\left(k_{2,2}+\frac{15}{2}\right)J_2^2\; . \eqlabel{quantexam}
\end{align}
For the anomaly to vanish without membranes we need a non-rational flux parameter $k_{2,2}=\pm\sqrt{145}/2$ and, hence, the quantisation condition cannot be satisfied. On the other hand, for any integer $k_{1,2}$ and any half-integer $k_{2,2}$ the coefficients on the right-hand-side of \eqref{anexam} are divisible by $24$ and, for sufficiently small flux integers, positive. Hence, the anomaly condition can be satisfied by inclusion of a membrane. 

Next, we consider the second example in \tabref{tab:cicy2}, the co-dimension one CICY five-folds in ${\cal A}=\mathbb{P}^3\otimes\mathbb{P}^3$ with configuration matrix
\begin{equation}
X\sim\left[\begin{array}{l}3\\3\end{array}\right|\left.\begin{array}{l}4\\4\end{array}\right]\; .\eqlabel{cicy44}
\end{equation}
With the flux parameterised as $g=k_{1,1}J_1^2+k_{1,2}J_1J_2+k_{2,2}J_2^2$ one finds for the right-hand-sides of the anomaly and quantisation condition~\eqref*{anomcond}
\begin{align}
 c_4(X)-12g^2 &= (3600-48k_{1,2}^2-96k_{1,1}k_{2,2}-96k_{1,2}k_{2,2})\tilde{J}^1\nonumber\\& +(3600-96k_{1,1}k_{1,2}-48k_{1,2}^2-96k_{1,1}k_{2,2})\tilde{J}^2\; ,\eqlabel{anexam44}\\
 g+\frac{1}{2}c_2(X) &= (k_{1,1}+3)J_1^2+(k_{1,2}+8)J_1J_2+(k_{2,2}+3)J_2^2\; . \eqlabel{quantexam44}
\end{align}
Again, without membranes, it can be checked that the anomaly condition cannot be satisfied for integers $k_{1,1}$, $k_{1,2}$ and $k_{2,2}$. However, as the right-hand-side of \eqref{anexam44} is divisible by $24$, a complete model can always be obtained be inclusion of membranes as long as the flux integers are not too large.

For the above examples, we have $h^{2,2}(X)=2$ or $3$ flux parameters and $h^{1,1}(X)=2$ equations from the anomaly condition, so it is perhaps not surprising that a rational solution without membranes cannot be found. In fact, a similar obstruction can be found for other simple CICYs defined in a product of two projective spaces. This suggests looking at more complicated examples in products of more than two projective spaces, so that $h^{2,2}(X)>h^{1,1}(X)$. To this end, we consider the CICY in the third row of \tabref{tab:cicy2}, defined in a product of three projective spaces and with configuration matrix
\begin{equation}
X\sim\left[\begin{array}{l}1\\2\\3\end{array}\right|\left.\begin{array}{l}2\\3\\4\end{array}\right]\; .
\end{equation}
Flux can be parameterized as $g=k_{1,2}J_1J_2+k_{1,3}J_1J_3+k_{2,2}J_2^2+k_{2,3}J_2J_3+k_{3,3}J_3^2$ and we find for the right-hand-sides of the anomaly and quantisation conditions
\begin{align}
 c_4(X)-12g^2 &= (130-4k_{1,3}k_{2,2}-4k_{1,2}k_{2,3}-3k_{1,3}k_{2,3}-k_{2,3}^2-3k_{1,2}k_{3,3}-2k_{2,2}k_{3,3})\tilde{J}^3\nonumber\\
 & +(114-4k_{1,3}k_{2,3}-4k_{1,2}k_{3,3}-3k_{1,3}k_{3,3}-2k_{2,3}k_{3,3})\tilde{J}^2\nonumber\\
 & +(84-2k_{2,3}^2-4k-{2,2}k_{3,3}-3k_{2,3}k_{3,3})\tilde{J}^1 \; , \\
 g-\frac{1}{2}c_2(X) &= (3+k_{1,2})J_1J_2+(4+k_{1,3})J_1J_3+\left(\frac{3}{2}+k_{2,2}\right)J_2^2\nonumber\\
 & +(6+k_{2,3})J_2J_3+(3+k_{3,3})J_3^2 \; .
\end{align}
A quick scan reveals that both conditions can be satisfied for the choice $(k_{1,2}, k_{1,3}, k_{2,2}, k_{2,3},$ $k_{3,3})=(1,3,7/2,0,6)$.

\vskip 0.4cm

In summary, we have seen that viable M-theory backgrounds based on CICY five-folds are not too hard to obtain by adding flux and membranes as well as membranes only. With some more effort, by exploring more complicated examples with $h^{2,2}(X)>h^{1,1}(X)$,  solutions with flux only can be found as well. Unfortunately, we have not managed to find CICY five-folds with holonomy ${\rm SU}(5)$ and $c_4(X)=0$ and such CICY five-folds may well not exist. However, an example with $c_4(X)=0$ which allows for a ``clean'' compactification without flux or membranes is still highly desirable and we, therefore, turn to another class of Calabi-Yau five-folds.

\subsubsection{Torus quotients}
The Chern classes of a torus vanish and it is, therefore, a promising starting point for the construction of Calabi-Yau five-folds with $c_4(X)=0$. Specifically, we start with a product $T=T^2\times\dots\times T^2$ of five two-tori, each with a complex coordinate $z_\mu$, where $\mu=1,\ldots ,5$, identified as $z_\mu\sim z_\mu+1$ and $z_\mu\sim z_\mu +i$. Then  we consider the symmetry $\mathbb{Z}_2^4$ defined by the four generators
\begin{align}
 \gamma_1(z_1,\ldots ,z_5) &= (-z_1+1/2,-z_2+i/2,z_3+1/2,z_4,z_5)\\
 \gamma_2(z_1,\dots ,z_5) &= (z_1,-z_2+1/2,-z_3+i/2,z_4+1/2,z_5)\\
 \gamma_3(z_1,\ldots ,z_5) &= (z_1,z_2,-z_3+1/2,-z_4+i/2,z_5+1/2)\\
 \gamma_4(z_1\ldots ,z_5) &= (z_1+1/2,z_2,z_3,-z_4+1/2,-z_5+i/2)\; .
\end{align}
It is straightforward to check that the $16$ elements of this group all act freely on $T$. Hence, the quotient $X=T/\mathbb{Z}_2^4$ is a manifold. Clearly, it inherits the property of vanishing Chern classes from the torus and, in particular, $c_4(X)=0$. The holonomy of $X$ is of course just $\mathbb{Z}_2^4$ but the four $\mathbb{Z}_2$ symmetries are still sufficient to reduce the number of supersymmetries by a factor of $1/16$.  Therefore, $X$ is a Calabi-Yau five-fold in this sense defined at the beginning of this appendix.

What are the properties of $X$? Clearly, $c_i(X)=0$ for $i=1,\ldots ,5$ and this implies that the Euler number, $\eta (X)$, also vanishes. The Hodge numbers are obtained by counting the number of $\mathbb{Z}_2^4$ invariant $(p,q)$ forms $dz_{\mu_1}\wedge\dots\wedge dz_{\mu_p}\wedge d\bar{z}_{\nu_1}\wedge\dots\wedge d\bar{z}_{\nu_q}$. This results in
\bea
 &h^{1,1}(X)=5\; ,\quad &h^{1,2}(X)=0\; ,\quad &h^{1,3}(X)=0\; ,\\ &h^{2,2}(X)=10\; ,\quad &h^{1,4}(X)=5\; ,\quad &h^{2,3}(X)=10\; .
\eea
Presumably five-folds from tori divided by other discrete symmetries can be constructed along similar lines. We will not pursue this explicitly, having shown the existence of Calabi-Yau five-folds with $c_4(X)=0$ by the simple example above. It remains an open question whether Calabi-Yau five-folds with full ${\rm SU}(5)$ holonomy and $c_4(X)=0$ exist. We are not aware of a general mathematical reason which rules this out and it would be interesting to look for such manifolds, for example among toric five-folds.

\subsection{Some differential geometry on five-folds}\applabel{cydg}

As discussed earlier, on a Calabi-Yau five-fold $X$ we have a spinor $\eta$, unique up to normalisation, which is invariant under the holonomy group ${\rm Hol}(X)$. This means, $\eta$ is covariantly constant with respect to the Levi-Civita connection associated to the Ricci-flat metric $g$. Its direction can be defined by imposing the five annihilation conditions~\footnote{For our conventions on $10$-dimensional gamma matrices and spinors, see Appendix~\ref{app:conv}.}
\begin{equation}
\gamma^{\bar{\mu}}\eta =0\; .
\end{equation}
Given the definition~\eqref*{gamma11} of the 10-dimensional chirality operator, $\eta$ has positive and $\eta^\star$ negative chirality, that is
\be\eqlabel{eta_chirality}
	\gamma^{(11)} \eta = \eta , \qquad \gamma^{(11)} \eta^\star = - \eta^\star \; .
\ee
As usual, we normalize $\eta$ such that
\begin{equation}
 \eta^\dagger\eta =1\; . \eqlabel{etanorm}
\end{equation}
It can be shown that $\eta$ satisfies the Fierz identity (see, for example, Ref.~\cite{Miemiec:2005ry}, Proposition 5, or Ref.~\cite{Naito:1986cr}, eq. (2.3)) 
\be\eqlabel{cy5_fierz_id}
	\eta^\star \eta^T = - \oneon{32} g_{\mu\bar\nu} \gamma^{\mu\bar\nu} ,
\ee
which will be useful in our reduction of the fermionic terms.
Apart from the normalisation~\eqref*{etanorm}, there exist two other non-zero spinor bilinears, namely the K\"ahler form $J$ and the holomorphic $(5,0)$ form $\Omega$ defined by
\be
	J_{\mu\bar{\nu}}= i \eta^\dagger \gamma_{\mu\bar{\nu}} \eta\; ,\quad
	\Omega_{\mu_1 \ldots \mu_5}=||\Omega || \eta^\dagger \gamma_{\mu_1 \ldots \mu_5} \eta^\star \;  , \eqlabel{JOdef} 
\ee
where $||\Omega ||=\Omega_{\mu_1\dots\mu_5}\bar{\Omega}^{\mu_1\dots\mu_5}/5!$. Apart from these expressions and their complex conjugates, all other spinor bilinears vanish. Both $J$ and $\Omega$ are covariantly constant as a direct consequence of $\eta$ being covariantly constant. The complex structure ${\cal J}$ is defined by the equation $J_{mn}={{\cal J}_m}^pg_{pn}$ and the metric $g$ is hermitian with respect to ${\cal J}$. The projection operators $P_\pm =(\id \mp i{\cal J})/2$ can be used to split tensors on $X$ into $(p,q)$ ``index types'' with $p$ holomorphic and $q$ anti-holomorphic indices. As usual, we will work in local complex coordinates such that ${{\cal J}_\mu}^\nu =i\delta_\mu^\nu$ and ${{\cal J}_{\bar{\mu}}}^{\bar{\nu}} =-i\delta_{\bar{\mu}}^{\bar{\nu}}$. In this basis, the $(2,0)$ and $(0,2)$ components of the metric and the K\"ahler form vanish and we have
\begin{equation}
J_{\mu\bar{\nu}}=ig_{\mu\bar{\nu}}\; .
\end{equation}
For a $(p,q)$ form $\omega^{(p,q)}$ with $p>0$ and $q>0$ we can define an associated $(p-1,q-1)$ form by contracting one holomorphic and one anti-holomorphic index of $\omega$ with the inverse metric $g^{\mu\bar{\nu}}$. In the following, it will be convenient to introduce the short-hand notation $\tilde{w}^{(p,q)}$ for this $(p-1,q-1)$ form. Note that  $\tilde{\omega}^{(p,q)}$ is harmonic if $\omega^{(p,q)}$ is, since the metric is covariantly constant. This short-hand notation for the contraction of forms is useful to write down explicit formul\ae for the Hodge duals of $(p,q)$ forms which are required in many physics applications. Straightforward but in part somewhat tedious component calculations show that
\begin{equation}\eqlabel{hodge_special}
\begin{aligned}
	&(0,1):\;	\ast \zeta 		= \frac{i}{4!} J^4 \wedge \zeta , \qquad & 
	&(1,1):\;	\ast \omega	= -\oneon{3!} J^3 \wedge \omega - \frac{i}{4!} \tilde{\omega}J^4 , \\
	&(2,1):\; 	\ast \nu 		= \frac{i}{2} J^2 \wedge \nu + \oneon{3!} J^3 \wedge \tilde{\nu} , \qquad & 
	&(3,1):\; 	\ast \varpi		= - J \wedge \varpi - \frac{i}{2} J^2 \wedge \tilde{\varpi} , \\
	&(4,1):\; 	\ast \chi 		= i \chi + J \wedge \tilde{\chi} , \qquad & 
	&(2,2):\; 	\ast \sigma	= J \wedge \sigma - \frac{i}{2} J^2 \wedge \tilde{\sigma} + \oneon{12} \tilde{\tilde{\sigma}}J^3, \\
	&(3,2):\; 	\ast \phi 		= -i \phi - J \wedge \tilde{\phi} - \frac{i}{12} J^2 \wedge \tilde{\tilde{\phi}}. & & 
\end{aligned}
\end{equation}
Some simplifications of these equations arise for harmonic $(p,q)$ forms. We recall that Calabi-Yau five-folds have vanishing Hodge numbers $h^{p,0}(X)=h^{0,p}(X)$ for $p=1,2,3,4$.  This means non-zero harmonic $(p,0)$ and $(0,p)$ forms do not exist and consequently
\begin{equation}
 \tilde{\omega}^{(p,1)}=\tilde{\omega}^{(1,p)}=0 \mbox{  for harmonic $(p,1)$ and $(1,p)$ forms with $p>1$.}
\end{equation}
Moreover, a harmonic $(0,0)$ form is a constant and, hence,
\begin{equation}
 \tilde{\omega}^{(1,1)}={\rm const.} \qquad \text{for harmonic $(1,1)$ forms.}
\end{equation} 
Combining these facts with the formul\ae~\eqref*{hodge_special}, one finds for the Hodge dual of harmonic $(p,q)$ forms on Calabi-Yau five-folds
\begin{equation}
\eqlabel{dualforms}
\begin{aligned}
         &(1,1):\;	\ast \omega	= -\oneon{3!} J^3 \wedge \omega - \frac{i}{4!} \tilde{\omega} J^4, &\\\
         &(2,1):\; 	\ast \nu 		= \frac{i}{2} J^2 \wedge \nu , \qquad &\\
         &(3,1):\; 	\ast \varpi		= - J \wedge \varpi , \qquad &
	 &(2,2):\; 	\ast \sigma	= J \wedge \sigma - \frac{i}{2} J^2 \wedge \tilde{\sigma} + \oneon{12}\tilde{\tilde{\sigma}}J^3, \\
 	 &(4,1):\; 	\ast \chi 		= i \chi , &
	 &(3,2):\; 	\ast \phi 		= -i \phi - J \wedge\tilde{\phi}\; ,
\end{aligned}
\end{equation}
where we should keep in mind that $\tilde{\omega}$ and $\tilde{\tilde{\sigma}}$ are constants and $\tilde{\sigma}$ is a harmonic $(1,1)$ form. The volume $V$ of the five-fold can be written as
\begin{equation}
 V\equiv\int_Xd^{10}x\,\sqrt{g} = \frac{1}{5!}\int_X J^5\; .
\end{equation}
Then, acting with $J\wedge$ on the $(1,1)$ part of \eqref{dualforms}, using that $J\wedge\ast w=-id^{10}x\,\sqrt{g}\,\tilde{w}$ and integrating over $X$ we learn that
\begin{equation}
 \tilde{w}=5i\frac{\int_XJ^4\wedge w}{\int_XJ^5}\; . \eqlabel{omegatilde}
\end{equation} 
A further useful relation for a Hodge dual is
\be
 \ast (\sigma\wedge J^2) = \tilde{\tilde{\sigma}} J - 2 i \tilde{\sigma}\; . \eqlabel{sigmaJJ}
\ee 
where $\sigma$ is a $(2,2)$ form. In the next sub-section, we will use this relation to explicitly compute $\tilde{\sigma}$ and $\tilde{\tilde{\sigma}}$.

\subsection{Five-fold moduli spaces}\applabel{modspace}

For Calabi-Yau three-folds the moduli space of Ricci-flat metrics is (locally) a direct product of a K\"ahler and complex structure moduli space which are associated to harmonic $(1,1)$ and $(2,1)$ forms, respectively. For Calabi-Yau five-folds the situation is analogous and we will naturally borrow from the literature for three-folds (in particular, see Ref.~\cite{Candelas:1990pi}, for an explicit description). Just as for three-folds, the K\"ahler deformations of a five-fold metric are associated to harmonic $(1,1)$ forms while the complex structure deformations can be described in terms of $(1,4)$ forms. All other harmonic forms on five-folds are unrelated to metric deformations but some of them still do play a r\^{o}le in M-theory compactifications. In particular, the $(2,1)$ forms determine the zero modes of the M-theory three-form field. No bosonic degrees of freedom can be associated with the $(1,3)$ forms but, as we discuss in the main part of the paper, they give rise to a set of fermionic zero modes. In summary, all harmonic $(p,1)$ (or, equivalently, $(1,p)$) forms, where $p=1,2,3,4$, are relevant for the zero-modes expansion of the M-theory fields.
In addition, harmonic $(2,2)$ forms play a r\^{o}le when flux is included in the compactification. It is useful to introduce sets of harmonic basis forms for these cohomologies as follows
\begin{align}
	&H^{(1,1)}(X): \quad  \{\omega_i\}_{i=1,\ldots,h^{1,1}(X)} , \\
	&H^{(2,1)}(X): \quad  \{\nu_p\}_{p=1,\ldots,h^{2,1}(X)} , \\
	&H^{(1,3)}(X): \quad  \{\varpi_x\}_{x=1,\ldots,h^{1,3}(X)} , \\
	&H^{(2,2)}(X): \quad  \{\sigma_e\}_{e=1,\ldots,h^{2,2}(X)} , \\
	&H^{(1,4)}(X): \quad  \{\chi_a\}_{a=1,\ldots,h^{1,4}(X)} , 
\end{align}
with $\omega_i$ and $\sigma_e$ real and all other forms complex. These forms can be used to construct various intersection numbers\footnote{The term ``intersection number'' is a slight misnomer in this context, as, in fact, all of these integrals, except $d_{i_1\ldots i_5}$, in general depend on the complex structure (due to the use of complex $(p,q)$-forms) and thus do not represent topological invariants.}
\begin{equation}
\begin{array}{lllllll}
	d_{i_1\ldots i_5} &=& \int_X \omega_{i_1} \wedge \cdots \wedge \omega_{i_5} ,&\quad&
	d_{p\bar{q}ij} &=& \int_X \nu_p \wedge \bar{\nu}_{\bar{q}} \wedge \omega_i \wedge \omega_j , \\
	d_{eijk}&=&\int_X\sigma_e\wedge \omega_i\wedge \omega_j\wedge \omega_k ,&\quad&
	d_{p\bar{q}e}&=&\int_X \nu_p \wedge \bar{\nu}_{\bar{q}} \wedge \sigma_e,\\
	d_{efi}&=&\int_X\sigma_e\wedge\sigma_f\wedge\omega_i,&\quad&
	d_{x\bar{y}i} &=& \int_X \varpi_x \wedge \bar{\varpi}_{\bar{y}} \wedge \omega_i .
\end{array}
\eqlabel{def_int_nos}
\end{equation}
which will play a r\^{o}le later on.

\vskip 0.4cm

We begin with the metric moduli. As usual, the basic requirement is that a variation $g_{mn}\rightarrow g_{mn}+\delta g_{mn}$ of the metric leaves the Ricci tensor zero at linear order in $\delta g$. Working this out in detail, reveals that the $(1,1)$ part of $\delta g$ can be expanded in terms of harmonic $(1,1)$ forms while the $(2,0)$ and $(0,2)$ parts can be expressed in terms of harmonic $(1,4)$ forms. Explicitly, one has
\begin{equation}
	\delta g_{\mu\bar{\nu}}=-iw_{i,\mu\bar{\nu}}\delta t^i\; ,\quad 
	\delta g_{\mu\nu}=-\frac{2}{4!||\Omega ||^2}{\Omega_{\mu}}^{\bar{\rho}_1\ldots\bar{\rho}_4}\chi_{a,\nu\bar{\rho}_1\ldots\bar{\rho}_4}\delta z^a\; ,\eqlabel{deltag}
\end{equation} 
with the variations $\delta t^i$ and $\delta z^a$ in the K\"ahler and complex structure moduli. The standard moduli space metric on the space of metric deformations is defined by
\begin{equation}
 {\cal G}(\delta g ,\widetilde{\delta g})=\frac{1}{4V}\int_X d^{10}x\,\sqrt{g}\,g^{mn}g^{pq}\delta g_{mp}\widetilde{\delta g}_{nq}\; .
 \eqlabel{wp}
\end{equation} 
This metric splits into a K\"ahler and a complex structure part which can be worked out separately. Let us first discuss the K\"ahler deformations. A straightforward calculation, inserting  the first \eqref{deltag} shows that
\begin{equation}
 {\cal G}^{(1,1)}_{ij}(\underline{t})=\frac{1}{2V}\int_X\omega_i\wedge\ast\omega_j\; .
\end{equation}
Using the expression in \eqref{dualforms} for the dual of $(1,1)$ forms together with \eqref{omegatilde}, this can be written in terms of topological integrals which involve $J$ and the forms $\omega_i$. Then, defining the K\"ahler moduli by
\begin{equation}
 J=t^i\omega_i \; ,
\end{equation}
one finds
\begin{equation}
{\cal G}^{(1,1)}_{ij}(\underline{t})=-10\frac{\kappa_{ij}}{\kappa}+\frac{25}{2}\frac{\kappa_i\kappa_j}{\kappa^2}\; , \eqlabel{metric11}
\end{equation}
where we have introduced the notation
\begin{align}
 \kappa &= \int_XJ^5=5!\, V=d_{i_1\dots i_5}t^{i_1}\dots t^{i_5} \; , \\
 \kappa_i&= \int_X\omega_i\wedge J^4=d_{ii_2\dots i_5}t^{i_2}\dots t^{i_5} \; , \\
 \kappa_{ij}&= \int_X\omega_i\wedge\omega_j\wedge J^3=d_{iji_1i_2i_3}t^{i_1}t^{i_2}t^{i_3} \; , \\
 & \qquad\qquad\quad\vdots
\end{align}
and so on for versions of $\kappa$ with more than two indices.  With this notation, \eqref{omegatilde} can be re-written as
\begin{equation}
 \tilde{\omega}_i=5i\frac{\kappa_i}{\kappa}\; .\eqlabel{omegat}
\end{equation} 
It is easy to check that the above moduli space metric~\eqref*{metric11} can be obtained from a ``K\"ahler potential'' $K^{(1,1)}$ as
\begin{equation}
 {\cal G}^{(1,1)}_{ij}=\partial_i\partial_j K^{(1,1)}, \qquad \text{where $K^{(1,1)}=-\frac{1}{2}\ln\kappa$} \; .
\end{equation} 
We can use the moduli space metric to define lower index moduli $t_i$ via $t_i={\cal G}^{(1,1)}_{ij}t^j$. From the explicit form~\eqref*{metric11} of the metric, it is easy to verify the useful relation
\be
 t_i=\frac{5\kappa_i}{2\kappa}\; . \eqlabel{tli}
\ee 
A further useful observation is related to ``metrics'' of the form
\be
 \tilde{\cal G}_{ij}={\cal G}^{(1,1)}_{ij}+c\frac{\kappa_i\kappa_j}{\kappa^2} \eqlabel{Gtilde}
\ee
for any real number $c$. A short calculation, using \eqref{tli} and $\kappa_it^i=\kappa$ repeatedly, shows that
\be
   \tilde{\cal G}_{ij}\left({\cal G}^{(1,1)jk}+\tilde{c}\frac{\kappa^j\kappa^k}{\kappa^2}\right)=\delta_i^k+\left(c+\tilde{c}+\frac{2}{5}c\tilde{c}\right)\frac{\kappa_i\kappa^k}{\kappa^2}\; , \eqlabel{GtGti}
\ee
where $\tilde{c}$ is an arbitrary real number. Here, the standard moduli space metric ${\cal G}^{(1,1)}_{ij}$ and its inverse ${\cal G}^{(1,1)ij}$ have been used to lower and raise indices. The above relation shows that for all $c\neq -5/2$ the metric~\eqref*{Gtilde} is invertible and that its inverse is given by
\be
 \tilde{\cal G}^{ij}={\cal G}^{(1,1)jk}+\tilde{c}\frac{\kappa^j\kappa^k}{\kappa^2}\; ,\qquad \tilde{c}=-\frac{5c}{5+2c}\; .\eqlabel{Gti}
\ee
These relations will be helpful when calculating the flux potential in the one-dimensional effective theory.

To summarise the main points, the K\"ahler moduli space for five-folds can be treated in complete analogy with the one for three-folds. The main difference is that the moduli space metric is now governed by a quintic pre-potential $\kappa$ instead of a cubic one for three-folds.

\vskip 0.4cm

We now move on to the complex structure moduli. Evaluating the standard moduli space metric~\eqref*{wp} for the $(2,0)$ variation of the metric in \eqref{deltag}, one finds
\be
 {\cal G}^{(1,4)}_{a\bar{b}} = \oneon{V||\Omega ||^2}\int_X\chi_a\wedge\ast\bar{\chi}_{\bar{b}}\; .
\ee
Using the result in \eqref{dualforms} for the Hodge dual of $(4,1)$ forms together with the relation $V||\Omega ||^2=i\int_X\Omega\wedge\bar\Omega$ then leads to the standard result
\be
	{\cal G}_{a\bar{b}}^{(1,4)} (\underline{z}, \underline{\bar{z}}) = \frac{\int_X \chi_a \wedge \bar{\chi}_{\bar{b}}}{\int_X \Omega\wedge\bar\Omega} \; . \eqlabel{metric41}
\ee
Kodaira's relation
\be\eqlabel{kodairas_relation}
 \frac{\partial\Omega}{\partial z^a}=k_a\Omega +\chi_a
\ee
can be shown exactly as in the case of Calabi-Yau three-folds~\cite{Candelas:1990pi}. It implies, via direct differentiation, that the moduli space metric~\eqref*{metric41} can be obtained from the K\"ahler potential $K^{(1,4)}$ as
\be
 {\cal G}^{(1,4)}_{a\bar{b}}=\partial_a\partial_{\bar{b}} K^{(1,4)}, \qquad\text{where $K^{(1,4)} = \ln \left[i \int_X \Omega \wedge \bar{\Omega}\right]$} \; . \eqlabel{K41}
\ee
In order to express $K^{(1,4)}$ more explicitly in terms of moduli, we introduce a symplectic basis $(A^{\cal A}, B_{\cal B})$ of five-cycles and a dual basis $(\alpha_{\cal A},\beta^{\cal B})$ of five-forms satisfying
\begin{equation}
	\int_{A^{\cal B}} \alpha_{\cal A} = \int_X \alpha_{\cal A} \wedge \beta^{\cal B} = \delta_{\cal A}^{\cal B}, \quad
	\int_{B_{\cal A}} \beta^{\cal B} = \int_X \beta^{\cal B} \wedge \alpha_{\cal A} = -\delta_{\cal A}^{\cal B} \; . \eqlabel{sympbasis}
\end{equation}
Then, the period integrals are defined in the usual way as
\begin{equation}\eqlabel{periods}
	{\cal Z}^{\cal A} = \int_{A^{\cal A}} \Omega , \qquad 	\mathcal{G}_{\cal A} = \int_{B_{\cal A}} \Omega.
\end{equation}
and the periods ${\cal G}_{\cal A}$ can be shown to be functions of ${\cal Z}^{\cal A}$, just as in the three-fold case.  In the dual basis $(\alpha_{\cal A},\beta^{\cal B})$ the $(5,0)$ form can then be expanded as $\Omega = {\cal Z}^{\cal A} \alpha_{\cal A} - \mathcal{G}_{\cal A} \beta^{\cal A}$ and inserting this into the expression~\eqref*{K41} for the K\"ahler potential yields
\begin{equation}
	K^{(1,4)} = \ln \left[i (\mathcal{G}_{\cal A} \bar{\cal Z}^{\cal A} - {\cal Z}^{\cal A} \bar{\mathcal{G}}_{\cal A})\right] \; .
\end{equation}
By virtue of Kodaira's relation, $\int_X\Omega\wedge\frac{\partial\Omega}{\partial {\cal Z}^{\cal A}}=0$ which immediately leads to ${\cal G}_{\cal A}=\frac{1}{2}\frac{\partial}{\partial {\cal Z}^{\cal A}}({\cal G}_{\cal B}{\cal Z}^{\cal B})$. Hence, the periods ${\cal G}_{\cal A}$ can be obtained as derivatives
\begin{equation}
 {\cal G}_{\cal A}=\frac{\partial {\cal G}}{\partial{\cal  Z}^{\cal A}}
\end{equation}
of a pre-potential ${\cal G}$ which is homogeneous of degree two in the projective coordinates ${\cal Z}^{\cal A}$. This is formally very similar to the three-fold case. However, an important difference is that the five-forms here contain not only $(5,0)$, $(0,5)$, $(4,1)$ and $(1,4)$ pieces but also $(3,2)$ and $(2,3)$ parts. That is, $\mathcal{A,B},\ldots = 0,1,\ldots, h^{1,4}+h^{2,3}$. As a consequence, the periods ${\cal Z}^{\cal A}$ do not simply serve as projective coordinates on the complex structure moduli space, though they can in principle be computed as functions of the $z^a$. However, their vast redundancy renders them much less useful as compared to the three-fold case.

\vskip 0.4cm

When flux is included, the one-dimensional effective theory depends on yet another set of moduli-dependent functions which arises from the contractions, $\tilde{\sigma}_e$ and $\tilde{\tilde{\sigma}}_e$, of the harmonic $(2,2)$ forms $\sigma_e$ which appear in the relation~\eqref*{dualforms} for the Hodge dual of $(2,2)$ forms. To explicitly compute these contractions, we note that $\tilde{\sigma}_e$ must be a harmonic $(1,1)$ form and can hence be expanded in terms of the basis $\omega_i$. Concretely, we write
\be
 \tilde{\sigma}_e=ik_e^i\omega_i
\ee
with some coefficients $k_e^i$ which, in general, depend on the $(1,1)$ moduli $t^i$.  Applying one more contraction to this relation and using \eqref{omegat} we learn that
\be 
 \tilde{\tilde{\sigma}}_e=-\frac{5}{\kappa}k_e^i\kappa_i\; .
\ee 
Hence, we can deal with all the contractions of harmonic $(2,2)$ forms if we are able to compute the coefficients $k_e^i$. This can be accomplished by multiplying \eqref{sigmaJJ} with $\omega_j$ and integrating over the Calabi-Yau five-fold $X$. This results in
\be
 k_e^i=\frac{1}{4V}\left({\cal G}^{(1,1)ij}-\frac{25}{6}\frac{\kappa^i\kappa^j}{\kappa^2}\right) d_{ejkl}t^kt^l\; , \eqlabel{kei}
\ee
where ${\cal G}^{(1,1)ij}$ is the inverse of ${\cal G}^{(1,1)}_{ij}$.

\subsubsection{Real vs. complex forms}\applabel{real_form_formalism}

For the purpose of disentangling and clarifying the intertwining of $(2,1)$- and $(1,3)$-modes with the complex structure moduli in the M-theory reduction, it turns out to be advantageous to revert to real harmonic 3- and 4-forms instead of their complex counterparts, namely harmonic $(2,1)$- and $(1,3)$-forms. In this subsection, we will investigate the relations between the two formulations.

Real harmonic 3-forms are naturally locked to 3-cycles and thus topologically invariant. The fact that $h^{3,0}(X)=0$ for Calabi-Yau five-folds ensures that a real 3-form\footnote{All differential forms occurring in this subsection are henceforth implicitly assumed to be harmonic.} is exclusively made up of a $(2,1)$- and a $(1,2)$-piece. However, the way in which a particular 3-form is split into $(2,1)$- and $(1,2)$-parts evidently depends on the choice of complex structure. We can parametrically represent this fact by introducing complex structure dependent linear maps $\mathfrak{A}$ and $\mathfrak{B}$ from real 3-forms to complex $(2,1)$-forms and vice versa. While this parametrization of the complex structure dependence in terms of unknown implicit maps turns out to be sufficient for the dimensional reduction we are carrying out in this paper, it would nonetheless be nice to find a way to calculate explicit expressions for these functions. However, this is beyond the scope of the present paper and will not be attempted here. 

For fixed bases, the linear maps have a matrix representation according to
\begin{align}
	\nu_p &= \mathfrak{A}_p {}^{\cal{Q}} N_{\cal{Q}} \qquad \qquad \qquad 
	\text{(and: $\bar{\nu}_{\bar{p}} = \bar{\mathfrak{A}}_{\bar{p}} {}^{\cal{Q}} N_{\cal{Q}}$)} \; ,\eqlabel{21_3_rel1}\\
	N_{\cal{P}} &= \mathfrak{B}_{\cal{P}} {}^q \nu_q + \bar{\mathfrak{B}}_{\cal{P}} {}^{\bar{q}} \bar{\nu}_{\bar{q}} \; ,\eqlabel{21_3_rel2}
\end{align}
where $\{ N_{\cal{P}} \}_{\mathcal{P}=1,\ldots,b^{3}(X)}$ is a real basis of $H^{3} (X)$ and $\{ \nu_p \}_{p=1,\ldots,h^{2,1}(X)}$ is a basis of $H^{(2,1)} (X)$. To avoid confusion with symbols defined elsewhere, we use Fraktur font letters to denote maps translating between real and complex forms and calligraphic letters for real form indices. Note that $\mathfrak{A}_p {}^{\cal{Q}}$ and $\mathfrak{B}_{\cal{P}} {}^q$ are complex and have dependence $\mathfrak{A}_p {}^{\cal{Q}} = \mathfrak{A}_p {}^{\cal{Q}} (\underline{z},\bar{\underline{z}})$, $\mathfrak{B}_{\cal{P}} {}^q = \mathfrak{B}_{\cal{P}} {}^q (\underline{z},\bar{\underline{z}})$, where $z^a$ and $\bar{z}^{\bar{a}}$ are the complex structure moduli of the Calabi-Yau five-fold. The equations above have two faces, for they can either be written in local real ten dimensional coordinates or in local (complex) Darboux coordinates. For example, \eqref{21_3_rel2} in real coordinates is
\be
	N_{\mathcal{P},m_1 m_2 m_3} 	= \mathfrak{B}_{\cal{P}} {}^q \nu_{q,m_1 m_2 m_3} 
								+ \bar{\mathfrak{B}}_{\cal{P}} {}^{\bar{q}} \bar{\nu}_{\bar{q},m_1 m_2 m_3} \; ,
\ee
whereas in Darboux coordinates it reads
\be\eqlabel{21_3_cplx_coords}
	N_{\mathcal{P},\mu_1 \mu_2 \bar{\nu}} = \mathfrak{B}_{\cal{P}} {}^q \nu_{q,\mu_1 \mu_2 \bar{\nu}} \; , \qquad \text{(and $\cc$)} \; ,
\ee
where forms with unnatural index types are to be translated manually using \eqref{real_complex_coords}. Inserting \eqref{21_3_rel1} into \eqref{21_3_rel2} and vice versa, we learn relations between the $\mathfrak{A}$ and $\mathfrak{B}$ maps:
\begin{align}
	& \mathfrak{A}_p {}^{\cal{Q}} \mathfrak{B}_{\cal{Q}} {}^q = \delta_p {}^q &\text{(and $\cc$)} \; , \eqlabel{AB_rel1} \\
	& \mathfrak{A}_p {}^{\cal{Q}} \bar{\mathfrak{B}}_{\cal{Q}} {}^{\bar{q}} = 0 &\text{(and $\cc$)} \; , \eqlabel{AB_rel2} \\
	& \mathfrak{B}_{\cal{P}} {}^q \mathfrak{A}_q {}^{\cal{Q}} + \bar{\mathfrak{B}}_{\cal{P}} {}^{\bar{q}} \bar{\mathfrak{A}}_{\bar{q}} {}^{\cal{Q}} = 
		\delta_{\cal{P}} {}^{\cal{Q}} \; . &\eqlabel{AB_rel3} 
\end{align}
For the complex structure dependence, one finds:
\begin{align}
	\partial_a N_{\cal{P}} &= 0 \; , &
	\partial_a \nu_p &= 
		\mathfrak{A}_p {}^{\cal{Q}} {}_{,a} \mathfrak{B}_{\cal{Q}} {}^q \nu_q + 
		\mathfrak{A}_p {}^{\cal{Q}} {}_{,a} \bar{\mathfrak{B}}_{\cal{Q}} {}^{\bar{q}} \bar{\nu}_{\bar{q}} \; , \\
	\partial_{\bar{a}} N_{\cal{P}} &= 0 \; , &
	\partial_{\bar{a}} \nu_p &= 
		\mathfrak{A}_p {}^{\cal{Q}} {}_{,\bar{a}} \mathfrak{B}_{\cal{Q}} {}^q \nu_q + 
		\mathfrak{A}_p {}^{\cal{Q}} {}_{,\bar{a}} \bar{\mathfrak{B}}_{\cal{Q}} {}^{\bar{q}} \bar{\nu}_{\bar{q}} \; .
\end{align}
Using \eqrangeref{21_3_rel1}{21_3_rel2} and \eqref{dualforms}, one can compute the Hodge star of the real 3-form $N_{\cal{P}}$:
\be\eqlabel{3form_hodgestar}
	\ast N_{\cal{P}} = \oneon{2} \Delta_{\cal{P}} {}^{\cal{Q}} N_{\cal{Q}} \wedge J^2 \; , 
\ee
where $\Delta_{\cal{P}} {}^{\cal{Q}} := i(\mathfrak{B}_{\cal{P}} {}^q \mathfrak{A}_q {}^{\cal{Q}} - \bar{\mathfrak{B}}_{\cal{P}} {}^{\bar{q}} \bar{\mathfrak{A}}_{\bar{q}} {}^{\cal{Q}})$. The linear map $\Delta$ provides a complex structure on the \emph{moduli space} of real 3-forms induced by the complex structure of the Calabi-Yau five-fold itself. It satisfies
\be\eqlabel{3form_cplx_struct_rel}
	\Delta_{\cal{P}} {}^{\cal{Q}} \Delta_{\cal{Q}} {}^{\cal{R}} = - \delta_{\cal{P}} {}^{\cal{R}} \; , \qquad\qquad
	(\Delta_{\cal{P}} {}^{\cal{Q}})^\ast = \Delta_{\cal{P}} {}^{\cal{Q}} \; , \qquad\qquad
	\tr \, \Delta = 0 \; .
\ee
Using the complex structure $\Delta$, we define projection operators
\be\eqlabel{3form_projectors_def}
	P_{\pm} {}_{\cal{P}} {}^{\cal{Q}} := \oneon{2} (\id \mp i \Delta)_{\cal{P}} {}^{\cal{Q}}
\ee
satisfying
\be\eqlabel{3form_projectors_rel}
	P_{\pm} {}_{\cal{P}} {}^{\cal{Q}} P_{\pm} {}_{\cal{Q}} {}^{\cal{R}} = P_{\pm} {}_{\cal{P}} {}^{\cal{R}} \; , \qquad
	P_{+} {}_{\cal{P}} {}^{\cal{Q}} P_{-} {}_{\cal{Q}} {}^{\cal{R}} = P_{-} {}_{\cal{P}} {}^{\cal{Q}} P_{+} {}_{\cal{Q}} {}^{\cal{R}} = 0 \; , \qquad
	(P_{\pm} {}_{\cal{P}} {}^{\cal{Q}})^\ast = P_{\mp} {}_{\cal{P}} {}^{\cal{Q}} \; .
\ee
In terms of the $\mathfrak{A}$ and $\mathfrak{B}$ maps, they are explicitly given by
\be
	P_{+} {}_{\cal{P}} {}^{\cal{Q}} = \mathfrak{B}_{\cal{P}} {}^q \mathfrak{A}_q {}^{\cal{Q}} \; , \qquad\qquad
	P_{-} {}_{\cal{P}} {}^{\cal{Q}} = \bar{\mathfrak{B}}_{\cal{P}} {}^{\bar{q}} \bar{\mathfrak{A}}_{\bar{q}} {}^{\cal{Q}} \; .
\ee
The standard metric on the moduli space of real 3-forms is
\be
	{\cal G}^{(3)}_{\cal{PQ}} = \int_X N_{\cal{P}} \wedge \ast N_{\cal{Q}} \; .
\ee
Using the expression for the Hodge star \eqref*{3form_hodgestar}, we can rewrite this so as to make the dependence on the moduli more explicit:
\be
	{\cal G}^{(3)}_{\cal{PQ}} (\underline{t},\underline{z},\underline{\bar{z}}) = 
		\oneon{2} \Delta_{\cal{\left(P\right.}} {}^{\cal{R}} d_{\left.\mathcal{Q}\right)\mathcal{R}ij} t^i t^j \; ,
\ee
where we have defined a new intersection number $d_{\mathcal{P}\mathcal{Q}ij} := \int_X N_{\cal{P}} \wedge N_{\cal{Q}} \wedge \omega_i \wedge \omega_j$, which is purely topological. Note that $d_{\mathcal{P}\mathcal{Q}ij} = - d_{\mathcal{Q}\mathcal{P}ij}$. The metric anti-commutes with the complex structure:
\be
	\Delta_{\cal{P}} {}^{\cal{Q}} \mathcal{G}^{(3)}_{\cal{QR}} + \mathcal{G}^{(3)}_{\cal{PQ}} \Delta_{\cal{R}} {}^{\cal{Q}} = 0 \; , 
\ee
which, in fact, becomes a Hermiticity condition on the metric $\mathcal{G}^{(3)}$:
\be\eqlabel{G3_hermitian}
	\mathcal{G}^{(3)}_{\cal{PQ}} = \Delta_{\cal{P}} {}^{\cal{R}} \Delta_{\cal{Q}} {}^{\cal{S}} \mathcal{G}^{(3)}_{\cal{RS}} \; .
\ee
Thus, the 3-form moduli space is a Hermitian manifold with $\mathcal{G}^{(3)}$ being a Hermitian metric.

\vskip 0.4cm

A real 4-form, which is topologically invariant, can be decomposed into the sum of $(1,3)$-, $(3,1)$- and $(2,2)$-forms using the complex structure of the Calabi-Yau five-fold $X$. In the same spirit as for the 3-forms, we introduce linear maps $\mathfrak{C}$, $\mathfrak{D}$, $\mathfrak{E}$ and $\mathfrak{F}$ to translate between real 4-forms and their $(1,3)$-, $(3,1)$- and $(2,2)$-pieces:
\begin{align}
	\varpi_x &= \mathfrak{C}_x {}^{\cal{X}} O_{\cal{X}} \qquad \qquad \qquad 
	\text{(and: $\bar{\varpi}_{\bar{x}} = \bar{\mathfrak{C}}_{\bar{x}} {}^{\cal{X}} O_{\cal{X}}$)} \; ,\eqlabel{31_4_rel1} \\
	\sigma_e &= \mathfrak{E}_e {}^{\cal{X}} O_{\cal{X}} \; , \eqlabel{31_4_rel2} \\
	O_{\cal{X}} &= \mathfrak{D}_{\cal{X}} {}^x \varpi_x + \bar{\mathfrak{D}}_{\cal{X}} {}^{\bar{x}} \bar{\varpi}_{\bar{x}} 
			    + \mathfrak{F}_{\cal{X}} {}^e \sigma_e \; , \eqlabel{31_4_rel3} 
\end{align}
where $\{ \varpi_x \}$ is a basis of $H^{(1,3)} (X)$, whereas $\{ \sigma_e \}$ and $\{ O_{\cal{X}} \}$ are real bases of $H^{(2,2)} (X)$ and $H^{4} (X)$, respectively. Unlike $\mathfrak{C}$ and $\mathfrak{D}$, $\mathfrak{E}$ and $\mathfrak{F}$ are real. All linear maps $\mathfrak{C}$, $\mathfrak{D}$, $\mathfrak{E}$ and $\mathfrak{F}$ \emph{a priori} depend on the complex structure moduli $z^a$ and $\bar{z}^{\bar{a}}$. By consecutively inserting \eqrangeref{31_4_rel1}{31_4_rel3} into each other, we learn relations among the linear maps
\begin{align}
	&\mathfrak{C}_x {}^{\cal{X}} \mathfrak{D}_{\cal{X}} {}^y = \delta_x {}^y \; , \qquad
	\bar{\mathfrak{C}}_{\bar{x}} {}^{\cal{X}} \bar{\mathfrak{D}}_{\cal{X}} {}^{\bar{y}} = \delta_{\bar{x}} {}^{\bar{y}} \; , \qquad
	\mathfrak{E}_e {}^{\cal{X}} \mathfrak{F}_{\cal{X}} {}^f = \delta_e {}^f \; , \eqlabel{CD_rel1} \\
	&\mathfrak{C}_x {}^{\cal{X}} \bar{\mathfrak{D}}_{\cal{X}} {}^{\bar{y}} = 
	\mathfrak{C}_x {}^{\cal{X}} \mathfrak{F}_{\cal{X}} {}^e =
	\mathfrak{E}_e {}^{\cal{X}} \mathfrak{D}_{\cal{X}} {}^x = 0 \; , \qquad\qquad\qquad \text{(and $\cc$)} \; , \eqlabel{CD_rel2} \\
	&\mathfrak{D}_{\cal{X}} {}^x \mathfrak{C}_x {}^{\cal{Y}} + 
	\bar{\mathfrak{D}}_{\cal{X}} {}^{\bar{x}} \bar{\mathfrak{C}}_{\bar{x}} {}^{\cal{Y}} +
	\mathfrak{F}_{\cal{X}} {}^e \mathfrak{E}_e {}^{\cal{Y}} = \delta_{\cal{X}} {}^{\cal{Y}} \; . \eqlabel{CD_rel3}
\end{align}

\vskip 0.4cm

The wedge product of two harmonic $(1,1)$-forms is a harmonic $(2,2)$-form. For the purpose of this paper, we will restrict attention to the case where all $(2,2)$-forms are obtained by wedging together two $(1,1)$-forms, that is we require\footnote{In the Calabi-Yau four-fold literature, the right hand side of \eqref{22_vertical} is often referred to as the vertical part, denoted $H^{(2,2)}_V$, of $H^{(2,2)}$ (see, for example, Ref.~\cite{Haack:2002tu}). The total space $H^{(2,2)}$ is given by $H^{(2,2)} = H^{(2,2)}_V \oplus H^{(2,2)}_H$, where $H^{(2,2)}_H$ comprises all $(2,2)$-forms that can \emph{not} be obtained by the product of two $(1,1)$-forms. In this terminology, we are considering Calabi-Yau five-folds $X$ for which $H^{(2,2)} (X) = H^{(2,2)}_V (X)$ and $H^{(2,2)}_H (X) = 0$.}
\be\eqlabel{22_vertical}
	H^{(2,2)}(X) = H^{(1,1)}(X) \wedge H^{(1,1)}(X) \; .
\ee
All examples of Calabi-Yau five-folds presented in \appref{cyexamples} satisfy \eqref{22_vertical}. The significance of this restriction is that, since the $(1,1)$-forms (being naturally locked to 2-cycles) are independent of the complex structure, so are the $(2,2)$-forms if they are entirely generated by the square of $(1,1)$-forms. This implies that $\sigma_e$, $\mathfrak{E}_e {}^{\cal{X}}$ and $\mathfrak{F}_{\cal{X}} {}^e$ are all independent of the complex structure moduli (or of any moduli fields, in fact). Since the left hand side and the last term on the right hand side of \eqref{31_4_rel3} are independent of the complex structure, the same must be true for the \emph{sum} of the first two terms on the right hand side. This observation allows us to treat the $(1,3)$ and $(3,1)$ part together in a complex structure independent way.

Let us now choose the basis $\{ O_{\cal{X}} \}$ such that the first $2 \, h^{1,3}(X)$ indices lie in the $(1,3)+(3,1)$ directions and the remaining indices lie in the $(2,2)$ direction, that is we divide the index range $\mathcal{X}=(\hat{\mathcal{X}},\tilde{\mathcal{X}})$, where $\hat{\mathcal{X}}=1,\ldots,2h^{1,3}(X)$ and $\tilde{\mathcal{X}}=1,\ldots,h^{2,2}(X)$. This rearrangement is also independent of the complex structure. \Eqrangeref{31_4_rel1}{31_4_rel3} then become
\begin{align}
	\varpi_x &= \mathfrak{C}_x {}^{\hat{\cal{X}}} O_{\hat{\cal{X}}} \qquad \qquad \qquad 
	\text{(and: $\bar{\varpi}_{\bar{x}} = \bar{\mathfrak{C}}_{\bar{x}} {}^{\hat{\cal{X}}} O_{\hat{\cal{X}}}$)} \; ,\eqlabel{31_4_new_rel1} \\
	\sigma_e &= \mathfrak{E}_e {}^{\tilde{\cal{X}}} O_{\tilde{\cal{X}}} \; , \eqlabel{31_4_new_rel2} \\
	O_{\hat{\cal{X}}} &= \mathfrak{D}_{\hat{\cal{X}}} {}^x \varpi_x + \bar{\mathfrak{D}}_{\hat{\cal{X}}} {}^{\bar{x}} \bar{\varpi}_{\bar{x}}, \qquad
	O_{\tilde{\cal{X}}} = \mathfrak{F}_{\tilde{\cal{X}}} {}^e \sigma_e \; , \eqlabel{31_4_new_rel3} 
\end{align}
where $O_{\hat{\cal{X}}}$, $O_{\tilde{\cal{X}}}$, $\mathfrak{F}_{\tilde{\cal{X}}} {}^e$, $\mathfrak{E}_e {}^{\tilde{\cal{X}}}$ and $\sigma_e$ are independent of the complex structure moduli, whereas all other objects are dependent on them. Instead of \eqrangeref{CD_rel1}{CD_rel3} we have
\begin{align}
	&\mathfrak{C}_x {}^{\hat{\cal{X}}} \mathfrak{D}_{\hat{\cal{X}}} {}^y = \delta_x {}^y \; , \qquad
	\bar{\mathfrak{C}}_{\bar{x}} {}^{\hat{\cal{X}}} \bar{\mathfrak{D}}_{\hat{\cal{X}}} {}^{\bar{y}} = \delta_{\bar{x}} {}^{\bar{y}} \; , \qquad
	\mathfrak{E}_e {}^{\tilde{\cal{X}}} \mathfrak{F}_{\tilde{\cal{X}}} {}^f = \delta_e {}^f \; , \eqlabel{CD_new_rel1} \\
	&\mathfrak{C}_x {}^{\hat{\cal{X}}} \bar{\mathfrak{D}}_{\hat{\cal{X}}} {}^{\bar{y}} = 0 \; , 
		\qquad\qquad\qquad \text{(and $\cc$)} \; , \eqlabel{CD_new_rel2} \\
	&\mathfrak{D}_{\hat{\cal{X}}} {}^x \mathfrak{C}_x {}^{\hat{\cal{Y}}} + 
	\bar{\mathfrak{D}}_{\hat{\cal{X}}} {}^{\bar{x}} \bar{\mathfrak{C}}_{\bar{x}} {}^{\hat{\cal{Y}}} = \delta_{\hat{\cal{X}}} {}^{\hat{\cal{Y}}} \; , \qquad
	\mathfrak{F}_{\tilde{\cal{X}}} {}^e \mathfrak{E}_e {}^{\tilde{\cal{Y}}} = \delta_{\tilde{\cal{X}}} {}^{\tilde{\cal{Y}}} \; . \eqlabel{CD_new_rel3}
\end{align}
The relations between $\mathfrak{C}_x {}^{\hat{\cal{X}}}$, $\mathfrak{D}_{\hat{\cal{X}}} {}^y$, $O_{\hat{\cal{X}}}$ and $\varpi_x$ are very similar to the relations between $\mathfrak{A}_p {}^{\cal{P}}$, $\mathfrak{B}_{\cal{P}} {}^q$, $N_{\cal{P}}$ and $\nu_p$ for the 3-form case discussed above. The complex structure dependence in the $(1,3)$-sector is parametrized by $\mathfrak{C}_x {}^{\hat{\cal{X}}}$ and $\mathfrak{D}_{\hat{\cal{X}}} {}^y$
\begin{align}
	\partial_a O_{\hat{\cal{X}}} &= 0 \; , &
	\partial_a \varpi_x &= 
		\mathfrak{C}_x {}^{\hat{\cal{Y}}} {}_{,a} \mathfrak{D}_{\hat{\cal{Y}}} {}^y \varpi_y + 
		\mathfrak{C}_x {}^{\hat{\cal{Y}}} {}_{,a} \bar{\mathfrak{D}}_{\hat{\cal{Y}}} {}^{\bar{y}} \bar{\varpi}_{\bar{y}} \; , \\
	\partial_{\bar{a}} O_{\hat{\cal{X}}} &= 0 \; , &
	\partial_{\bar{a}} \varpi_x &= 
		\mathfrak{C}_x {}^{\hat{\cal{Y}}} {}_{,\bar{a}} \mathfrak{D}_{\hat{\cal{Y}}} {}^y \varpi_y + 
		\mathfrak{C}_x {}^{\hat{\cal{Y}}} {}_{,\bar{a}} \bar{\mathfrak{D}}_{\hat{\cal{Y}}} {}^{\bar{y}} \bar{\varpi}_{\bar{y}} \; .
\end{align}
Using \eqref{31_4_new_rel1,31_4_new_rel3,dualforms}, one can compute the Hodge star of the real 4-form $O_{\hat{\cal{X}}}$:
\be\eqlabel{4form_hodgestar}
	\ast O_{\hat{\cal{X}}} = - O_{\hat{\cal{X}}} \wedge J \; .
\ee
Whenever we use the forms $O_{\hat{\cal{X}}}$ to describe $(1,3)$- and $(3,1)$-forms we will refer to it as the $\hat{4}$-form formulation. The standard metric on the moduli space of real $\hat{4}$-forms is given by
\be\eqlabel{4form_metric}
	{\cal G}^{(\hat{4})}_{\cal{\hat{X}\hat{Y}}} = \int_X O_{\hat{\cal{X}}} \wedge \ast O_{\hat{\cal{Y}}} \; .
\ee
Using the expression for the Hodge star \eqref*{4form_hodgestar}, we can rewrite this so as to make the dependence on the moduli more explicit:
\be
	{\cal G}^{(\hat{4})}_{\cal{\hat{X}\hat{Y}}} (\underline{t}) = 
		- d_{\mathcal{\hat{X}}\mathcal{\hat{Y}}i} t^i \; ,
\ee
where we have defined a new intersection number $d_{\mathcal{\hat{X}}\mathcal{\hat{Y}}i} := \int_X O_{\hat{\cal{X}}} \wedge O_{\hat{\cal{Y}}} \wedge \omega_i$, which is purely topological. Note that $d_{\mathcal{\hat{X}}\mathcal{\hat{Y}}i} = d_{\mathcal{\hat{Y}}\mathcal{\hat{X}}i}$.

Similarly to the 3-form case, there is a complex structure $\Delta_{\hat{\cal{X}}} {}^{\hat{\cal{Y}}}$ on the $\hat{4}$-form moduli space inherited from the complex structure of the Calabi-Yau five-fold and given by
\be
	\Delta_{\hat{\cal{X}}} {}^{\hat{\cal{Y}}} = i (\mathfrak{D}_{\hat{\cal{X}}} {}^x \mathfrak{C}_x {}^{\hat{\cal{Y}}} - 
	\bar{\mathfrak{D}}_{\hat{\cal{X}}} {}^{\bar{x}} \bar{\mathfrak{C}}_{\bar{x}} {}^{\hat{\cal{Y}}}) \; .
\ee
It satisfies the relations of \eqref{3form_cplx_struct_rel}. The projection operators are
\be\eqlabel{4form_projectors_def}
	P_{\pm} {}_{\hat{\cal{X}}} {}^{\hat{\cal{Y}}} := \oneon{2} (\id \mp i \Delta)_{\hat{\cal{X}}} {}^{\hat{\cal{Y}}} \; ,
\ee
which satisfy \eqref{3form_projectors_rel} and are explicitly given by
\be
	P_{+} {}_{\hat{\cal{X}}} {}^{\hat{\cal{Y}}} = \mathfrak{D}_{\hat{\cal{X}}} {}^y \mathfrak{C}_y {}^{\hat{\cal{Y}}} \; , \qquad\qquad
	P_{-} {}_{\hat{\cal{X}}} {}^{\hat{\cal{Y}}} = \bar{\mathfrak{D}}_{\hat{\cal{X}}} {}^{\bar{y}} \bar{\mathfrak{C}}_{\bar{y}} {}^{\hat{\cal{Y}}} \; .
\ee
Note, however, that unlike in the 3-form case, the standard $\hat{4}$-form metric~\eqref*{4form_metric} is not Hermitian with respect to the complex structure $\Delta_{\hat{\cal{X}}} {}^{\hat{\cal{Y}}}$.

\section{$\susyno=2$ supersymmetry in one dimension}\applabel{superspace}

In this appendix we will review and develop one-dimensional ${\cal N}=2$ supersymmetry to the level necessary for the theories which arise from our M-theory reductions. One-dimensional supersymmetry has previously been discussed in the literature (see, for example,~\cite{Coles:1990hr,vanHolten:1995qt,Machin:2002zn} and references therein), notably in the context of black hole moduli spaces~\cite{Gibbons:1997iy}. However, to describe the effective actions which arise from M-theory reduction on Calabi-Yau five-folds a number of generalisations and extensions of the one-dimensional ${\cal N}=2$ theories studied in the literature are required. For example, we find that we require theories in which the two main types of multiplets, the $2a$ and $2b$ multiplets, are coupled. Some of the five-fold zero modes fall into fermionic ($2b$) multiplets so we need to introduce and develop these multiplets properly. Even though gravity in one dimension is non-dynamical, it leads to constraints which cannot be ignored. This means we need to consider one-dimensional {\em local} supersymmetry. Finally, when we include M-theory four-form flux we need to incorporate a potential and an associated superpotential into the $2a$ sector of the theory. All those features have not been fully worked out in the literature. We have, therefore, opted for a systematic exposition of one-dimensional ${\cal N}=2$ global and local supersymmetry, in order to develop a solid base for our application to M-theory.

\subsection{Global $\susyno=2$ supersymmetry}

Before turning to one-dimensional $\susyno=2$ \emph{curved} superspace, we will briefly recapitulate the case of global ${\cal N}=2$ supersymmetry in one dimension~\cite{Coles:1990hr}. One-dimensional superspace (\emph{supertime}) is most easily obtained by dimensional reduction from $d=2$, which has attracted a lot of attention in view of formulating superstring actions in superspace~\cite{Howe:1977xz,Martinec:1983um}. In $d=2$, there are Majorana, Weyl and Majorana-Weyl spinors and hence the same amount of supersymmetry can be realized by different choices of spinorial representation for the supercharges (see, for example, Ref.~\cite{West:1990tg}). For $\susyno=2$, the two options are $(1,1)$ and $(2,0)$ supersymmetry.

Upon reduction to one dimension, these two choices for two-dimensional ${\cal N}=2$ supersymmetry lead to two different one-dimensional ${\cal N}=2$ super multiplets, referred to as $2a$ (descending from two-dimensional $(1,1)$ supersymmetry) and $2b$ (descending from two-dimensional $(2,0)$ supersymmetry) multiplets. These two multiplets will play a central r\^{o}le in our discussion. Off-shell, the $2a$ multiplet contains a real scalar as its lowest component plus a complex fermion and a real scalar auxiliary field while the $2b$ multiplets contains a complex scalar as its lowest component, accompanied by a complex fermion. The $2b$ multiplet does not contain an auxiliary field.  Other off-shell multiplets, not obtained from a standard toroidal reduction, are the fermionic $2a$ and $2b$ multiplets and the non-linear multiplet~\cite{vanHolten:1995qt}. From those we will only need and discuss in detail the fermionic $2b$ multiplet. It has a complex fermion as its lowest component which is balanced by a complex scalar at the next level. 

Flat $\susyno=2$ supertime, $\R^{1|2}$, is  parametrised by coordinates $\{x^0=\tau;\theta,\bar\theta \}$, where $\theta$ is a complex one-dimensional spinor. In the following, we use indices $A,B,\ldots = 0,\theta,\bar{\theta}$ to label supertime tensors. The supersymmetry algebra is generated by two supercharges $Q$ and $\bar Q$ defined as
\be\eqlabel{sspace_flat_superchrg_defn}
	Q = \partial_\theta - \frac{i}{2} \bar\theta \partial_0 , \qquad
	\bar{Q} = - \partial_{\bar\theta} + \frac{i}{2} \theta \partial_0
\ee	
where $\partial_\theta = \frac{\partial}{\partial \theta}$, $\partial_{\bar\theta} = \frac{\partial}{\partial \bar\theta} = - \left(\partial_\theta\right)^\ast$, $\partial_0 = \frac{\partial}{\partial x^0} = \frac{\partial}{\partial\tau}$. Using the conventions for one-dimensional spinors summarised in \appref{conv} it is easy to verify that they satisfy the algebra
\be	
  \{Q, \bar{Q} \} = i \partial_0 = H , \quad \{ Q,Q \} = 0, \quad \{ \bar{Q},\bar{Q} \} = 0 \; .
\ee
Supersymmetry transformations of ${\cal N}=2$ supertime are parameterised by a complex one-di\-men\-sio\-nal spinor $\epsilon$ and act as
\be
 \delta_\epsilon = i \epsilon Q\; ,\qquad \delta_{\bar{\epsilon}} = i \bar{\epsilon}\bar{Q}\; . \eqlabel{susytransform}
\ee
This choice ensures that the total supersymmetry variation $\delta_{\epsilon,\text{tot.}} = \delta_\epsilon + \delta_{\bar{\epsilon}}$ is real. As usual, we introduce the associated covariant derivatives $D$ and $\bar{D}$ which anti-commute with the supercharges, that is $\{D,Q\} = \{D,\bar{Q}\} = \{\bar{D},Q\} = \{\bar{D},\bar{Q}\} = 0$, and are explicitly given by
\be\eqlabel{sspace_flat_cov_deriv_defn}
  D = \partial_\theta + \frac{i}{2} \bar\theta \partial_0 , \qquad \bar{D} = - \partial_{\bar\theta} - \frac{i}{2} \theta \partial_0\; .
\ee
They satisfy the anti-commutation relations
\be	
  \{D,\bar{D}\} = - i \partial_0 = - H, \quad \{D,D\} = 0,\quad \{\bar{D},\bar{D}\} = 0\; .
\ee
Although not really required for the global case it is useful for comparison with local supersymmetry later on to develop the geometry of flat supertime. To this end, we introduce the notation $(\partial_A)=(\partial_0,\partial_\theta ,\partial_{\bar{\theta}})$ for the partial derivatives and similarly for the covariant derivatives, $(D_\UA)=(D_0,D_\theta ,D_{\bar{\theta}})$. These two types of derivatives are generally related by
\be 
 D_\UA = E_\UA {}^B \partial_B\; , \eqlabel{Dpartial}
\ee
where $E_\UA {}^B$ is the inverse of the supervielbein $E_B{}^\UA$, that is $E_\UA {}^C E_C {}^\UB = \delta_\UA {}^\UB$. For flat supertime we have $D_{\underline{0}}=\partial_0$, $D_\theta = D$ and $D_{\bar{\theta}}=\bar{D}$ with $D$ and $\bar{D}$ given in \eqref{sspace_flat_cov_deriv_defn}. A short computation using \eqref{Dpartial} then shows that the supervielbein of flat supertime is given by
\bea\eqlabel{sspace_flat_seinbein}
	&E_0 {}^{\underline{0}} = 1 , 				&\quad &  E_0 {}^\Uth = 0 , 			&\quad &  E_0 {}^\BUth = 0 , \\
	&E_\theta {}^\UZ = -\frac{i}{2} \bar\theta , 	&\quad &  E_\theta {}^\Uth = 1 , 		&\quad & E_\theta {}^\BUth = 0 , \\
	&E_{\bar{\theta}} {}^\UZ = -\frac{i}{2} \theta , 	&\quad &  E_{\bar{\theta}} {}^\Uth = 0 ,	&\quad &  E_{\bar{\theta}} {}^\BUth = -1 .
\eea
The torsion tensor $T_{\UA\UB} {}^{\underline{C}}$ and curvature tensor $R_{\UA\UB}{}^{rs}$ can be obtained from the general relation~\cite{West:1990tg,Wess:1992cp}
\be\eqlabel{sspace_flat_torsion_curvature_defn}
	\left[ D_\UA, D_\UB \right\} = 
		- T_{\UA\UB} {}^{\underline{C}} D_{\underline{C}} - R_{\UA\UB}{}^{rs} M_{rs} , 
\ee
where $M_{rs}$ are the Lorentz generators. In $d=1$, the Lorentz indices only run over one value and hence the single Lorentz generator and the curvature tensor vanish. To compute the torsion tensor of flat superspace we use the flat superspace covariant derivatives~\eqref*{sspace_flat_cov_deriv_defn} in the above relation~\eqref*{sspace_flat_torsion_curvature_defn} for the torsion tensor. One finds that the only non-vanishing component is
\be\eqlabel{sspace_flat_torsion_value}
	T_{\Uth\BUth} {}^{\underline{0}} = i \; .
\ee
Finally, we find for the super-determinant of the flat supervielbein~\eqref*{sspace_flat_seinbein}
\be
  \mathrm{sdet} E_A{}^{\underline{B}} = -1\; .
\ee

\vskip .4cm

Now we need to introduce superfields. One-dimensional ${\cal N}=2$ superfields are functions of the supertime coordinates $\tau$, $\theta$ and $\bar{\theta}$. As usual, their component field content can be worked out by expanding in $\theta$ and $\bar{\theta}$. Since $\theta^2=\bar{\theta}^2=0$, only the terms proportional to $\theta$, $\bar{\theta}$ and $\theta\bar{\theta}$ arise, in addition to the lowest, $\theta$-independent component. Different types of irreducible superfields can be obtained by imposing constraints on this general superfield. We now discuss these various types in turn.

A $2a$ superfield $\phi =\phi(\tau,\theta ,\bar{\theta})$ is a real superfield, that is, a superfield satisfying the constraint $\phi =\phi^\dagger$. A short calculation shows that the most general component expansion consistent with this constraint is
\be\eqlabel{sspace_flat_2a_comp}
 \phi = \varphi + i \theta\psi + i\bar\theta \bar\psi + \oneon{2} \theta\bar\theta f\; ,
\ee
where $\varphi$ and $f$ are real scalars and $\psi$ is a complex fermion. The highest component $f$ will turn out to be an auxiliary field so that a $2a$ superfield contains one real physical scalar field. From \eqref{sspace_flat_superchrg_defn,susytransform}, the supersymmetry transformations of these components are given by
\begin{align}
 &\delta_\epsilon \varphi = - \epsilon \psi, \quad \delta_\epsilon \psi = 0, \quad 
					\delta_\epsilon \bar\psi = \frac{i}{2} \epsilon \dot{\varphi} - \oneon{2} \epsilon f , \quad
					\delta_\epsilon f = - i \epsilon \dot{\psi}, \eqlabel{sspace_flat_transf_2a} \\
 &\delta_{\bar\epsilon} \varphi = \bar\epsilon \bar\psi, \quad \delta_{\bar\epsilon} \bar\psi = 0, \quad
                                             \delta_{\bar\epsilon} \psi = - \frac{i}{2} \bar\epsilon\dot{\varphi} - \oneon{2} \bar\epsilon f,\quad
					\delta_{\bar\epsilon} f = - i\bar\epsilon \dot{\bar\psi} \; . \eqlabel{sspace_flat_transf_2a_bar}
\end{align}
For a set, $\{\phi^i\}$, of $2a$ superfields the most general non-linear sigma model\footnote{For an introduction to supersymmetric non-linear sigma models in one and two dimensions, see, for example, Ref.~\cite{Machin:2002zn}.} can be written in superspace as~\cite{Coles:1990hr,Gibbons:1997iy,Hull:1999ng}
\be
S_{2a} = \oneon{4} \int d\tau d^2\theta \left\{ (G(\underline{\phi})+B(\underline{\phi}))_{ij} D \phi^i \bar{D} \phi^j + L_{ij}(\underline{\phi}) D \phi^i D \phi^j+ M_{ij}(\underline{\phi}) \bar{D} \phi^i \bar{D} \phi^j + {\cal W}(\underline{\phi}) \right\}\; , \eqlabel{nlsm_2a} 
\ee
where $G_{ij}$ is symmetric, $B_{ij}$, $L_{ij}$, $M_{ij}$ are anti-symmetric and ${\cal W}$ is an arbitrary function of $\phi^i$. The component version of ${\cal W}(\underline{\phi})$ is obtained by a Taylor expansion about $\varphi^i$: 
\be
	{\cal W}(\underline{\phi}) = {\cal W}(\underline{\varphi}) + i\theta\psi^i {\cal W}_{,i}(\underline{\varphi})
		+ i\bar\theta\bar\psi^i {\cal W}_{,i}(\underline{\varphi}) 
		+ \oneon{2} \theta\bar\theta ({\cal W}_{,i}(\underline{\varphi}) f^i + 2 {\cal W}_{,ij}(\underline{\varphi}) \psi^i \bar\psi^j ) .
\ee
The $,i$ notation denotes the ordinary derivative with respect to $\varphi^i$. From this and the other formul\ae given in this appendix it is straightforward to work out the component action of this superspace action. Here, we will not present the most general result but focus on the first and last term in \eqref{nlsm_2a} which are the only ones relevant to our M-theory reduction. One finds
\begin{align}
 S_{2a} &= \oneon{4} \int d\tau d^2\theta \left\{ G_{ij}(\underline{\phi}) D \phi^i \bar{D} \phi^j + {\cal W}(\underline{\phi}) \right\}\\
 &= \oneon{4}\int d\tau\,\left\{ \oneon{4} G_{ij}(\underline{\varphi}) \dot{\varphi}^i \dot{\varphi}^j
			- \frac{i}{2} G_{ij}(\underline{\varphi}) (\psi^i \dot{\bar\psi}^j - \dot{\psi}^i\bar{\psi}^j)
			+ \oneon{4}G_{ij}(\underline{\varphi}) f^i f^j\right. \nonumber\\
			&- \oneon{2}G_{ij,k}(\underline{\varphi}) (\psi^i\bar{\psi}^j f^k - \psi^k\bar{\psi}^j f^i - \psi^i\bar{\psi}^k f^j)
			+ \frac{i}{2} G_{ij,k}(\underline{\varphi}) (\psi^k\bar{\psi}^i + \bar{\psi}^k\psi^i) \dot{\varphi}^j \\
			&\left.- G_{ij,kl}(\underline{\varphi}) \psi^i\bar{\psi}^j\psi^k\bar{\psi}^l-\frac{1}{2}{\cal W}_{,i}(\underline{\varphi})f^i- {\cal W}_{,ij}(\underline{\varphi})\psi^i\bar{\psi}^j\right\}\nonumber .
\end{align} 
Apart from the standard kinetic terms we have Pauli terms (coupling two fermions and the time derivative of a scalar), Yukawa couplings and four-fermi terms. We also see that the highest components $f^i$ are indeed auxiliary field. The $f^i$ equation of motion can be solved explicitly and leads to
\be
 f^i =  G^{ij}{\cal W}_{,j}+\dots
\ee
where $G^{ij}$ is the inverse of $G_{ij}$.  The dots indicate fermion bilinear terms which we have not written down explicitly. Using this solution to integrating out the $f^i$ produces additional four-fermi terms and  the scalar potential
\be
	S_{2a,{\rm pot}} = - \oneon{8} \int d\tau\, {\cal U}\; ,\qquad {\cal U} = \frac{1}{2}G^{ij}{\cal W}_{,i}{\cal W}_{,j}\; .
\ee 

\vskip 0.4cm

The other major type of multiplet is the $2b$ multiplet $Z=Z(\tau,\theta,\bar{\theta})$ which is defined by the constraint $\bar{D}Z=0$. Working out its most general component expansion one finds
\be 
 Z = z + \theta\kappa + \frac{i}{2} \theta\bar\theta \dot{z} , \eqlabel{sspace_flat_sfield_2b}
\ee
where $z$ is a complex scalar and $\kappa$ is a complex fermion. We note that, unlike for the $2a$ multiplet, the highest component is not an independent field but simply $\dot{z}$. Hence, a $2b$ multiplet contains a complex physical scalar field and no auxiliary field. This difference in physical bosonic field content in comparison with the $2a$ multiplet will be quite useful when it comes to identifying which supermultiplets arise from our M-theory reduction. \Eqref{sspace_flat_superchrg_defn,susytransform} lead to the component supersymmetry transformations
\begin{align}
&\delta_\epsilon z = i\epsilon \kappa, \quad\delta_\epsilon \bar{z} = 0, \quad\delta_\epsilon \kappa = 0, \quad
					\delta_\epsilon \bar\kappa = \epsilon \dot{\bar{z}}, \eqlabel{sspace_flat_transf_2b} \\
&\delta_{\bar\epsilon} \bar{z} = i\bar\epsilon \bar\kappa, \quad\delta_{\bar\epsilon} z = 0,\quad
                                              \delta_{\bar\epsilon} \bar\kappa = 0,\quad\delta_{\bar\epsilon} \kappa = \bar\epsilon \dot{z}\; .
                                              \eqlabel{sspace_flat_transf_2b_bar}
\end{align}
A general non-linear sigma model for a set, $\{ Z^a\}$, of $2b$ multiplets has the form~\cite{Coles:1990hr,Gibbons:1997iy,Hull:1999ng}
\be
  S_{2b} = \oneon{4} \int d\tau d^2\theta \left\{ G_{a{\bar{b}}}(\underline{Z},\underline{\bar{Z}}) DZ^a \bar{D}\bar{Z}^{\bar{b}} 
		+ \left[\oneon{2} B_{ab}(\underline{Z},\underline{\bar{Z}}) DZ^a DZ^b + \cc\right] +
		  F(\underline{Z},\bar{\underline{Z}})\right\} , \eqlabel{nlsm_2b}
\ee
where $G_{a\bar{b}}$ is hermitian, $B_{ab}$ is anti-symmetric and $F$ is an arbitrary real function. The component version of $F(\underline{Z},\bar{\underline{Z}})$ is obtained by a Taylor expansion about $z^a$ and $\bar{z}^{\bar{a}}$: 
\begin{multline}
	F(\underline{Z},\bar{\underline{Z}}) = 
	F(\underline{z},\bar{\underline{z}})
	+ \theta \kappa^a F_{,a}(\underline{z},\bar{\underline{z}})
	- \bar\theta \bar{\kappa}^{\bar{a}} F_{,\bar{a}}(\underline{z},\bar{\underline{z}}) \\ 
	+ \oneon{2} \theta\bar\theta \left\{ i F_{,a}(\underline{z},\bar{\underline{z}}) \dot{z}^a 
		- i F_{,\bar{a}}(\underline{z},\bar{\underline{z}}) \dot{\bar{z}}^{\bar{a}}
		+ 2 F_{,a\bar{b}}(\underline{z},\bar{\underline{z}}) \kappa^a \bar{\kappa}^{\bar{b}} \right\} .
\end{multline}
The component form of the action~\eqref*{nlsm_2b} can again be worked out straightforwardly from the above formal\ae~but we will not pursue this here. Instead, we focus on a slightly different superspace action which is better adapted to what we need in the context of our M-theory reduction. First, we drop the term proportional to $B_{ab}$ which does not arise from M-theory. Secondly, we introduce a slight generalisation in that we allow the sigma model metric $G_{a\bar{b}}$ to also depend on $2a$ superfields $\phi^i$, in addition to the $2b$ superfields $Z^a$ and their complex conjugates. A multi-variable Taylor expansion of a function $G(\underline{\phi},\underline{Z},\underline{\bar{Z}})$ depending on $2a$ as well as $2b$ superfields yields the component form:
\bea
	G(\underline{\phi},\underline{Z},\underline{\bar{Z}}) = 
	  G(\underline{\varphi},\underline{z},\underline{\bar{z}})
	+ \theta [i\psi^i G_{,i}(\underline{\varphi},\underline{z},\underline{\bar{z}}) 
		+ \kappa^a G_{,a}(\underline{\varphi},\underline{z},\underline{\bar{z}})]
	+ \bar\theta [i\bar\psi^i G_{,i}(\underline{\varphi},\underline{z},\underline{\bar{z}}) 
		\\ - \bar{\kappa}^{\bar{a}} G_{,\bar{a}}(\underline{\varphi},\underline{z},\underline{\bar{z}})] 
	+ \theta\bar\theta \left[
		   \oneon{2} G_{,i}(\underline{\varphi},\underline{z},\underline{\bar{z}}) f^i
		+ G_{,ij}(\underline{\varphi},\underline{z},\underline{\bar{z}}) \psi^i \bar\psi^j \right.\\ \left.
		+ i G_{,ia}(\underline{\varphi},\underline{z},\underline{\bar{z}}) \bar\psi^i \kappa^a
		+ i G_{,i\bar{a}}(\underline{\varphi},\underline{z},\underline{\bar{z}}) \psi^i \bar{\kappa}^{\bar{a}} 
		+ G_{,a\bar{b}}(\underline{\varphi},\underline{z},\underline{\bar{z}}) \kappa^a \bar{\kappa}^{\bar{b}} \right.\\ \left.
		+ \frac{i}{2} G_{,a}(\underline{\varphi},\underline{z},\underline{\bar{z}}) \dot{z}^a
		- \frac{i}{2} G_{,\bar{a}}(\underline{\varphi},\underline{z},\underline{\bar{z}}) \dot{\bar{z}}^{\bar{a}}
		\right] .
\eea
The relevant action is
\begin{align}
 S_{2b} &= \frac{1}{4}\int d\tau\,d^2\theta\,\left\{G_{a{\bar{b}}}(\underline{\phi},\underline{Z},\underline{\bar{Z}}) DZ^a \bar{D}\bar{Z}^{\bar{b}} +  F(\underline{Z},\bar{\underline{Z}})\right\} \nonumber\\
 &= \frac{1}{4}\int d\tau\,\left\{ 
 	G_{a\bar{b}}(\underline{\varphi}, \underline{z}, \underline{\bar{z}}) \dot{z}^a \dot{\bar{z}}^{\bar{b}}
	- \frac{i}{2} G_{a\bar{b}}(\underline{\varphi}, \underline{z}, \underline{\bar{z}}) 
		(\kappa^a \dot{\bar{\kappa}}^{\bar{b}} - \dot{\kappa}^a \bar{\kappa}^{\bar{b}}) \right. \\  & \left.
	- \frac{i}{2} G_{a\bar{b}, c}(\underline{\varphi}, \underline{z}, \underline{\bar{z}}) 
		(\kappa^a \bar{\kappa}^{\bar{b}} \dot{z}^c - 2\kappa^c \bar{\kappa}^{\bar{b}} \dot{z}^a ) 
	+ \frac{i}{2} G_{a\bar{b}, \bar{c}}(\underline{\varphi}, \underline{z}, \underline{\bar{z}}) 
		(\kappa^a \bar{\kappa}^{\bar{b}} \dot{\bar{z}}^{\bar{c}} + 2\kappa^a \bar{\kappa}^{\bar{c}} \dot{\bar{z}}^{\bar{b}} )   \right. \\  & \left.
	- G_{a\bar{b}, c\bar{d}}(\underline{\varphi}, \underline{z}, \underline{\bar{z}}) 
		\kappa^a \bar{\kappa}^{\bar{b}} \kappa^c \bar{\kappa}^{\bar{d}}
	- \oneon{2} G_{a\bar{b}, i}(\underline{\varphi}, \underline{z}, \underline{\bar{z}}) \kappa^a \bar{\kappa}^{\bar{b}} f^i
	- G_{a\bar{b}, ij}(\underline{\varphi}, \underline{z}, \underline{\bar{z}}) \kappa^a \bar{\kappa}^{\bar{b}} \psi^i \bar{\psi}^j   \right. \\  & \left.
	- i G_{a\bar{b}, ic}(\underline{\varphi}, \underline{z}, \underline{\bar{z}}) \kappa^a \bar{\kappa}^{\bar{b}} \bar{\psi}^i \kappa^c 
	- i G_{a\bar{b}, i\bar{c}}(\underline{\varphi}, \underline{z}, \underline{\bar{z}}) \kappa^a \bar{\kappa}^{\bar{b}} \psi^i \bar{\kappa}^{\bar{c}}
	- G_{a\bar{b}, i}(\underline{\varphi}, \underline{z}, \underline{\bar{z}}) 
		\psi^i \bar{\kappa}^{\bar{b}} \dot{z}^a  \right. \\  & \left.
	+ G_{a\bar{b}, i}(\underline{\varphi}, \underline{z}, \underline{\bar{z}}) 
		\bar{\psi}^i \kappa^a \dot{\bar{z}}^{\bar{b}}
	- \frac{i}{2} (F_{,a}\dot{z}^a-F_{,\bar{b}}\dot{\bar{z}}^{\bar{b}})-F_{,a\bar{b}}\kappa^a\bar{\kappa}^{\bar{b}}\right\} .
\end{align} 
Note that the function $F$ gives rise to a Chern-Simons type term (and fermion mass terms) but not to a scalar potential.

\vskip 0.4cm

The $2a$ and $2b$ superfields introduced above are bosonic superfields in the sense that their lowest components are bosons. However, for both types of multiplets there also exists a fermionic version, satisfying the same constraint as their bosonic counterparts but starting off with a fermion as the lowest component. In our context, we will only need fermionic $2b$ superfields so we will focus on them. The details for fermionic $2a$ superfields can be worked out analogously.

Fermionic $2b$ superfields $R=R(\tau ,\theta,\bar{\theta})$ have a spinorial lowest component and are defined by the constraint $\bar{D}R=0$. Their general component expansion reads
\be
 R = \rho + \theta h + \frac{i}{2} \theta\bar\theta \dot{\rho}\; , \eqlabel{2bfexp}
\ee
where $\rho$ is a complex fermion and $h$ is a complex scalar. For its component supersymmetry transformations one finds
\begin{align}
	&\delta_\epsilon \rho = i\epsilon h, \quad\delta_\epsilon \bar\rho = 0, \quad\delta_\epsilon h = 0, \quad
						\delta_\epsilon \bar{h} = - \epsilon \dot{\bar{\rho}}, \eqlabel{sspace_flat_transf_2bf} \\
	&\delta_{\bar\epsilon} \bar\rho = -i\bar\epsilon \bar{h}, \quad\delta_{\bar\epsilon} \rho = 0, \quad
	                                             \delta_{\bar\epsilon} \bar{h} = 0.\quad\delta_{\bar\epsilon} h = \bar\epsilon \dot{\rho}\; .
           \eqlabel{sspace_flat_transf_2bf_bar}
\end{align}
A set, $\{R^x\}$, of fermionic $2b$ superfields can be used to build non-linear sigma models where only fermions are propagating. A class of such models is given by
\begin{align}
	S_{2b,{\rm F}} &= \oneon{4} \int d\tau\, d^2\theta\, G_{x\bar{y}}(\underline{\phi}) R^x \bar{R}^{\bar{y}} \\
	                        &=\frac{1}{4}\int d\tau \, \left\{
		   \frac{i}{2} G_{x\bar{y}}(\underline{\varphi}) ( \rho^x \dot{\bar{\rho}}^{\bar{y}} - \dot{\rho}^x \bar{\rho}^{\bar{y}} ) 
		- G_{x\bar{y}}(\underline{\varphi}) h^x \bar{h}^{\bar{y}}
		- \oneon{2} G_{x\bar{y},i}(\underline{\varphi}) \rho^x \bar{\rho}^{\bar{y}} f^i  \right. \\ & \left.
		- i G_{x\bar{y},i}(\underline{\varphi}) ( \psi^i \rho^x \bar{h}^{\bar{y}} + \bar{\psi}^i \bar{\rho}^{\bar{y}} h^x ) 
		- G_{x\bar{y},ij}(\underline{\varphi}) \rho^x \bar{\rho}^{\bar{y}} \psi^i \bar{\psi}^j  \right\} .
\end{align}
Here, we have allowed the sigma model metric to depend on a set, $\{\phi^i\}$, of $2a$ moduli, a situation which will arise from M-theory reductions.  Note that the bosons $h^x$ are indeed auxiliary fields and only the fermions $\rho^x$ have kinetic terms.

\subsection{Local $\susyno=2$ supersymmetry} \applabel{localsusy}

The goal of this subsection is to develop one-dimensional $\susyno=2$ \emph{curved} superspace to an extent that will allow us to write down actions over this superspace and compare their component expansion with our result from dimensional reduction of M-theory on Calabi-Yau five-folds. Eventually, we are using the results of this subsection to write our one-dimensional effective action in full one-dimensional $\susyno=2$ curved superspace thereby making the residual supersymmetry manifest. 

The on-shell one-dimensional $\susyno=2$ supergravity multiplet comprises the lapse function (or ``einbein'') $N$, which is a real scalar, and the ``lapsino'' $\psi_0$, which is a one-component complex spinor. In all expressions provided in this sub-section, flat superspace (and thus the equations of the previous subsection) can be recovered by gauge fixing the supergravity fields to $N=1$ and $\psi_0 = 0$. From a more geometric viewpoint, curved $\susyno=2$ supertime looks locally like flat $\susyno=2$ supertime $\R^{1|2}$.

The well-known case of $\susyno=1$ in four dimensions~\cite{Wess:1977fn,Grimm:1977kp,Wess:1978bu,Stelle:1978ye,Ferrara:1978em,Sohnius:1981tp} and supergravity theories in two-dimensions~\cite{Howe:1978ia,Ertl:2001sj} will guide us in constructing our curved supertime here. Modulo some subtleties, which are explained below, many textbook formul\ae\footnote{We shall closely follow Refs.~\cite{West:1990tg,Wess:1992cp}, here.} carry over to the case of $\susyno=2$ supertime with the index ranges adjusted appropriately.

The geometrical description of curved superspace follows ordinary Riemannian geometry, however with the range of indices extended to include the spinorial coordinates.  In particular, certain (super-)tensors, such as the supervielbein $E_A {}^\UB$, super-spin-connection $\Omega_{A\UB}{}^{\underline{C}}$, supertorsion $T_{AB} {}^{\underline{C}}$ and supercurvature $R_{AB\underline{C}}{}^{\underline{D}}$, play an important r\^{o}le when working with curved superspace. As in the previous subsection, the indices $A,B,\ldots = 0,\theta,\bar{\theta}$ are used to label supertime tensors and underlined versions $\UA, \UB, \ldots$ correspond to local Lorentz indices. As local coordinates, we choose $\{x^0=\tau;\theta,\bar\theta \}$, where $\theta$ is a complex one-dimensional spinor. The supervielbein can be used to convert curved to flat indices and vice versa, so that
\be
	V_A = E_A {}^\UB V_\UB , \qquad\qquad 
	V_\UA = E_\UA {}^B V_B\; .
\ee
In the second relation the inverse of the supervielbein has been used, which is defined via
\be\eqlabel{sspace_curved_supervielbein_inverse_rel}
	E_A {}^\UB E_\UB {}^C = \delta_A {}^C , \qquad\qquad 
	E_\UA {}^B E_B {}^{\underline{C}} = \delta_\UA {}^{\underline{C}} .
\ee
Note that one may use the superdifferential $dz^A$ together with the graded wedge-product $\wedge$ to write all the aforementioned supertensors as super-differential forms, for example
\be
	E^\UA = dz^B E_B {}^\UA, \qquad\qquad
	T^\UA = \oneon{2} dz^B \wedge dz^C T_{CB} {}^\UA \; .
\ee
The supertorsion is defined as covariant derivative of the supervielbein:
\be\eqlabel{sspace_curved_def_torsion}
	T^\UA = d E^\UA + E^\UB \wedge \Omega_\UB {}^\UA .
\ee

The r\^{o}le of the local Lorentz indices is rather subtle in $\susyno=2$ supertime. In order to recover flat supertime, these indices are taken to be valued in the {bosonic} Lorentz group $\SO(1)$, which is just the trivial group, and not in the full super-Lorentz group $\SO(1|2)$. Since there is no Lie algebra for the trivial group, there are no Lorentz generators in one dimension and the local Lorentz indices $\UA, \UB, \ldots$ do not transform under any group action but should merely thought of as labels. They label the two different representations of $\mathrm{Spin}(1)$, namely for $\UA=\UZ$ the ``vector'' representation, which is nothing but the real numbers in one dimension, and for $\UA=\Uth$ the spinor representation which are real Grassmann numbers. In addition, the fact that we want to realize $\susyno$-extended supersymmetry (with $\susyno > 1$) means we need in principle another index, say $i,j=1,\ldots,\susyno$ on the $\UA=\Uth$ components to label the $\susyno$-extendedness of the spinorial components (cf. the notation used in four dimensional $\susyno=2$ superspace~\cite{Breitenlohner:1979np,Castellani:1980cu,Stelle:1980uw,Gates:1980ky}). Here, $\susyno=2$ and hence $\UA,\UB,\ldots = \UZ, \Uth_{1}, \Uth_{2}$. For ease of notation, we combine the two $\Uth_i$ into a combination of one complex index $\Uth = \Uth_{1} + i \Uth_{2}$ (and similarly $\BUth = \Uth_{1} - i \Uth_{2}$) thereby suppressing the additional $\susyno$-extension index $i$. After this step, the local Lorentz indices $\UA, \UB, \ldots$ range over $\UZ$, $\Uth$ and $\BUth$. Note that this coincides precisely with the notation used for curved indices except for the additional underline added for distinction. In summary, even though the local Lorentz indices can take on three different values, there is no group acting on them. Objects carrying an anti-symmetrized combination of two or more local Lorentz indices vanish identically, since the Lorentz generator in each representation of $\mathrm{Spin}(1)$ is zero and there are no representation-mixing Lorentz transformations. This immediately implies $\Omega_{A\UB} {}^{\underline{C}} = 0$ and $R_{AB\underline{C}} {}^{\underline{D}} = 0$, which profoundly simplifies the further discussion.

Since the on-shell supergravity multiplet contains only one real scalar, we take the geometrical supertime tensors to be $2a$ superfields, which means they comprise four component fields when expanded out in powers of $\theta$ and $\bar{\theta}$ (see~\eqref{sspace_flat_2a_comp}). The supervielbein $E_A {}^\UB$, in general, consists of a set of $3\times 3=9$ $2a$ superfields, which totals to $9 \times 4 = 36$ component (that is, off-shell) fields, and is expanded as 
\be\eqlabel{sspace_curved_supervielbein_gen_expansion}
	E_A {}^\UB = E_{A(0)} {}^\UB + i \theta E_{A(1)} {}^\UB + i \bar\theta E_{A(\bar{1})} {}^\UB + \oneon{2} \theta\bar\theta E_{A(2)} {}^\UB\; .
\ee
This is a large number of apparently independent fields given that on-shell, we just have three, namely $N$, $\psi_0$ and $\bar{\psi}_0$. In order to not obscure the physical content and to formulate supertime theories in the most efficient way, it is important to find a formulation with the minimum number of component fields. This can be achieved by imposing covariant constraints on the supervielbein and by gauging away some components using the super-general coordinate transformations
\be\eqlabel{sspace_curved_inf_sgct_supervielbein}
	\delta E_A {}^\UB = \xi^C (\partial_C E_A {}^\UB) + (\partial_A \xi^C) E_C {}^\UB ,
\ee
with infinitesimal parameters $\xi^A$, which comprise a set of three four-component $2a$ superfields (that is, 12 component fields in total). The lowest component of $\xi^0 | = \zeta$ is the infinitesimal parameter of worldline reparametrizations, whereas the lowest components of the spinorial parameters $\xi^\theta | = i\epsilon$ and $\xi^{\bar\theta} | = i\bar\epsilon$ correspond to the infinitesimal local $\susyno=2$ supersymmetry parameters. The notation $\phi |$ is a shorthand for $\phi |_{\theta=\bar\theta = 0}$, that is denoting the lowest component of the superfield $\phi$.
An infinitesimal local $\susyno=2$ supersymmetry transformations with parameters $\epsilon$ and $\bar\epsilon$ on a general superfield $\phi$ can be written by means of the supercharges $Q$ and $\bar{Q}$ as
\be\eqlabel{sspace_curved_gen_sfield_transf}
	\delta_\epsilon \phi = i \epsilon Q \phi, \qquad\qquad \delta_{\bar\epsilon} \phi = i \bar\epsilon \bar{Q} \phi .
\ee
If we use the following general component expansion for $\phi$:
\be\eqlabel{sspace_curved_gen_cov_sfield_expansion}
	\phi = \phi | + \theta (\sderiv\phi |) - \bar\theta (\bar\sderiv\phi |) + \oneon{2} \theta\bar\theta ( [\sderiv, \bar\sderiv]\phi | ) ,
\ee
then the components of $\phi$ transform as
\be\eqlabel{sspace_curved_gen_sfield_comp_transf}
	\delta_\epsilon (\phi |) = i \epsilon Q \phi | , \quad
	\delta_\epsilon (\sderiv \phi |) = i \epsilon Q \sderiv \phi | , \quad
	\delta_\epsilon (\bar\sderiv \phi |) = i \epsilon Q \bar\sderiv \phi | , \quad
	\delta_\epsilon ( [\sderiv, \bar\sderiv]\phi | ) = i \epsilon Q [\sderiv, \bar\sderiv] \phi | .
\ee
Both~\eqref*{sspace_curved_gen_cov_sfield_expansion,sspace_curved_gen_sfield_comp_transf} are manifestly super-covariant expressions since we used the tangentized covariant super-derivative of curved supertime $\sderiv_\UA =E_\UA {}^B \partial_B$ for building them. Note that, similarly to $D$ and $\bar{D}$ in the flat case, the tangentized, spinorial super-covariant derivatives are abbreviated as $\sderiv := \sderiv_\Uth = E_\Uth {}^A \partial_A$ and $\bar\sderiv := \sderiv_\BUth = E_\BUth {}^A \partial_A$. From the general fact that $Q | = \sderiv | = \partial_\theta$, it follows that one may replace $Q$s by $\sderiv$s everywhere in~\eqref*{sspace_curved_gen_sfield_comp_transf} and hence knowing the component expansion of $\sderiv$ is enough for working out the entire component version of \eqref*{sspace_curved_gen_sfield_transf}, namely:
\be\eqlabel{sspace_curved_gen_sfield_comp_transf_final}
	\delta_\epsilon (\phi |) = i \epsilon \sderiv \phi | , \quad
	\delta_\epsilon (\sderiv \phi |) = i \epsilon \sderiv^2 \phi | = 0 , \quad
	\delta_\epsilon (\bar\sderiv \phi |) = i \epsilon \sderiv \bar\sderiv \phi | , \quad
	\delta_\epsilon ( [\sderiv, \bar\sderiv]\phi | ) = - i \epsilon \sderiv \bar\sderiv \sderiv \phi | ,
\ee
and similarly for the $\bar\epsilon$-transformations. In the second and fourth equation in~\eqref*{sspace_curved_gen_sfield_comp_transf_final}, we used the property $\sderiv^2 = 0$.

\vskip 0.4cm

Now continuing our quest for finding the minimal formulation of off-shell $\susyno=2$, $d=1$ supergravity, we have here opted for the analogue of the Wess-Zumino gauge in $d=4$ and the way to formulate it in the present case will be explained in the following. Since we have three physical components in the supergravity multiplet, we shall use $9=12-3$ components out of $\xi^A$ to gauge fix 9 out of the 36 components of $E_A {}^\UB$, namely
\begin{align}
	&E_\theta {}^\UZ | = E_{\bar{\theta}} {}^\UZ | =
	E_\theta {}^\BUth | = E_{\bar{\theta}} {}^\Uth | =
	\sderiv E_{\bar{\theta}} {}^\Uth | = 
	\sderiv E_{\bar{\theta}} {}^\BUth | = 0 , \eqlabel{sspace_curved_gauge_fix1}\\
	&\bar\sderiv E_\theta {}^\UZ | = \frac{i}{2} , \quad
	E_\theta {}^\Uth | = 1, \quad E_{\bar{\theta}} {}^\BUth | = -1 . \eqlabel{sspace_curved_gauge_fix2}
\end{align}
The three remaining parameters in $\xi^A$ act on the three physical fields $N$, $\psi_0$ and $\bar\psi_0$, which we choose to identify in the following way:
\be\eqlabel{sspace_curved_gauge_fix3}
	E_0 {}^\UZ | = N , \qquad
	E_0 {}^\Uth | = \psi_0 , \qquad
	E_0 {}^\BUth | = - \bar\psi_0 .
\ee

We will now discuss our choice of covariant constraints. Usually, they are imposed on certain components of the tangentized supertorsion $T_{\underline{A}\underline{B}} {}^{\underline{C}}$. ``Trial and error'' and ``educated guesses'' eventually lead to a combination of constraints that yield the minimum number of fields in the $\theta$-expansion of the supervielbein $E_A {}^\UB$. The main idea is to take the system of constraints from $\susyno=1$, $d=4$ and restrict the index ranges appropriately. Doing this, we obtain the following torsion constraints:
\begin{align}
	 T_{\Uth\BUth} {}^\UZ &= i , \qquad &
	T_{\Uth\BUth} {}^\Uth &= 0 , \qquad &
	&\text{(conventional constraints)}\eqlabel{sspace_curved_conv_constr} , \\
	 T_{\BUth\BUth} {}^\UZ &= 0 , \qquad &
	T_{\BUth\BUth} {}^\Uth &= 0 , \qquad &
	&\text{(representation preserving constraints)}\eqlabel{sspace_curved_rep_pres_constr} , \\
	& &T_{\Uth\Uth} {}^\Uth &= 0 , \qquad &
	&\text{(``type 3'' constraint)}\eqlabel{sspace_curved_type3_constr} , 
\end{align}
and their complex conjugates, of course. We are equating superfields to superfields here and hence each of the above relations is manifestly (super-)covariant. The first line is the analogue of the conventional constraints in $\susyno=1$, $d=4$ and are characterized by being algebraically solvable. In the absence of $R_\UA {}^\UB$, the torsion is directly related to the graded commutator of two super-covariant derivatives via
\be
	[ \sderiv_\UA , \sderiv_\UB \} = - T_{\UA\UB} {}^{\underline{C}} \sderiv_{\underline{C}} .
\ee
The conventional constraints now stem from imposing (cf. \eqref{sspace_flat_cov_deriv_defn})
\be
	\{ \sderiv, \bar\sderiv \} = - i \sderiv_\UZ ,
\ee
which guarantees that the \emph{tangentized} covariant super-derivatives of curved superspace, $\sderiv$ and $\bar\sderiv$, satisfy the \emph{flat} algebra. A $2b$ superfield $Z$ by definition satisfies $\bar\sderiv Z = 0$. The representation preserving constraints listed in~\eqref*{sspace_curved_rep_pres_constr} follow from the corresponding integrability condition, that is from
\be
	\{ \bar\sderiv, \bar\sderiv \} Z = 0 \qquad\qquad \forall \;\;\text{$2b$ superfields $Z$}.
\ee
For the constraint in~\eqref*{sspace_curved_type3_constr}, we do not have a direct motivation from a one-dimensional viewpoint, so we impose it purely by analogy to the conformal constraint of $\susyno=1$ in $d=4$.

In general superspace theory, the torsion and curvature tensors satisfy the two Bianchi identities (BIs)
\begin{align}
	\nabla T^\UA &= E^\UB \wedge R_\UB {}^\UA , \\
	\nabla R_\UA{}^\UB &= 0 ,
\end{align}
where $\nabla = d + \Omega\wedge$. Specializing to $\susyno=2$ supertime, the second BI identically vanishes due to $R_\UA {}^\UB = 0$ and the first BI becomes
\be\eqlabel{sspace_curved_1d_BI}
	d T^\UA = 0  \qquad\Leftrightarrow\qquad
	\sderiv_{\left[\UA\right.} T_{\left.\UB\underline{C}\right\}} {}^{\underline{D}} + 
	T_{\left[\UA\UB\right.} {}^{\underline{E}} T_{\left.|\underline{E}|\underline{C}\right\}} {}^{\underline{D}} = 0.
\ee
In the presence of constraints, consistency requires that the BIs are sill obeyed and this needs to be checked by explicit calculation. In this respect, the BIs become ``contentful'' (rather than being genuine identities) when constraints are present and then the BIs must be \emph{imposed}. For the case at hand, one learns from the BI~\eqref*{sspace_curved_1d_BI} that all remaining torsion components which are not already fixed by the constraints~\eqrangeref*{sspace_curved_conv_constr}{sspace_curved_type3_constr} must be zero.

From the definition of the supertorsion~\eqref*{sspace_curved_def_torsion}, the choice of gauge fixing~\eqrangeref*{sspace_curved_gauge_fix1}{sspace_curved_gauge_fix3} and torsion constraints~\eqrangeref*{sspace_curved_conv_constr}{sspace_curved_type3_constr} and the imposition of the BI~\eqref*{sspace_curved_1d_BI}, all 36 components in the supervielbein expansion~\eqref*{sspace_curved_supervielbein_gen_expansion} are fixed uniquely to
\begin{align}
	&E_0 {}^\UZ = N + i \theta \bar\psi_0 + i \bar\theta \psi_0 , \eqlabel{sspace_curved_seinbein1} \\
	&E_0 {}^\Uth = \psi_0 , \qquad
	E_0 {}^\BUth = - \bar\psi_0 , \eqlabel{sspace_curved_seinbein2} \\
	&E_\theta {}^\UZ = -\frac{i}{2} \bar\theta , \qquad
	E_{\bar{\theta}} {}^\UZ = -\frac{i}{2} \theta , \eqlabel{sspace_curved_seinbein3} \\
	&E_\theta {}^\Uth = 1 , \quad
	E_\theta {}^\BUth = 0 , \quad
	E_{\bar{\theta}} {}^\Uth = 0 , \quad
	E_{\bar{\theta}} {}^\BUth = -1 \eqlabel{sspace_curved_seinbein4}.
\end{align}
Note that the minimal set of fields of off-shell pure $\susyno=2$, $d=1$ supergravity does not comprise any auxiliary fields. From \eqref{sspace_curved_supervielbein_inverse_rel} we compute the component expansion of the inverse supervielbein
\begin{align}
	&E_\UZ {}^0 = N^{-1} - \frac{i}{2} \theta N^{-2} \bar\psi_0 - \frac{i}{2} \bar\theta N^{-2} \psi_0
		- \oneon{2} \theta\bar\theta N^{-3} \psi_0 \bar\psi_0 , \\
	&E_\UZ {}^\theta = - N^{-1} \psi_0 - \frac{i}{2} \theta N^{-2} \psi_0 \bar\psi_0 , \qquad
	E_\UZ {}^{\bar\theta} = - N^{-1} \bar\psi_0 + \frac{i}{2} \bar\theta N^{-2} \psi_0 \bar\psi_0 , \\
	&E_\Uth {}^0 = \frac{i}{2} \bar\theta N^{-1} - \oneon{4} \theta\bar\theta N^{-2} \bar\psi_0 , \quad\quad
	E_\BUth {}^0 = - \frac{i}{2} \theta N^{-1} - \oneon{4} \theta\bar\theta N^{-2} \psi_0 , \\
	&E_\Uth {}^\theta = 1 - \frac{i}{2} \bar\theta N^{-1} \psi_0 - \oneon{4} \theta\bar\theta N^{-2} \psi_0 \bar\psi_0 ,
		\qquad\qquad
	E_\Uth {}^{\bar\theta} = -\frac{i}{2} \bar\theta N^{-1} \bar\psi_0 , \\
	&E_\BUth {}^{\bar\theta} = -1 + \frac{i}{2} \theta N^{-1} \bar\psi_0 + \oneon{4} \theta\bar\theta N^{-2} \psi_0 \bar\psi_0 , 
		\qquad\qquad
	E_\BUth {}^\theta = \frac{i}{2} \theta N^{-1} \psi_0 .
\end{align}
Since $\sderiv_\UA = E_\UA {}^B \partial_B$, the above expressions allow us to write down the component expansion of the tangentized, spinorial super-covariant derivative
\begin{equation}\eqlabel{sspace_curved_sderiv_comp_exp}
	\sderiv = \left( 1 - \frac{i}{2}N^{-1}\bar\theta\psi_0 - \oneon{4}N^{-2}\theta\bar\theta\psi_0\bar\psi_0 \right) \partial_\theta
		+ \left( \frac{i}{2}N^{-1}\bar\theta - \oneon{4}N^{-2}\theta\bar\theta\bar\psi_0 \right) \partial_0 
		- \frac{i}{2}N^{-1}\bar\theta\bar\psi_0 \partial_{\bar\theta} ,
\end{equation}
and similarly for $\bar\sderiv$. By comparing the component expansion of \eqref{sspace_curved_inf_sgct_supervielbein} with \eqrangeref{sspace_curved_seinbein1}{sspace_curved_seinbein4}, we learn how the supergravity fields transform under local $\susyno=2$ supersymmetry
\be
	\delta_\epsilon N = - \epsilon\bar\psi_0 , \quad
	\delta_{\bar\epsilon} N = \bar\epsilon\psi_0 , \qquad 
	\delta_\epsilon\psi_0 = i\dot{\epsilon} , \quad
	\delta_\epsilon\bar\psi_0 = 0 , \quad
	\delta_{\bar\epsilon}\psi_0 = 0 , \quad
	\delta_{\bar\epsilon}\bar\psi_0 = -i\dot{\bar\epsilon}.
\ee
In order to build curved superspace actions that are manifestly invariant under local $\susyno=2$ supersymmetry, we need the analogue of $\sqrt{-g}$ to construct an invariant volume form. It turns out that this is given by the super-determinant of the supervielbein, denoted simply by $\mathcal{E}$, and defined, in general, as
\be
	\mathcal{E} := \mathrm{sdet} E_A {}^\UB = (\det E_a {}^{\underline{b}})
		( \det [E_\alpha {}^{\underline{\beta}} 
		- E_\alpha {}^{\underline{c}} (E_d {}^{\underline{c}})^{-1} E_d {}^{\underline\beta}] )^{-1} ,
\ee
where $a,b,\ldots$ and $\alpha, \beta, \ldots$ denote vector and spinor indices, respectively. Specializing to $\susyno=2$ supertime and inserting \eqrangeref{sspace_curved_seinbein1}{sspace_curved_seinbein4}, one finds for the super-determinant of the supervielbein
\be
	\mathcal{E} = - N - \frac{i}{2} \theta\bar\psi_0 - \frac{i}{2} \bar\theta\psi_0 .
\ee
Since there is no $\theta\bar\theta$-component in this expression, it follows that the canonical action of pure supergravity vanishes as expected, that is
\be
	S_{\text{pure sugra}} = \int d\tau d^2\theta \, \mathcal{E} = 0 .
\ee
Also, as an additional consistency check, one may verify that $\mathcal{E}$ is super-covariantly constant, so that
\be
	\int d\tau d^2\theta \, \sderiv \mathcal{E} = \int d\tau d^2\theta \, \bar\sderiv \mathcal{E} = \text{(total derivative)} = 0\; .
\ee
This allows us to use the partial-integration rule for superspace.

\vskip 0.4cm

In analogy to the flat superspace case, we will now present the different irreducible multiplets. We begin with the $2a$ multiplet, defined by the constraint $\phi=\phi^\dagger$. The general solution to this constraint leads to the component expansion
\begin{equation}
  \phi = \varphi + i \theta\psi + i\bar\theta \bar\psi + \oneon{2} \theta\bar\theta f \; , \eqlabel{sspace_curved_sfield_2a} 
\end{equation}
where the component fields are labelled as in \eqref{sspace_flat_2a_comp}. This can also be written in a manifestly super-covariant fashion as
\be
	\phi = \phi | + \theta (\sderiv\phi |) - \bar\theta (\bar\sderiv\phi |) + \oneon{2} \theta\bar\theta ( [\sderiv, \bar\sderiv]\phi | ) .
\ee
For the supersymmetry transformations of the $2a$ component fields one finds
\begin{align}
	        \eqlabel{sspace_curved_transf_2a} \begin{aligned} &
		\delta_\epsilon \varphi = - \epsilon \psi, \quad 
		\delta_\epsilon \psi = 0, \quad 
		\delta_\epsilon \bar\psi = 	\frac{i}{2} N^{-1} \epsilon \dot{\varphi} - \oneon{2} \epsilon f 
							+ \oneon{2} N^{-1} \epsilon (\psi_0\psi + \bar{\psi}_0\bar\psi), \\& 
		\delta_\epsilon f = 	- i N^{-1} \epsilon \dot{\psi} + \frac{i}{2} N^{-2} \epsilon\bar{\psi}_0 \dot{\varphi} 
						+ \oneon{2} N^{-1} \epsilon\bar{\psi}_0 f - \oneon{2} N^{-2} \epsilon\psi\psi_0\bar{\psi}_0, \end{aligned} \\
	         \eqlabel{sspace_curved_transf_2a_bar}  \begin{aligned} &
		\delta_{\bar\epsilon} \varphi = \bar\epsilon \bar\psi, \quad 
		\delta_{\bar\epsilon} \psi = 	- \frac{i}{2} N^{-1} \bar\epsilon \dot{\varphi} - \oneon{2} \bar\epsilon f 
								- \oneon{2} N^{-1} \bar\epsilon (\psi_0\psi + \bar{\psi}_0\bar\psi), \quad 
		\delta_{\bar\epsilon} \bar\psi = 0, \\& 
		\delta_{\bar\epsilon} f = 	- i N^{-1} \bar\epsilon \dot{\bar\psi} + \frac{i}{2} N^{-2} \bar\epsilon\psi_0 \dot{\varphi} 
							- \oneon{2} N^{-1} \bar\epsilon\psi_0 f + \oneon{2} N^{-2} \bar\epsilon\bar\psi\psi_0\bar{\psi}_0 \; .
							\end{aligned}
\end{align}
This is obtained by plugging in the component expansions~\eqref*{sspace_curved_sderiv_comp_exp,sspace_curved_sfield_2a} into the general formula~\eqref*{sspace_curved_gen_sfield_comp_transf_final}. A standard kinetic term of a single $2a$ superfield $\phi$ and its associated component action are given by
\bea
	S_{\rm 2a,kin} &= -\oneon{4} \int d\tau\, d^2\theta\, \mathcal{E}\, \sderiv \phi \bar\sderiv \phi = \oneon{4} \int d\tau\, \mathcal{L}_{\rm 2a,kin}, \\
	\mathcal{L}_{\rm 2a,kin} &= \oneon{4}N^{-1}\dot{\varphi}^2 - \frac{i}{2}(\psi\dot{\bar\psi} - \dot\psi\bar\psi) + \oneon{4} N f^2
			+ \frac{i}{2}N^{-1}(\psi\psi_0+\bar\psi\bar\psi_0)\dot\varphi 
			+ \oneon{2}N^{-1}\psi_0\bar\psi_0\psi\bar\psi .
\eea
In the context of M-theory five-fold compactifications we need to consider more general actions, representing non-linear sigma models for a set of $2a$ fields $\phi^i$ which also include a (super)-potential term. The superspace and component forms for such actions read
\bea\eqlabel{sspace_curved_2a_action}
	S_{\rm 2a} &= -\oneon{4} \int d\tau\, d^2\theta\, \mathcal{E}\, \{ G_{ij}(\underline{\phi}) \sderiv \phi^i \bar\sderiv \phi^j + {\cal W}(\underline{\phi}) \} 
			    = \oneon{4} \int d\tau\, \mathcal{L}_{\rm 2a}, \\
	\mathcal{L}_{\rm 2a} 	&= \oneon{4} N^{-1} G_{ij}(\underline{\varphi}) \dot{\varphi}^i \dot{\varphi}^j
			- \frac{i}{2} G_{ij}(\underline{\varphi}) (\psi^i \dot{\bar\psi}^j - \dot{\psi}^i\bar{\psi}^j)
			+ \oneon{4} N G_{ij}(\underline{\varphi}) f^i f^j \\
			&+ \frac{i}{2} N^{-1} G_{ij}(\underline{\varphi}) (\psi^i\psi_0 + \bar{\psi}^i\bar\psi_0) \dot{\varphi}^j
			+ \oneon{2} N^{-1} G_{ij}(\underline{\varphi}) \psi_0\bar\psi_0 \psi^i \bar{\psi}^j \\
			&- \oneon{2} N G_{ij,k}(\underline{\varphi}) (\psi^i\bar{\psi}^j f^k - \psi^k\bar{\psi}^j f^i - \psi^i\bar{\psi}^k f^j)
			+ \frac{i}{2} G_{ij,k}(\underline{\varphi}) (\psi^k\bar{\psi}^i + \bar{\psi}^k\psi^i) \dot{\varphi}^j \\
			&- N G_{ij,kl}(\underline{\varphi}) \psi^i\bar{\psi}^j\psi^k\bar{\psi}^l 
			- \oneon{2} N {\cal W}_{,i}(\underline{\varphi}) f^i 
			- N {\cal W}_{,ij}(\underline{\varphi}) \psi^i \bar{\psi}^j
			- \oneon{2} {\cal W}_{,i}(\underline{\varphi}) (\psi^i \psi_0 - \bar{\psi}^i \bar{\psi}_0) ,
\eea
with a sigma model metric $G_{ij}(\underline{\phi})$ and a superpotential ${\cal W}(\underline{\phi})$ . Here, $G_{\ldots,i}$  denotes differentiation with respect to the bosonic fields $\varphi^i$. Note that the fields $f^i$ are indeed auxiliary. Solving their equations of motion leads to
\begin{equation}
	f^i = G^{ij}{\cal W}_j + G^{ij} G_{kl,j} \psi^k \bar{\psi}^l - G^{ij} G_{jk,l} (\psi^k \bar{\psi}^l + \psi^l \bar{\psi}^k) \; ,
\end{equation}	
where $G^{ij}$ is the inverse of $G_{ij}$ and ${\cal W}_i = {\cal W}_{,i} = \frac{\partial{\cal W}}{\partial\varphi^i}$. Inserting this back into the component action leads, among other terms, to the scalar potential
\be
	S_{2a,{\rm pot}} = - \oneon{8} \int d\tau\, N\, {\cal U}\; ,\qquad {\cal U} = \oneon{2} G^{ij} {\cal W}_i {\cal W}_j\; ,
\ee
for the scalars $\varphi^i$ in the $2a$ multiplets. We will also need a slight generalization of \eqref{sspace_curved_2a_action}, namely an action for a set of $2a$ superfields $X^p$ coupling to a set of other $2a$ superfields $\phi^i$ and to a set of $2b$ superfields $Z^a$ via the sigma model metric $G_{pq}(\underline{\phi},\underline{Z},\bar{\underline{Z}})$:
\bea\eqlabel{sspace_curved_2a_action_3form}
	S_{\rm 2a,gen.} &= -\oneon{4} \int d\tau\, d^2\theta\, \mathcal{E}\, \{ 
		G_{pq}(\underline{\phi},\underline{Z},\bar{\underline{Z}}) \sderiv X^p \bar\sderiv X^q\} 
			    = \oneon{4} \int d\tau\, \mathcal{L}_{\rm 2a,gen.}, \\
	\mathcal{L}_{\rm 2a,gen.} &= \oneon{4} N^{-1} G_{pq}(\underline{\varphi},\underline{z},\bar{\underline{z}}) \dot{x}^p \dot{x}^q
			- \frac{i}{2} G_{pq}(\underline{\varphi},\underline{z},\bar{\underline{z}}) 
				(\lambda^p \dot{\bar\lambda}^q - \dot{\lambda}^p \bar{\lambda}^q)
			+ \oneon{4} N G_{pq}(\underline{\varphi},\underline{z},\bar{\underline{z}}) g^p g^q \\
			&+ \frac{i}{2} N^{-1} G_{pq}(\underline{\varphi},\underline{z},\bar{\underline{z}}) 
				(\lambda^p\psi_0 + \bar{\lambda}^p\bar\psi_0) \dot{x}^q
			+ \oneon{2} N^{-1} G_{pq}(\underline{\varphi},\underline{z},\bar{\underline{z}}) \psi_0\bar\psi_0 \lambda^p \bar{\lambda}^q \\
			&- \oneon{2} N G_{pq,i}(\underline{\varphi},\underline{z},\bar{\underline{z}}) 
				(\lambda^p\bar{\lambda}^q f^i - \psi^i\bar{\lambda}^p g^q + \bar{\psi}^i \lambda^p g^q)
			+ \frac{i}{2} G_{pq,i}(\underline{\varphi},\underline{z},\bar{\underline{z}}) (\psi^i\bar{\lambda}^p + \bar{\psi}^i\lambda^p) \dot{x}^q \\
			&- N G_{pq,ij}(\underline{\varphi},\underline{z},\bar{\underline{z}}) \lambda^p\bar{\lambda}^q\psi^i\bar{\psi}^j 
			- \frac{i}{2} G_{pq,a}(\underline{\varphi},\underline{z},\bar{\underline{z}}) \lambda^p \bar\lambda^q (\dot{z}^a - \psi_0 \kappa^a) \\
			&+ \frac{i}{2} G_{pq,\bar{a}}(\underline{\varphi},\underline{z},\bar{\underline{z}}) 
				\lambda^p \bar\lambda^q (\dot{\bar{z}}^{\bar{a}} + \bar\psi_0 \bar\kappa^{\bar{a}}) 
			- \frac{i}{2} N G_{pq,a}(\underline{\varphi},\underline{z},\bar{\underline{z}}) \kappa^a \bar\lambda^p g^q \\
			&- \frac{i}{2} N G_{pq,\bar{a}}(\underline{\varphi},\underline{z},\bar{\underline{z}}) \bar\kappa^{\bar{a}} \lambda^p g^q 
			+ \oneon{2} G_{pq,a}(\underline{\varphi},\underline{z},\bar{\underline{z}}) \kappa^a \bar\lambda^p \dot{x}^q
			- \oneon{2} G_{pq,\bar{a}}(\underline{\varphi},\underline{z},\bar{\underline{z}}) \bar\kappa^{\bar{a}} \lambda^p \dot{x}^q \\ 
			&- N G_{pq,a\bar{b}}(\underline{\varphi},\underline{z},\bar{\underline{z}}) \lambda^p \bar\lambda^q \kappa^a \bar\kappa^{\bar{b}}
			- i N G_{pq,ia}(\underline{\varphi},\underline{z},\bar{\underline{z}}) \lambda^p \bar\lambda^q \bar\psi^i \kappa^a \\
			&- i N G_{pq,i\bar{a}}(\underline{\varphi},\underline{z},\bar{\underline{z}}) \lambda^p \bar\lambda^q \psi^i \bar\kappa^{\bar{a}} \; .
\eea

\vskip 0.4cm

\begin{sloppypar} 
Next we turn to $2b$ multiplets. They are defined by the constraint $\bar{\cal D}Z=0$ which leads to the component expansion
\begin{equation}
  Z = z + \theta\kappa + \frac{i}{2} N^{-1} \theta\bar\theta (\dot{z} - \psi_0 \kappa) \; . \eqlabel{sspace_curved_sfield_2b}
\end{equation}
Here, $N$ and $\psi_0$ are the components of the supergravity multiplet and the other fields are labelled in analogy with the globally supersymmetric case \eqref*{sspace_flat_sfield_2b}. 
Expression~\eqref*{sspace_curved_sfield_2b} is equivalent to the manifestly super-covariant version:
\be
	Z = Z | + \theta (\sderiv Z |) - \oneon{2} \theta\bar\theta ( \bar\sderiv \sderiv Z | ) .
\ee
By plugging in the component expansions~\eqref*{sspace_curved_sderiv_comp_exp, sspace_curved_sfield_2b} into the general formula~\eqref*{sspace_curved_gen_sfield_comp_transf_final}, the component field supersymmetry transformations are derived and read
\begin{align}
                              \delta_\epsilon z = i \epsilon \kappa, \quad
					\delta_\epsilon \bar{z} = 0, \quad
					\delta_\epsilon \kappa = 0, \quad
					\delta_\epsilon \bar\kappa = N^{-1} \epsilon (\dot{\bar{z}} + \bar{\psi}_0 \bar\kappa), 
\eqlabel{sspace_curved_transf_2b}\\
	                                     \delta_{\bar\epsilon} z = 0, \quad
					\delta_{\bar\epsilon} \bar{z} = i \bar\epsilon \bar\kappa, \quad
					\delta_{\bar\epsilon} \kappa = N^{-1} \bar\epsilon (\dot{z} - \psi_0 \kappa), \quad
					\delta_{\bar\epsilon} \bar\kappa = 0. \eqlabel{sspace_curved_transf_2b_bar}
\end{align}
A standard kinetic term for a single $2b$ multiplet $Z$ can be written and expanded into components as
\bea
	S_{\rm 2b,kin} &= - \oneon{4} \int d\tau\, d^2\theta\, \mathcal{E}\, \sderiv Z \bar\sderiv \bar{Z} = \oneon{4} \int d\tau\, \mathcal{L}_{\rm 2b,kin}, \\
	\mathcal{L}_{\rm 2b,kin} &= N^{-1}\dot{z}\dot{\bar{z}} - \frac{i}{2}(\kappa\dot{\bar\kappa} - \dot\kappa\bar\kappa) 
			- N^{-1}(\psi_0\kappa\dot{\bar{z}} - \bar\psi_0\bar\kappa\dot{z}) + N^{-1}\psi_0\bar\psi_0\kappa\bar\kappa.
\eea
The generalization to a non-linear sigma model for a set, $\{ Z^a \}$, of $2b$ multiplets is given by
\bea\eqlabel{sspace_curved_2b_action}
	S_{\rm 2b} &= - \oneon{4} \int d\tau\, d^2\theta\, \mathcal{E}\, G_{a\bar{b}}(\underline{Z}, \underline{\bar{Z}}) 
		\sderiv Z^a \bar\sderiv \bar{Z}^{\bar{b}} = \oneon{4} \int d\tau\, \mathcal{L}_{\rm 2b}, \\
	\mathcal{L}_{\rm 2b} &= N^{-1} G_{a\bar{b}}(\underline{z}, \underline{\bar{z}}) \dot{z}^a \dot{\bar{z}}^{\bar{b}}
		- \frac{i}{2} G_{a\bar{b}}(\underline{z}, \underline{\bar{z}}) (\kappa^a \dot{\bar{\kappa}}^{\bar{b}} - \dot{\kappa}^a \bar{\kappa}^{\bar{b}})
		- N^{-1} G_{a\bar{b}}(\underline{z}, \underline{\bar{z}}) (\psi_0 \kappa^a \dot{\bar{z}}^{\bar{b}} - \bar\psi_0 \bar{\kappa}^{\bar{b}} \dot{z}^a) \\ &
		+ N^{-1} G_{a\bar{b}}(\underline{z}, \underline{\bar{z}}) \psi_0 \bar\psi_0 \kappa^a \bar{\kappa}^{\bar{b}}
		- \frac{i}{2} G_{a\bar{b}, c}(\underline{z}, \underline{\bar{z}}) 
			(\kappa^a \bar{\kappa}^{\bar{b}} (\dot{z}^c - 2\psi_0 \kappa^c) - 2\kappa^c \bar{\kappa}^{\bar{b}} \dot{z}^a ) \\ &
		+ \frac{i}{2} G_{a\bar{b}, \bar{c}}(\underline{z}, \underline{\bar{z}}) 
			(\kappa^a \bar{\kappa}^{\bar{b}} (\dot{\bar{z}}^{\bar{c}} + 2\bar{\psi}_0 \bar{\kappa}^{\bar{c}}) 
			- 2\kappa^a \bar{\kappa}^{\bar{c}} \dot{\bar{z}}^{\bar{b}} ) 
		- N G_{a\bar{b}, c\bar{d}}(\underline{z}, \underline{\bar{z}}) \kappa^a \bar{\kappa}^{\bar{b}} \kappa^c \bar{\kappa}^{\bar{d}} .
\eea
Here, $G_{\ldots,a}$ means differentiation with respect to the bosonic fields $z^a$. In our application to M-theory, we need a variant of this action where the sigma model metric $G_{a\bar{b}}$ is also allowed to depend on a set of $2a$ multiplets $\phi^i$ in addition to $Z^a$ and $\bar{Z}^{\bar{b}}$. This leads to a coupling between $2a$ and $2b$ multiplets. The action for this case reads
{\allowdisplaybreaks
\begin{align}
	S_{\rm 2b} &= - \oneon{4} \int d\tau\, d^2\theta\, \mathcal{E}\, G_{a\bar{b}}(\underline{\phi}, \underline{Z}, \underline{\bar{Z}}) 
		\sderiv Z^a \bar\sderiv \bar{Z}^{\bar{b}} = \oneon{4} \int d\tau\, \mathcal{L}_{\rm 2b}, \nonumber \\ \allowdisplaybreaks
	\mathcal{L}_{\rm 2b} &= N^{-1} G_{a\bar{b}}(\underline{\varphi}, \underline{z}, \underline{\bar{z}}) \dot{z}^a \dot{\bar{z}}^{\bar{b}}
		- \frac{i}{2} G_{a\bar{b}}(\underline{\varphi}, \underline{z}, \underline{\bar{z}}) 
			(\kappa^a \dot{\bar{\kappa}}^{\bar{b}} - \dot{\kappa}^a \bar{\kappa}^{\bar{b}})
		- N^{-1} G_{a\bar{b}}(\underline{\varphi}, \underline{z}, \underline{\bar{z}}) 
			(\psi_0 \kappa^a \dot{\bar{z}}^{\bar{b}} \nonumber \\ \allowdisplaybreaks & - \bar\psi_0 \bar{\kappa}^{\bar{b}} \dot{z}^a) 
		+ N^{-1} G_{a\bar{b}}(\underline{\varphi}, \underline{z}, \underline{\bar{z}}) \psi_0 \bar\psi_0 \kappa^a \bar{\kappa}^{\bar{b}}
		- \frac{i}{2} G_{a\bar{b}, c}(\underline{\varphi}, \underline{z}, \underline{\bar{z}}) 
			(\kappa^a \bar{\kappa}^{\bar{b}} (\dot{z}^c - 2\psi_0 \kappa^c) - 2\kappa^c \bar{\kappa}^{\bar{b}} \dot{z}^a ) \nonumber \\ \allowdisplaybreaks &
		+ \frac{i}{2} G_{a\bar{b}, \bar{c}}(\underline{\varphi}, \underline{z}, \underline{\bar{z}}) 
			(\kappa^a \bar{\kappa}^{\bar{b}} (\dot{\bar{z}}^{\bar{c}} + 2\bar{\psi}_0 \bar{\kappa}^{\bar{c}}) 
			- 2\kappa^a \bar{\kappa}^{\bar{c}} \dot{\bar{z}}^{\bar{b}} ) 
		- N G_{a\bar{b}, c\bar{d}}(\underline{\varphi}, \underline{z}, \underline{\bar{z}}) \kappa^a \bar{\kappa}^{\bar{b}} \kappa^c \bar{\kappa}^{\bar{d}} \nonumber \\ \allowdisplaybreaks &
		- \oneon{2} N G_{a\bar{b}, i}(\underline{\varphi}, \underline{z}, \underline{\bar{z}}) \kappa^a \bar{\kappa}^{\bar{b}} f^i
		- N G_{a\bar{b}, ij}(\underline{\varphi}, \underline{z}, \underline{\bar{z}}) \kappa^a \bar{\kappa}^{\bar{b}} \psi^i \bar{\psi}^j
		- i N G_{a\bar{b}, ic}(\underline{\varphi}, \underline{z}, \underline{\bar{z}}) \kappa^a \bar{\kappa}^{\bar{b}} \bar{\psi}^i \kappa^c \nonumber \\ \allowdisplaybreaks &
		- i N G_{a\bar{b}, i\bar{c}}(\underline{\varphi}, \underline{z}, \underline{\bar{z}}) \kappa^a \bar{\kappa}^{\bar{b}} \psi^i \bar{\kappa}^{\bar{c}}
		- G_{a\bar{b}, i}(\underline{\varphi}, \underline{z}, \underline{\bar{z}}) 
			\psi^i \bar{\kappa}^{\bar{b}} (\dot{z}^a - \oneon{2} \psi_0 \kappa^a) \nonumber \\ \allowdisplaybreaks &
		+ G_{a\bar{b}, i}(\underline{\varphi}, \underline{z}, \underline{\bar{z}}) 
			\bar{\psi}^i \kappa^a (\dot{\bar{z}}^{\bar{b}} + \oneon{2} \bar{\psi}_0 \bar{\kappa}^{\bar{b}}) .  \eqlabel{sspace_curved_2b_action_coupled}
\end{align}}
This result can be readily specialized to $G_{a\bar{b}}(\underline{\phi}, \underline{Z}, \underline{\bar{Z}}) = f(\underline{\phi}) G_{a\bar{b}}(\underline{Z}, \underline{\bar{Z}})$, for a real function $f=f(\underline{\phi})$, which is the case relevant to M-theory compactifications.
\end{sloppypar}

\vskip 0.4cm

Finally, we need to discuss fermionic $2b$ multiplets, that is, super-multiplets $R$ with a fermionic lowest component and  satisfying $\bar{\cal D}R=0$. Their component expansion is given by
\be\eqlabel{sspace_curved_sfield_2bf}
	 R = \rho + \theta h + \frac{i}{2} N^{-1} \theta\bar\theta (\dot{\rho} - \psi_0 h),
\ee
where the notation for the component fields is completely analogous to the globally supersymmetric case~\eqref*{2bfexp}. The component supersymmetry transformations follow from plugging in the component expansions~\eqref*{sspace_curved_sderiv_comp_exp, sspace_curved_sfield_2bf} into the general formula~\eqref*{sspace_curved_gen_sfield_comp_transf_final} and are given by
\begin{align}
	\delta_\epsilon \rho = i \epsilon h, \quad
						\delta_\epsilon \bar\rho = 0, \quad
						\delta_\epsilon h = 0, \quad
						\delta_\epsilon \bar{h} = - N^{-1} \epsilon (\dot{\bar{\rho}} - \bar{\psi}_0 \bar{h}), \eqlabel{sspace_curved_transf_2bf} \\
		\delta_{\bar\epsilon} \rho = 0, \quad
						\delta_{\bar\epsilon} \bar\rho = - i \bar\epsilon \bar{h}, \quad
						\delta_{\bar\epsilon} h = N^{-1} \bar\epsilon (\dot{\rho} - \psi_0 h), \quad
						\delta_{\bar\epsilon} \bar{h} = 0. \eqlabel{sspace_curved_transf_2bf_bar}
\end{align}
A simple kinetic term for a single fermionic $2b$ superfield $R$ takes the form
\bea
	S_{\rm 2b-f,kin} &= - \oneon{4} \int d\tau\, d^2\theta\, \mathcal{E}\, R \bar{R} = \oneon{4} \int d\tau\, \mathcal{L}_{\rm 2b-f,kin}, \\
	\mathcal{L}_{\rm 2b-f,kin} &= \frac{i}{2} (\rho\dot{\bar{\rho}} - \dot{\rho}\bar{\rho}) - N h \bar{h} \; .
\eea
Note that the only bosonic field, $h$, in this multiplet is auxiliary and, hence, we are left with only fermionic physical degrees of freedom. This observation will be crucial for writing down a superspace version of the effective one-dimensional theories obtained from M-theory. As for the other types of multiplets, we need to generalise to a sigma model for a set, $\{R^x\}$, of fermionic $2b$ multiplets. The sigma model metric $G_{x\bar{y}}=G_{x\bar{y}}(\underline{\phi})$ should be allowed to depend on $2a$ multiplets $\phi^i$. Such an action takes the form
\bea\eqlabel{sspace_curved_2bf_action}
	S_{\rm 2b-f} 	&= - \oneon{4} \int d\tau\, d^2\theta\, \mathcal{E}\, G_{x\bar{y}}(\underline{\phi}) R^x \bar{R}^{\bar{y}} 
				  = \oneon{4} \int d\tau\, \mathcal{L}_{\rm 2b-f}, \\
	\mathcal{L}_{\rm 2b-f} &= \frac{i}{2} G_{x\bar{y}}(\underline{\varphi}) ( \rho^x \dot{\bar{\rho}}^{\bar{y}} - \dot{\rho}^x \bar{\rho}^{\bar{y}} ) 
		- N G_{x\bar{y}}(\underline{\varphi}) h^x \bar{h}^{\bar{y}}
		- i N G_{x\bar{y},i}(\underline{\varphi}) ( \psi^i \rho^x \bar{h}^{\bar{y}} + \bar{\psi}^i \bar{\rho}^{\bar{y}} h^x ) \\ &
		- \oneon{2} N G_{x\bar{y},i}(\underline{\varphi}) \rho^x \bar{\rho}^{\bar{y}} f^i
		- N G_{x\bar{y},ij}(\underline{\varphi}) \rho^x \bar{\rho}^{\bar{y}} \psi^i \bar{\psi}^j
		+ \oneon{2} G_{x\bar{y},i}(\underline{\varphi}) \rho^x \bar{\rho}^{\bar{y}} ( \psi_0 \psi^i - \bar{\psi}_0 \bar{\psi}^i ) \; .
\eea
%


\bibliographystyle{JHEP}
\bibliography{mcy5v2}

\providecommand{\href}[2]{#2}\begingroup\raggedright\begin{thebibliography}{10}

\bibitem{Candelas:1985en}
P.~Candelas, G.~T. Horowitz, A.~Strominger, and E.~Witten, {\it Vacuum
  configurations for superstrings},  {\em Nucl. Phys.} {\bf B258} (1985)
  46--74.

\bibitem{Brunner:1996pk}
I.~Brunner and R.~Schimmrigk, {\it {F-theory on Calabi-Yau fourfolds}},  {\em
  Phys. Lett.} {\bf B387} (1996) 750--758,
  [\href{http://xxx.lanl.gov/abs/hep-th/9606148}{{\tt hep-th/9606148}}].

\bibitem{Hull:1994ys}
C.~M. Hull and P.~K. Townsend, {\it Unity of superstring dualities},  {\em
  Nucl. Phys.} {\bf B438} (1995) 109--137,
  [\href{http://xxx.lanl.gov/abs/hep-th/9410167}{{\tt hep-th/9410167}}].

\bibitem{Witten:1995ex}
E.~Witten, {\it String theory dynamics in various dimensions},  {\em Nucl.
  Phys.} {\bf B443} (1995) 85--126,
  [\href{http://xxx.lanl.gov/abs/hep-th/9503124}{{\tt hep-th/9503124}}].

\bibitem{Becker:1996gj}
K.~Becker and M.~Becker, {\it {M}-theory on eight-manifolds},  {\em Nucl.
  Phys.} {\bf B477} (1996) 155--167,
  [\href{http://xxx.lanl.gov/abs/hep-th/9605053}{{\tt hep-th/9605053}}].

\bibitem{Haack:2002tu}
M.~Haack, {\it Calabi-{Y}au fourfold compactifications in string theory},  {\em
  Fortsch. Phys.} {\bf 50} (2002) 3--99.

\bibitem{Kumar:1996zx}
A.~Kumar and C.~Vafa, {\it {U-manifolds}},  {\em Phys. Lett.} {\bf B396} (1997)
  85--90, [\href{http://xxx.lanl.gov/abs/hep-th/9611007}{{\tt
  hep-th/9611007}}].

\bibitem{Curio:1998bv}
G.~Curio and D.~Lust, {\it New {N} = 1 supersymmetric 3-dimensional superstring
  vacua from {U}-manifolds},  {\em Phys. Lett.} {\bf B428} (1998) 95--104,
  [\href{http://xxx.lanl.gov/abs/hep-th/9802193}{{\tt hep-th/9802193}}].

\bibitem{Lu:2004ng}
H.~Lu, C.~N. Pope, K.~S. Stelle, and P.~K. Townsend, {\it {String and M-theory
  deformations of manifolds with special holonomy}},  {\em JHEP} {\bf 07}
  (2005) 075, [\href{http://xxx.lanl.gov/abs/hep-th/0410176}{{\tt
  hep-th/0410176}}].

\bibitem{Coles:1990hr}
R.~A. Coles and G.~Papadopoulos, {\it The geometry of the one-dimensional
  supersymmetric nonlinear sigma models},  {\em Class. Quant. Grav.} {\bf 7}
  (1990) 427--438.

\bibitem{Julia:1980gr}
B.~Julia, {\it Group disintegrations}, . Invited paper presented at Nuffield
  Gravity Workshop, Cambridge, Eng., Jun 22 - Jul 12, 1980.

\bibitem{Damour:2002cu}
T.~Damour, M.~Henneaux, and H.~Nicolai, {\it {E(10) and a 'small tension
  expansion' of M theory}},  {\em Phys. Rev. Lett.} {\bf 89} (2002) 221601,
  [\href{http://xxx.lanl.gov/abs/hep-th/0207267}{{\tt hep-th/0207267}}].

\bibitem{Ooguri:2005vr}
H.~Ooguri, C.~Vafa, and E.~P. Verlinde, {\it {Hartle-Hawking wave-function for
  flux compactifications}},  {\em Lett. Math. Phys.} {\bf 74} (2005) 311--342,
  [\href{http://xxx.lanl.gov/abs/hep-th/0502211}{{\tt hep-th/0502211}}].

\bibitem{Hartle:1983ai}
J.~B. Hartle and S.~W. Hawking, {\it {Wave Function of the Universe}},  {\em
  Phys. Rev.} {\bf D28} (1983) 2960--2975.

\bibitem{Sethi:1996es}
S.~Sethi, C.~Vafa, and E.~Witten, {\it {Constraints on low-dimensional string
  compactifications}},  {\em Nucl. Phys.} {\bf B480} (1996) 213--224,
  [\href{http://xxx.lanl.gov/abs/hep-th/9606122}{{\tt hep-th/9606122}}].

\bibitem{Candelas:1990pi}
P.~Candelas and X.~de~la Ossa, {\it {Moduli space of Calabi-Yau manifolds}},
  {\em Nucl. Phys.} {\bf B355} (1991) 455--481.

\bibitem{Gibbons:1997iy}
G.~W. Gibbons, G.~Papadopoulos, and K.~S. Stelle, {\it {HKT} and {OKT}
  geometries on soliton black hole moduli spaces},  {\em Nucl. Phys.} {\bf
  B508} (1997) 623--658, [\href{http://xxx.lanl.gov/abs/hep-th/9706207}{{\tt
  hep-th/9706207}}].

\bibitem{Bilal:2003es}
A.~Bilal and S.~Metzger, {\it Anomaly cancellation in {M}-theory: {A} critical
  review},  {\em Nucl. Phys.} {\bf B675} (2003) 416--446,
  [\href{http://xxx.lanl.gov/abs/hep-th/0307152}{{\tt hep-th/0307152}}].

\bibitem{Miemiec:2005ry}
A.~Miemiec and I.~Schnakenburg, {\it Basics of {M}-theory},  {\em Fortschr.
  Phys.} {\bf 54} (2006) 5--72,
  [\href{http://xxx.lanl.gov/abs/hep-th/0509137}{{\tt hep-th/0509137}}].

\bibitem{Cremmer:1978km}
E.~Cremmer, B.~Julia, and J.~Scherk, {\it Supergravity theory in 11
  dimensions},  {\em Phys. Lett.} {\bf B76} (1978) 409--412.

\bibitem{Duff:1995wd}
M.~J. Duff, J.~T. Liu, and R.~Minasian, {\it Eleven-dimensional origin of
  string / string duality: A one-loop test},  {\em Nucl. Phys.} {\bf B452}
  (1995) 261--282, [\href{http://xxx.lanl.gov/abs/hep-th/9506126}{{\tt
  hep-th/9506126}}].

\bibitem{Green:1997di}
M.~B. Green and P.~Vanhove, {\it D-instantons, strings and {M}-theory},  {\em
  Phys. Lett.} {\bf B408} (1997) 122--134,
  [\href{http://xxx.lanl.gov/abs/hep-th/9704145}{{\tt hep-th/9704145}}].

\bibitem{Green:1997as}
M.~B. Green, M.~Gutperle, and P.~Vanhove, {\it One loop in eleven dimensions},
  {\em Phys. Lett.} {\bf B409} (1997) 177--184,
  [\href{http://xxx.lanl.gov/abs/hep-th/9706175}{{\tt hep-th/9706175}}].

\bibitem{Russo:1997mk}
J.~G. Russo and A.~A. Tseytlin, {\it One-loop four-graviton amplitude in
  eleven-dimensional supergravity},  {\em Nucl. Phys.} {\bf B508} (1997)
  245--259, [\href{http://xxx.lanl.gov/abs/hep-th/9707134}{{\tt
  hep-th/9707134}}].

\bibitem{deAlwis:1996ez}
S.~P. de~Alwis, {\it A note on brane tension and {M}-theory},  {\em Phys.
  Lett.} {\bf B388} (1996) 291--295,
  [\href{http://xxx.lanl.gov/abs/hep-th/9607011}{{\tt hep-th/9607011}}].

\bibitem{deAlwis:1996hr}
S.~P. de~Alwis, {\it Anomaly cancellation in {M}-theory},  {\em Phys. Lett.}
  {\bf B392} (1997) 332--334,
  [\href{http://xxx.lanl.gov/abs/hep-th/9609211}{{\tt hep-th/9609211}}].

\bibitem{Schwarz:1982jn}
J.~H. Schwarz, {\it {Superstring Theory}},  {\em Phys. Rept.} {\bf 89} (1982)
  223--322.

\bibitem{Witten:1996md}
E.~Witten, {\it On flux quantization in {M}-theory and the effective action},
  {\em J. Geom. Phys.} {\bf 22} (1997) 1--13,
  [\href{http://xxx.lanl.gov/abs/hep-th/9609122}{{\tt hep-th/9609122}}].

\bibitem{Candelas:1987kf}
P.~Candelas, A.~M. Dale, C.~A. Lutken, and R.~Schimmrigk, {\it {Complete
  Intersection Calabi-Yau Manifolds}},  {\em Nucl. Phys.} {\bf B298} (1988)
  493.

\bibitem{Cadavid:1995bk}
A.~C. Cadavid, A.~Ceresole, R.~D'Auria, and S.~Ferrara, {\it Eleven-dimensional
  supergravity compactified on calabi-yau threefolds},  {\em Phys. Lett.} {\bf
  B357} (1995) 76--80, [\href{http://xxx.lanl.gov/abs/hep-th/9506144}{{\tt
  hep-th/9506144}}].

\bibitem{DeWitt:1967yk}
B.~S. DeWitt, {\it {Quantum Theory of Gravity. 1. The Canonical Theory}},  {\em
  Phys. Rev.} {\bf 160} (1967) 1113--1148.

\bibitem{Pope:www}
C.~N. Pope, {\it lecture notes}, .
  \href{http://faculty.physics.tamu.edu/pope/ihplec.ps}{http://faculty.physics%
.tamu.edu/pope/ihplec.ps}.

\bibitem{Roest:2004pk}
D.~Roest, {\it M-theory and gauged supergravities},  {\em Fortsch. Phys.} {\bf
  53} (2005) 119--230, [\href{http://xxx.lanl.gov/abs/hep-th/0408175}{{\tt
  hep-th/0408175}}].

\bibitem{Gillard:2004xq}
J.~Gillard, U.~Gran, and G.~Papadopoulos, {\it {The spinorial geometry of
  supersymmetric backgrounds}},  {\em Class. Quant. Grav.} {\bf 22} (2005)
  1033--1076, [\href{http://xxx.lanl.gov/abs/hep-th/0410155}{{\tt
  hep-th/0410155}}].

\bibitem{Hubsch:1992nu}
T.~Hubsch, {\it Calabi-{Y}au manifolds: {A} {B}estiary for physicists}, .
  Singapore, Singapore: World Scientific (1992) 362 p.

\bibitem{Candelas:1987is}
P.~Candelas, {\it Lectures on complex manifolds}, . In *Trieste 1987,
  Proceedings, Superstrings '87* 1-88.

\bibitem{Green:1987cr}
P.~S. Green, T.~Hubsch, and C.~A. Lutken, {\it {All Hodge Numbers of All
  Complete Intersection Calabi-Yau Manifolds}},  {\em Class. Quant. Grav.} {\bf
  6} (1989) 105--124.

\bibitem{Naito:1986cr}
S.~Naito, K.~Osada, and T.~Fukui, {\it Fierz identities and invariance of
  11-dimensional supergravity action},  {\em Phys. Rev.} {\bf D34} (1986)
  536--552.

\bibitem{vanHolten:1995qt}
J.~W. van Holten, {\it $d = 1$ supergravity and spinning particles},
  \href{http://xxx.lanl.gov/abs/hep-th/9510021}{{\tt hep-th/9510021}}.

\bibitem{Machin:2002zn}
W.~Machin, {\it {Supersymmetric sigma models, gauge theories and vortices}},
  \href{http://xxx.lanl.gov/abs/hep-th/0311126}{{\tt hep-th/0311126}}.

\bibitem{Howe:1977xz}
P.~S. Howe, {\it {Superspace and the Spinning String}},  {\em Phys. Lett.} {\bf
  B70} (1977) 453.

\bibitem{Martinec:1983um}
E.~J. Martinec, {\it {Superspace geometry of fermionic strings}},  {\em Phys.
  Rev.} {\bf D28} (1983) 2604.

\bibitem{West:1990tg}
P.~C. West, {\it Introduction to supersymmetry and supergravity}, . Singapore,
  Singapore: World Scientific (1990) 425 p.

\bibitem{Wess:1992cp}
J.~Wess and J.~Bagger, {\it Supersymmetry and supergravity}, . Princeton, USA:
  Univ. Pr. (1992) 259 p.

\bibitem{Hull:1999ng}
C.~M. Hull, {\it The geometry of supersymmetric quantum mechanics},
  \href{http://xxx.lanl.gov/abs/hep-th/9910028}{{\tt hep-th/9910028}}.

\bibitem{Wess:1977fn}
J.~Wess and B.~Zumino, {\it {Superspace Formulation of Supergravity}},  {\em
  Phys. Lett.} {\bf B66} (1977) 361--364.

\bibitem{Grimm:1977kp}
R.~Grimm, J.~Wess, and B.~Zumino, {\it {Consistency Checks on the Superspace
  Formulation of Supergravity}},  {\em Phys. Lett.} {\bf B73} (1978) 415.

\bibitem{Wess:1978bu}
J.~Wess and B.~Zumino, {\it {Superfield Lagrangian for Supergravity}},  {\em
  Phys. Lett.} {\bf B74} (1978) 51.

\bibitem{Stelle:1978ye}
K.~S. Stelle and P.~C. West, {\it {Minimal Auxiliary Fields for Supergravity}},
   {\em Phys. Lett.} {\bf B74} (1978) 330.

\bibitem{Ferrara:1978em}
S.~Ferrara and P.~van Nieuwenhuizen, {\it {The Auxiliary Fields of
  Supergravity}},  {\em Phys. Lett.} {\bf B74} (1978) 333.

\bibitem{Sohnius:1981tp}
M.~F. Sohnius and P.~C. West, {\it {An Alternative Minimal Off-Shell Version of
  N=1 Supergravity}},  {\em Phys. Lett.} {\bf B105} (1981) 353.

\bibitem{Howe:1978ia}
P.~S. Howe, {\it Super {W}eyl transformations in two-dimensions},  {\em J.
  Phys.} {\bf A12} (1979) 393--402.

\bibitem{Ertl:2001sj}
M.~F. Ertl, {\it Supergravity in two spacetime dimensions},
  \href{http://xxx.lanl.gov/abs/hep-th/0102140}{{\tt hep-th/0102140}}.

\bibitem{Breitenlohner:1979np}
P.~Breitenlohner and M.~F. Sohnius, {\it {Superfields, auxiliary fields, and
  tensor calculus for N=2 extended supergravity}},  {\em Nucl. Phys.} {\bf
  B165} (1980) 483.

\bibitem{Castellani:1980cu}
L.~Castellani, P.~van Nieuwenhuizen, and S.~J. Gates, {\it {The constraints for
  N=2 superspace from extended supergravity in ordinary space}},  {\em Phys.
  Rev.} {\bf D22} (1980) 2364.

\bibitem{Stelle:1980uw}
K.~S. Stelle and P.~C. West, {\it {Algebraic derivation of N=2 supergravity
  constraints}},  {\em Phys. Lett.} {\bf B90} (1980) 393.

\bibitem{Gates:1980ky}
S.~J. Gates, {\it {Supercovariant derivatives, super weyl groups, and N=2
  supergravity}},  {\em Nucl. Phys.} {\bf B176} (1980) 397.

\end{thebibliography}\endgroup

\end{document}